\documentclass[usenatbib,twocolumn]{mn2e}
\usepackage{graphicx}
\usepackage{amsfonts}
\usepackage{epsfig,rotating,natbib,pstricks}
\usepackage{color}
\usepackage{amssymb,latexsym}
\usepackage{amsmath,wasysym}
\usepackage{verbatim}

\def\beq{\begin{equation}}
\def\eeq{\end{equation}}
\def\baq{\begin{eqnarray}}
\def\eaq{\end{eqnarray}}

\def\p3m{P$^3$M}
\def\ap3m{AP$^3$M}

\def\h1{H\/I}
\def\omegah1{\Omega_{\h1}}
\def\ph1{P_{_{\h1}}}
\def\ph1k{P_{_{\h1}}(k)}
\def\dh1k{\Delta^2_{_{\h1}}(k)}

\def\lbol{L_{\mathrm{bol}}}
\def\msun{M_{\odot}}

\def\hmpc{h^{-1}\mathrm{Mpc}}

\def\mdm{m_{_{\mathrm{DM}}}}
\def\mgas{m_{_{\mathrm{gas}}}}
\def\mbh{M_{_{\mathrm{BH}}}}

\def\mstar{M_{*}}
\def\mhalo{M_{\mathrm{halo}}}
\def\mblack2{{MassiveBlack-II }}
\def\mblack{{MassiveBlack }}
\def\lsim{\apprle}

\newcommand{\be}{\begin{equation}}
\newcommand{\e}{\end{equation}}

\def\apj{ApJ}
\def\mnras{MNRAS}
\def\apjl{ApJL}
\def\physrep{Physics Reports}
\def\apjs{ApJS}
\def\pasp{PASP}

\title[The MassiveBlack-II Simulation]{The MassiveBlack-II Simulation: The Evolution of Halos and Galaxies to $z \sim 0$}

\author[]{\parbox{18cm}{Nishikanta Khandai$^1$,
    Tiziana Di Matteo$^2$,
    Rupert Croft$^2$,
    Stephen Wilkins$^3$,\\
    Yu Feng$^2$,
    Evan Tucker$^2$,
    Colin DeGraf$^4$,
    Mao-Sheng Liu$^2$}
  \vspace{0.3cm}\\
  $^1$ {Brookhaven National Laboratory, Department of Physics, Bldg 510, Upton, NY 11973, USA}\\
  $^2$ {McWilliams Center for Cosmology, Carnegie Mellon University, 5000 Forbes Avenue, Pittsburgh, PA 15213, USA}\\
  $^3$ {Astronomy Centre, Department of Physics and Astronomy, University of Sussex, Brighton, BN1 9QH, U.K.}\\
  $^4$ {Racah Institute of Physics, The Hebrew University, Jerusalem 91904, Israel}}
\pubyear{2014}
\label{firstpage}

\def\LaTeX{L\kern-.36em\raise.3ex\hbox{a}\kern-.15em
    T\kern-.1667em\lower.7ex\hbox{E}\kern-.125emX}

\pagerange{\pageref{firstpage}--\pageref{lastpage}}

\begin{document}

\maketitle

\begin{abstract}
We investigate the properties and clustering
of halos, galaxies and blackholes to $z=0$ in the high resolution
hydrodynamical simulation MassiveBlack-II (MBII).
MBII evolves a $\Lambda$CDM cosmology in  a cubical comoving volume
of $V_{box} = (100 {\rm Mpc/h})^3$ and is able to resolve halos of
mass $\mhalo = 10^9\msun$/h.
It is the highest resolution simulation of this size which includes a
self-consistent model for star formation, black hole accretion and
associated feedback.
We provide a simulation browser web application which enables
interactive search and tagging of halos, subhalos and their properties
and publicly release our galaxy catalogs to the scientific community.
Our analysis of the halo mass function in MBII reveals that
baryons have strong effects, with changes in the halo abundance
of  20-35\% below the knee of the mass function
($\mhalo \lsim 10^{13.2}\msun$/h at $z=0$)
when compared to fits based on dark matter only simulations.
We provide a fitting function for the halo mass function valid for the full
range of halo masses in MBII out to redshift $z=11$ and discuss how the 
onset of non-universal behavior in the mass function limits the accuracy of our fit. 
We examine  the halo occupation distribution of satellite galaxies
and present results valid over 5 orders of magnitude in host halo mass.
We study the clustering of galaxies, and in particular the
evolution and scale dependence of stochasticity and bias. Comparison with
observational data for these quantities for samples with different stellar
mass thresholds yields reasonable agreement.
Using population synthesis, we find that the shape of the cosmic spectral
energy distribution predicted by MBII is consistent with observations,
but lower in amplitude.
The Galaxy Stellar Mass Function (GSMF) function is broadly consistent
with observations at $z \ge 2$.
At $z < 2$, observations probe deeper into the faint end and the
population of passive low mass (for $M_{*}<10^{9} \msun$) galaxies
in the simulation makes the GSMF too steep.
At the high mass end ($M_{*}>10^{11} \msun$) galaxies
hosting bright AGN make significant contributions to the GSMF.
The quasar bolometric luminosity function is also largely consistent
with observations. We note however that more efficient AGN feedback
(beyond simple thermal coupling used here) is likely necessary for the
largest, rarest objects/clusters at low redshifts.

\end{abstract}


\begin{keywords}
methods: numerical -- cosmology: theory -- cosmology: large-scale structure of Universe -- galaxies: formation 
-- galaxies: evolution -- quasars: general
\end{keywords}

\section{Introduction}

The Cold Dark Matter model with a cosmological constant ($\Lambda$CDM)
 is well established enough (see e.g.,  \citealt{2013arXiv1303.5076P},\citealt{2013ApJS..208...19H},
\citealt{2014arXiv1401.1389L},\citealt{2013arXiv1311.0840C}) that
individual  large-scale simulation efforts can be carried out that focus on just this one cosmology.
We have also reached the point at which supercomputers enable
numerical modeling of cosmological volumes with enough
resolution to study the properties of individual galaxies. In this paper
we report on a {\small P-GADGET} hydrodynamic simulation of $100 \hmpc$ cubic volume,
the MassiveBlackII simulation. It has $\sim 10^{6}\msun$ mass resolution,
cooling, star formation, black holes and feedback, and represents the
evolution of a $\Lambda$CDM universe to redshift $z=0$.

Numerical simulations  (see reviews by \citealt{2008SSRv..134..229D},
\citealt{2012AN....333..515S}, \citealt{1998ARA&A..36..599B})
are the tool of choice to address many questions in
cosmology, as galaxy formation is a complex non-linear problem. Two criteria
which must be satisfied for accurate results are:

(1) A large enough simulation volume that Fourier density modes
on the largest scales are evolving independently. The volume simulated must
be a representative region of the Universe, otherwise inferences drawn
from it regarding such questions as clustering and the mass function of
objects will be incorrect (\citealt{2005MNRAS.358.1076B},
\citealt{2006MNRAS.370..993B}, \citealt{2013arXiv1312.5256O}). When $\Lambda$CDM
models are evolved to redshift $z=0$, then a volume of at least $\sim 100 \hmpc$ on a side
becomes necessary to predict the overall star formation rate,
for example (\citealt{2003MNRAS.339..312S}).

(2) High enough mass and spatial
resolution that the properties of the objects of interest
have converged. This requires many resolution elements
(particles or grid cells). If we focus on particle based simulations
relevant to the current work, at the very least for identification of objects
there must enough particles to overcome shot noise.
If we require detailed properties of the objects such
as galaxy spectra, or angular momenta then this can require many more (
e.g., \citealt{2007MNRAS.374.1479G}).

These criteria of large simulation volume and high resolution are more
straightforward to address in the context of more restricted
physical modeling.
As a result,  dark matter and gravity-only simulations have long been
used to make cosmological predictions that cover both large and small scales
in the same volume (e.g., \citealt{2005Natur.435..629S},
\citealt{2009MNRAS.398.1150B}, \citealt{2011ApJ...740..102K}).
Semi-analytic modeling
has been used to process dark matter simulations, resulting in many
studies of the galaxies and their properties in the $\Lambda$CDM model
(see e.g., \citealt{2006RPPh...69.3101B}, \citealt{2012MNRAS.419.3200H} and
references therein).

Baryonic physics including hydrodynamics obviously plays an important
role in the formation of luminous objects and structure. This has lead to
the inclusion of the relevant equations in simulation codes in many forms,
Smoothed Particle Hydrodynamics (SPH, \citealt{1992ARAA..30..543M}),
Eulerian grid solvers (e.g., \citealt{1992ApJS...78..341C},
\citealt{1997ASPC..123..363B})
and hybrid Lagrangian/Eulerian schemes (e.g., \citealt{2010MNRAS.401..791S}).
Although previous
work has not simultaneously reached the combination of large volume and
high resolution that we present here, research has progressed using many
methods, including making use
of zoom simulations of smaller volumes inside a representative one
(\citealt{1993ApJ...412..455K}, \citealt{2012MNRAS.423.1726S}),
 simulations that stop at high redshifts before large-scale
modes become non-linear (e.g., \citealt{2012ApJ...745L..29D}),
or by tackling problems
which require lower mass resolution (e.g., \citealt{2012ApJ...758...74B}).

The advent of large-scale computing facilities with
100,000 compute cores or more (such as the Cray XT5, ``Kraken'' on
which the current simulation was run) and the development of
highly efficient  distributed
memory simulation codes  (such as {\small GADGET2}, \citealt{2005MNRAS.364.1105S})
means that simulations
which satisfy both criteria (1) and (2) are now possible. We have run one
such simulation as part of the NSF Petascale Applications in Cosmology
program, using the code {\small P-GADGET}
(see e.g., \citealt{2012ApJ...745L..29D}).

 Our aim was to simulate and analyze a large, representative volume of
the $\Lambda$CDM model with the most important physical processes previously
included in zoom runs or simulations with smaller boxes. These are
hydrodynamics (using SPH), cooling, a subgrid multiphase
model for star formation (\citealt{2003MNRAS.339..289S}) and subgrid black hole
modeling (\citealt{2005MNRAS.361..776S},\citealt{2005Natur.433..604D}), both
with feedback. Our use of
the physical modeling and algorithms used in
previous work such as \cite{2008ApJ...676...33D}, \cite{2009MNRAS.400...43C},
\cite{2010MNRAS.402.1927D} and \cite{2012ApJ...745L..29D}
enables continuity and therefore comparison with this previous work. Our aim is to see what this ``fiducial''
model ($\Lambda$CDM + {\small GADGET} SPH + the particular
subgrid  algorithms employed) predicts about
the properties of galaxies, their halos and their clustering at redshifts
extending down to the present day. We have not adjusted methods,
algorithms and parameters used in previous work (e.g.,
\citealt{2012MNRAS.424.1892D}) to try to tune
to observational results. Our goal is to see how this model performs,
now that there is a large volume at high resolution. We naturally expect both
regions of agreement and disagreement with observations and we aim
that our work  will offer guidance to
future work to address the problems.

In this paper, we make the first use of the MassiveBlackII simulation
evolved to $z=0$, and  use
it to explore some topics in structure formation. Here we choose to focus on topics
relevant to galaxy and AGN formation and large-scale structure, including
mass functions, galaxy and halo properties and clustering. Our emphasis
is on lower redshifts; the simulation at redshifts $z>5$
has been explored by \cite{2013MNRAS.430.2885W}, \cite{2013MNRAS.429.2098W},
and \cite{2013MNRAS.435.2885W}. Topics that we leave to future work include
the intergalactic medium, absorption lines, galaxy clusters and X-ray emission.

Our plan for the paper is as follows. In Section~\ref{sec_methods} we briefly describe
the numerical methods and algorithms used to run the simulation, select
galaxies and carry out stellar population synthesis. In Section~\ref{sec_vis}, we
describe visualization of the simulation.
In Sections \ref{sec_massfn} and \ref{sec_hod}
respectively we present the mass function  and halo occupation
distribution and in Section \ref{sec_gal_clustering}
we examine the clustering of dark matter and galaxies.
The properties of galaxies and supermassive black holes are examined in Sections~\ref{sec_galaxy}  and \ref{sec_bh}
and we derive some conclusions from our work in Section~\ref{sec_conc}.

\section{Methods: }
\label{sec_methods}

\subsection{Numerical Code}
We have used {\small P-GADGET}, an upgraded version of {\small GADGET3}
(see \cite{2005MNRAS.364.1105S}
for an earlier version) which we are developing for use at
upcoming Petascale supercomputer facilities. This code was also
used to run the MassiveBlack (MB) simulation (\cite{2012ApJ...745L..29D}.
Both MB and MBII are  cosmological simulation of a $\Lambda$CDM cosmology.
The major differences between MB and MBII are resolution and volume.
However there are minor differences in cosmology between the two.

The initial conditions for MBII
were generated with the CMBFAST transfer function at
$z=159$ and the simulation was evolved to $z=0$.  The
cosmological parameters used were:
amplitude of mass fluctuations, $\sigma_8=0.816$,
spectral index, $n_s = 0.968$,
cosmological constant parameter $\Omega_{\Lambda}= 0.725$,
mass density parameter $\Omega_m = 0.275$ , baryon density
parameter $\Omega_b = 0.046$ and $h=0.701$ (Hubble's constant in units of
{$100 \mathrm{km\: s}^{-1} \mathrm{Mpc}^{-1}$)}. These are
consistent with the WMAP7 cosmology \citep{2011ApJS..192...18K}.

\begin{table}
      \begin{center}
        \begin{tabular}{c|c|c|c|c}
          \hline
          $L_{\mathrm{box}}$ & $N_{\mathrm{part}}$ & $\mdm$ & $\mgas$ & $\epsilon$ \\
          $\left(h^{-1}\mathrm{Mpc}\right)$& & $\left(h^{-1}\msun\right)$ & $\left(h^{-1}\msun\right)$ & $\left(h^{-1}\mathrm{kpc}\right)$ \\
          \hline
          100 & $2\times 1792^3$ & 1.1$\times 10^7$ & 2.2$\times 10^6$ & 1.85  \\
         \hline
        \end{tabular}
      \end{center}
\caption{Basic simulation parameters for the simulation. The columns
  list the size of the simulation box, $L_{\mathrm{box}}$, the
  number of  particles (dark matter + gas) used in the simulation,
  $N_{\mathrm{part}}$,
  the mass of a single dark matter particle, $\mdm$,
  the initial mass of a gas particle, $\mgas$,
  and the gravitational softening length,
  $\epsilon$.  All length scales are in comoving units.}
\label{table_simparam}
\end{table}

\subsection{Halo And Subhalo Identification}
We identify halos with the friends-of-friends
(FOF) procedure  \citep{1985ApJ...292..371D} applied to dark matter particles
with a linking length of $b=0.2$ times the mean inter-particle separation.
Gas, stars and BHs are then associated to their nearest dark matter particles.
The subhalo finder SUBFIND \citep{2001MNRAS.328..726S} was then used, working
with particles in the FOF halo and computing a local density for each particle.
Starting from isolated density peaks within the FOF halo, additional particles
with decreasing density are attached to it.  Whenever a saddle point, which connects two disjoint overdensities
is reached, the smaller of the two is treated as a substructure candidate followed by merging of the two regions.
Eventually all particles within a substructure are checked for self-boundedness and only those
particles are retained which have a total negative energy.

\subsection{Subgrid Model for Star Formation BH growth and associated feedback}
The subgrid models for star formation, BH growth and associated feedback processes are identical
to that employed in the MB simulation. We briefly describe them here and refer the reader to the MB simulation (e.g., \cite{2012ApJ...745L..29D})
for a more detailed description.

We adopt the multiphase model for star forming gas developed by \citet{2003MNRAS.339..289S}.
This has two principal ingredients: (1) a star formation prescription and
(2) an effective equation of state (EOS). (1) is
motivated by observations and given by the Schmidt-Kennicutt Law
\citep{1989ApJ...344..685K}, where the star formation rate is proportional to the density of cold clouds ($\rho_{\rm SFR} \propto \rho_{gas}^{N}$ and $ N =1.5$.
Star particles are created from gas particles probabilistically according to their star formation rates.
(2) encapsulates the self-regulated nature of star formation due to supernovae feedback in a
simple model for a multiphase ISM.
In this model, a thermal instability is assumed to operate above a critical density threshold $\rho_{\rm th}$, producing a
two phase medium consisting of cold clouds embedded in a tenuous gas at pressure equilibrium.
Stars form from  the cold clouds, and short-lived stars supply an energy of $10^{51}\,{\rm ergs}$
to the surrounding gas as supernovae. This energy heats the
diffuse phase of the ISM and evaporates cold clouds, thereby establishing a self-regulation cycle for star formation.
$\rho_{\rm th}$ is determined self-consistently in the model by requiring that the EOS is
continuous at the onset of star formation. The cloud evaporation process and the
cooling function of the gas then determine the temperatures and the  mass fractions of the two 'hot and cold' phases of the ISM,
such that the EOS of the model can be directly
computed as a function of density. In addition, a parametrization of stellar winds is also used \citep[see][for further details]{2003MNRAS.339..289S}.

In MBII BHs are modeled as collisionless sink particles within newly
collapsing halos, which are identified by the FOF halofinder called on the
fly at regular time intervals.  A seed BH of mass $M_{\mathrm{seed}} =
5\times10^{5}h^{-1}\msun$ is inserted into a halo with mass $\mhalo \geq
5\times10^{10}h^{-1}\msun$ if it does not already contain a BH.
Once seeded, BHs grow by accreting gas in its surrounding region or by merging with other BHs.
Gas is accreted with an accretion rate $\dot{M}_{\mathrm{BH}} =
\frac{4\pi G^2 \mbh^2 \rho}{\left(c_s^2 + v_{\mathrm{BH}}^2\right)^{3/2}}$,
 where $v_{\mathrm{BH}}$ is the velocity of the black hole relative to the surrounding gas,
$\rho$ and $c_s$ are the density and sound speed of the hot
and cold phase of the ISM gas (which when taken into account
appropriately as in \cite{2007ApJ...665..107P} eliminates
the need for a correction factor $\alpha$ previously introduced).
We allow the accretion rate to be mildly super-Eddington
but limit it to a maximum allowed value equal to $2\times$Eddington rate ($\dot{M}_{\mathrm{Edd}}$)
to prevent artificially high
values, consistent with \cite{2006MNRAS.370..289B,2006ApJ...650..669V}.
The BH radiates with a bolometric luminosity which is
proportional to the accretion rate, $\lbol = \eta\dot{M}_{\mathrm{BH}}c^2$
\citep{1973A&A....24..337S}, where $\eta$ is the radiative efficiency and its
standard value of 0.1 is kept throughout, and $c$ is the speed of light.
In the simulation 5\% of the radiated energy couples thermally
to the surrounding gas  and this energy is deposited isotropically on gas particles
that are within the BH
kernel (64 nearest neighbors) and acts as a form of feedback \citep{2005Natur.433..604D}.
The value of 5\% is the only free parameter in the
model and was set using galaxy merger simulations \citep{2005Natur.433..604D}
to match the normalization in the observed $\mbh - \sigma$ relation.
BHs also grow by merging once one BH comes within the kernel of another
with a relative velocity below the local gas sound speed.

This model for the growth of BHs has been developed by
\cite{2005Natur.433..604D,2005MNRAS.361..776S}.  It has been implemented and
studied extensively in cosmological simulations
\citep{2007MNRAS.380..877S,2007ApJ...665..187L,2008MNRAS.387.1163C,2008ApJ...676...33D,
  2009MNRAS.400...43C,2009MNRAS.398...53B,2009MNRAS.400..100S,2010MNRAS.402.1927D,
  2011MNRAS.413.1383D,2011MNRAS.416.1591D,2012MNRAS.419.2657C}, successfully
reproducing basic properties of BH growth, the observed $\mbh -
\sigma$ relation and the BH mass function \citep{2008ApJ...676...33D},
the quasar luminosity function \citep{2010MNRAS.402.1927D} and the
clustering of quasars \citep{2011MNRAS.413.1383D}.

\subsection{Stellar Population Synthesis}
The spectral energy distribution (SED) of a galaxy is generated by summing the SEDs of each star particle in the galaxy.
The SED of the star particles is generated using the Pegase.2 stellar
population synthesis (SPS) code \citep{1997A&A...326..950F, 1999astro.ph.12179F}
by considering their ages, mass and  metallicities and assuming a
Salpeter IMF.
Nebula (continuum and line) emission is also added to each star particle SED. We also apply a correction for absorption in the intergalactic medium (IGM)
using the standard \citet{1996MNRAS.283.1388M}
prescription. We finally sum the SED of each galaxy and convolve
with given filters (see bottom panels of figure~\ref{fig_csed}) to finally obtain  the broadband photometry, hence the CSED.

\subsection{Public Release of MBII Galaxy Catalogs}
We release the MBII galaxy catalogs to the scientific community.
Some of the properties included in these catalogs are position, velocity, mass, 
mass by particle type (such as gas, dark matter, stars and BHs), 
circular velocity and rest-frame luminosities. 
We encourage the community to use these galaxy catalogs which can be 
accessed from \emph{http://mbii.phys.cmu.edu/data/}. A more detailed
description and sample codes can also be found at the above URL.
 
\section{Visualization}
\label{sec_vis}
\begin{figure*}
  \includegraphics[width=\textwidth]{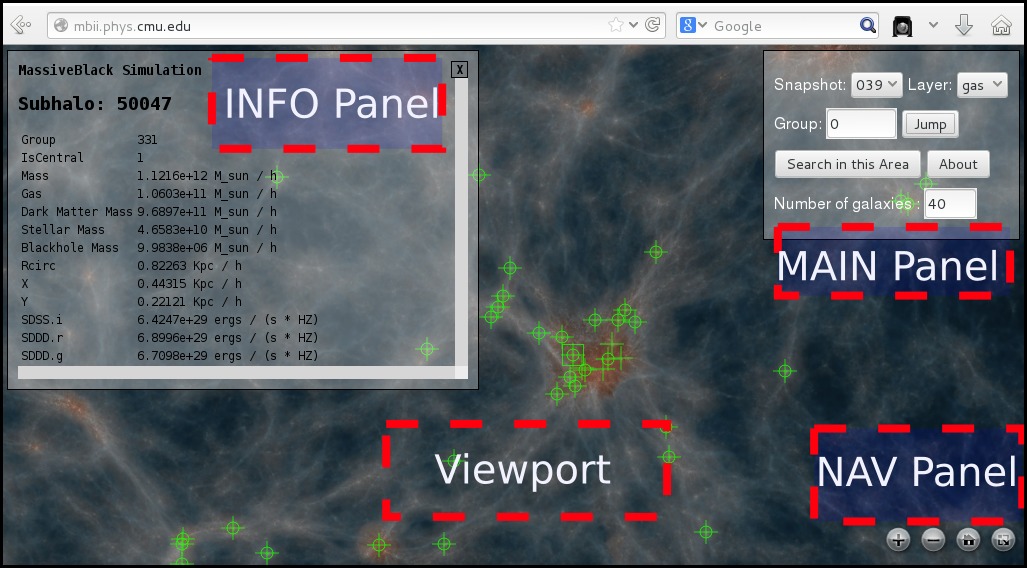}
  \caption{Interface for the interactive simulation browser.}
  \label{fig:interactive}
\end{figure*}

\begin{figure*}
\includegraphics[height=0.90\textheight]{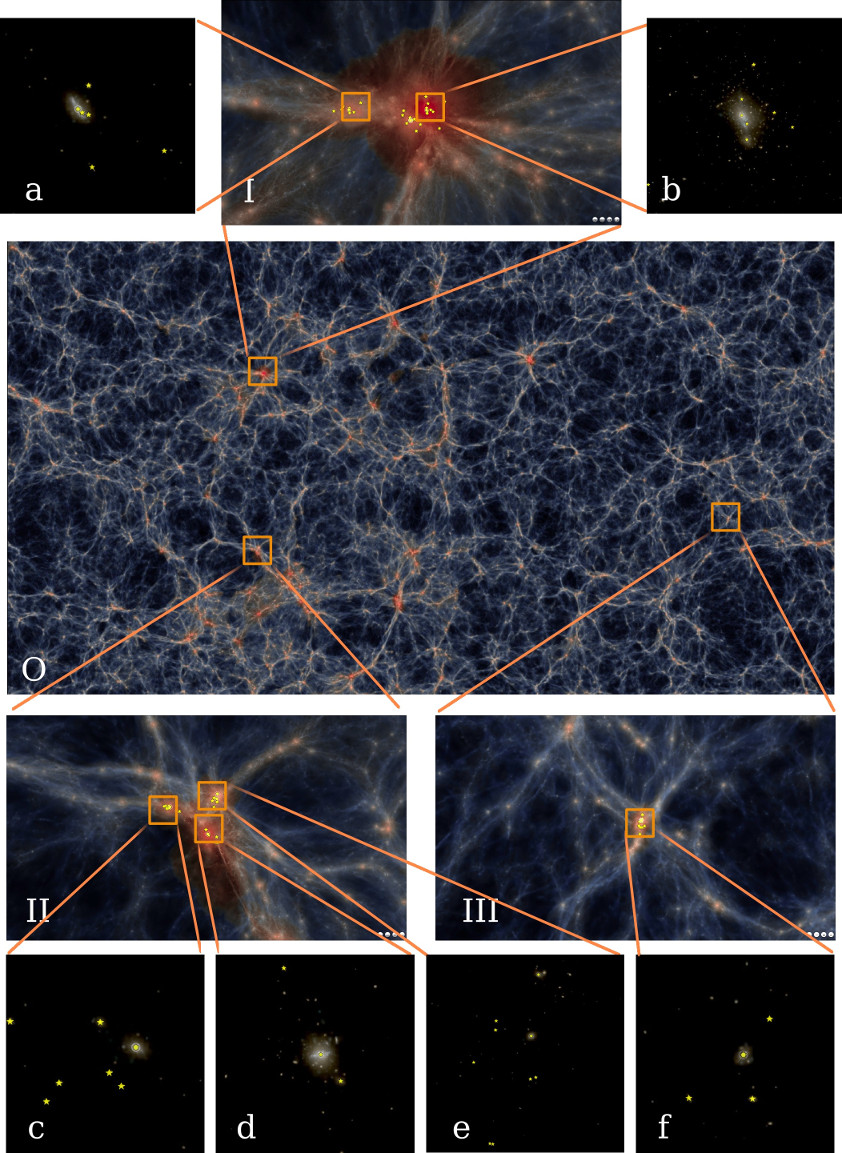}
\caption{Visualization of MBII simulation. The central panel, O shows the
full simulation box: the $z=1.0$ snapshot is mapped into a
$8\, \mathrm{h^{-1}Mpc}$ thick slice. Panels I, II, and III show the gaseous
environment of three FOF groups. Panel a to f show the stellar
component of the subhalos. The central  subhalos are marked with dots,
and 10 of the brightest subhalos are marked with stars. Please see
text for a description of the color scheme. The interactive simulation
browser is available at http://mbii.phys.cmu.edu .}
\label{fig:gigapan}
\end{figure*}

To enable easy visual exploration of the large dataset represented
by the \mblack2 simulation,
we have developed an interactive simulation browser web-application.
The browser allows real-time zooming, panning in the simulation, and
enables searching and locating of halos and subhalos in the
simulation.
The application is built upon existing web technology. Two main
libraries used are
Gigapan\footnote{http://www.gigapan.org}, and
the Microsoft Seadragon library
\footnote{http://gallery.expression.microsoft.com/SeadragonAjax}.

Figure \ref{fig:interactive} shows the interface for the interactive
browser. The browser can be accessed from the URL
\emph{http://mbii.phys.cmu.edu} .
It consists of a viewport and three floating control panels:
the MAIN panel, located at the top-right corner of the interface;
the INFOrmation panel, located at the left side of the interface;
and the NAVigation panel, located at the bottom right corner of the
interface.

The Gigapan image of the selected snapshot is
displayed in the viewport, where subhalos are also marked with green crosses.
In addition, central subhalos ($M_\mathrm{subhalo} > 0.1 M_\mathrm{group}$)
are labeled with an additional circle. Interactive zooming and panning
in the viewport is implemented via mouse clicking and dragging.

The MAIN panel provides the following functionalities:
\begin{enumerate}
\item selecting an epoch from the snapshot number;
\item switching between the gas and stellar image layer;
\item jumping among FOF groups;
\item querying subhalos in the current view.
\end{enumerate}

The INFO panel displays the properties of the currently selected
subhalo or group. In Figure \ref{fig:interactive}, for example,
the panels shows the property of the currently selected subhalo
(marked with a rectangle).

The NAV panel provides zoom-in and zoom-out controls, and a switch to
toggle the visibility of other control panels.

Figure \ref{fig:gigapan} shows a collage of images extracted from the
browser.
In this example
we have selected three halos in the simulation at redshift $z=1.0$: (I) at
a major confluence of filaments; (II) a moderately sized halo
with three main subhalos; (III) a relatively isolated halo. For each of
the halo we show the stellar component in their subhalos, embedded
in their surrounding gas environment.

The gigapan images used in the browser are high resolution 2-D images of the
full simulation
rendered with the visualization software Gaepsi
\citep{2011ApJS..197...18F}.
The gas images (panels O, I, II, III) are rendered with the divergent
Cool-Warm color-map introduced by \cite{Moreland...2009DCM}.
The density information is encoded in the brightness of the pixels:
brighter pixels have higher column density, and voids are represented
with black (zero-brightness).
The temperature of gas is encoded in the hue of the pixels,
blue represents low temperature ($T < 10^{3.5}\,\mathrm{K}$), and red
represents high temperature ($T > 10^{7.5}\,\mathrm{K}$).
The stellar images (panels a, b, c, d, e, and f) are composed from the
simulated i, r, and g band luminosity. This band definition follows the
convention used by the Sloan Digital Sky Survey
\citep[see the procedure described in][]{2004PASP..116..133L}.

\section{Mass Function of Halos}
\label{sec_massfn}

Given that dark matter halos represent the locations where gas
can cool and form stars and galaxies it is important
to predict their abundances - the halo mass function - accurately.
The halo or subhalo mass function, is one of the fundamental
quantities in structure formation. It is an important ingredient
in a diverse set of tools used for making theoretical predictions in cosmology.
At low redshifts the tail of the mass function which probes
the abundance of clusters is extremely sensitive to cosmological parameters.
It is also a key component in studying the clustering of galaxies as the
halo-halo term  \citep[see][]{2002PhR...372....1C} depends on the mass function.
At higher redshifts the mass function is used in modeling the sources of
reionization which reside in dark matter halos like PopIII stars,
early galaxies and quasars.
Any significant deviation in the mass function as predicted
by the $\Lambda$CDM model would therefore create some tension in our current
understanding of structure formation.

Traditionally dark matter simulations have been used to compute
the abundance of halos for a given cosmology. A key component of these
analyses is the halo definition. The FOF definition identifies regions
bounded by an isodensity contour whereas the Spherical Overdensity (SO)
definition identifies an artificial spherical region centered on
a density maximum such that the density within it is
at a given density threshold. A dark matter halo is never perfectly spherical making the SO definition artificial.
The FOF definition is prone to artificially
bridging two (or more) nearby halos connected by a filament.

In this section we will not look at how the halo definition affects
the mass function predicted by MBII since much work has been
done on this subject \citep{1994MNRAS.271..676L, 2001MNRAS.321..372J, 2002ApJS..143..241W, 2008ApJ...688..709T, 2013MNRAS.433.1230W}.
We will rather choose a halo definition and see how baryonic
effects affect the mass function and compare our results
with  fitting functions based on dark matter only simulations.

We generate two catalogues of halos based on the FOF and SUBFIND halofinders.
These catalogues contain the total number of particles by type
(e.g. gas, dark matter, stars and BHs),
and the total mass by type amongst other important halo properties.
For the analysis in this section we consider the smallest
halo to have a resolution limit of 40 particles by type.
E.g. An object is considered to be a halo if its mass satisfies
$\mhalo \geq 40 \times (m_{dm}+m_{gas})$, where $m_{dm}$ and $m_{gas}$
are given in table~\ref{table_simparam}. If we are interested only
in the dark matter component of the halo then the above condition is relaxed
such that $\mhalo \geq 40\times m_{dm}$. Note that these criteria affect
the statistics and counts of only the smallest halos.

In the case of dark matter simulations it has been shown
\citep{2006ApJ...646..881W} that halos with small particle counts
have a mass which is systematically overestimated.
Corrections have been proposed to alleviate this
\citep{2006ApJ...646..881W,2009ApJ...692..217L,2011ApJ...732..122B,2011ApJS..195....4M}. We choose to ignore for this effect for two reasons.
(1) It is not clear how such corrections
apply to each particle type in MBII. Providing
a similar correction to  halo masses
for hydrodynamical simulations is beyond the scope of this paper.
(2) In section~\ref{subsec_massfn} we will show that the baryonic effects
already show up in the halo mass function at lower masses at the 10-35\% level
when compared to dark matter simulations.
A correction to the mass of the halo at smaller masses will only
enhance the discrepancy in the mass function.

The largest mode that any cosmological simulation can sample
is governed by the physical size of the simulation volume.
Large scale modes $k< 2\pi/L_{box}$ are not sampled in the simulation
and lead to a suppression of structure formation
and hence the mass function. This is a well known effect
\citep{2005MNRAS.358.1076B,2005ApJ...634..728S,2006MNRAS.370..993B}
and  masses can be corrected by accounting for the missing power
\citep{2007MNRAS.374....2R,2013MNRAS.433.1230W}.
\citet{2005MNRAS.358.1076B} point out
that a boxsize $L_{box} = 100 \hmpc$
is sufficient to obtain reasonably reliable mass functions
to halo masses
$\mhalo \lsim 10^{14}\msun$/h for the $\Lambda$CDM model at $z=0$;
the requirement for large boxes becomes less stringent at higher redshift.
Our focus will in any case be for smaller masses,
which are less affected by effects of finite volume.
We therefore choose not to make any corrections to the mass function
due to missing large scale power.

\subsection{The baryon fraction of halos}
\label{subsec_barfrac}

\begin{figure}
  \begin{tabular}{c}
    \includegraphics[width=3.2truein]{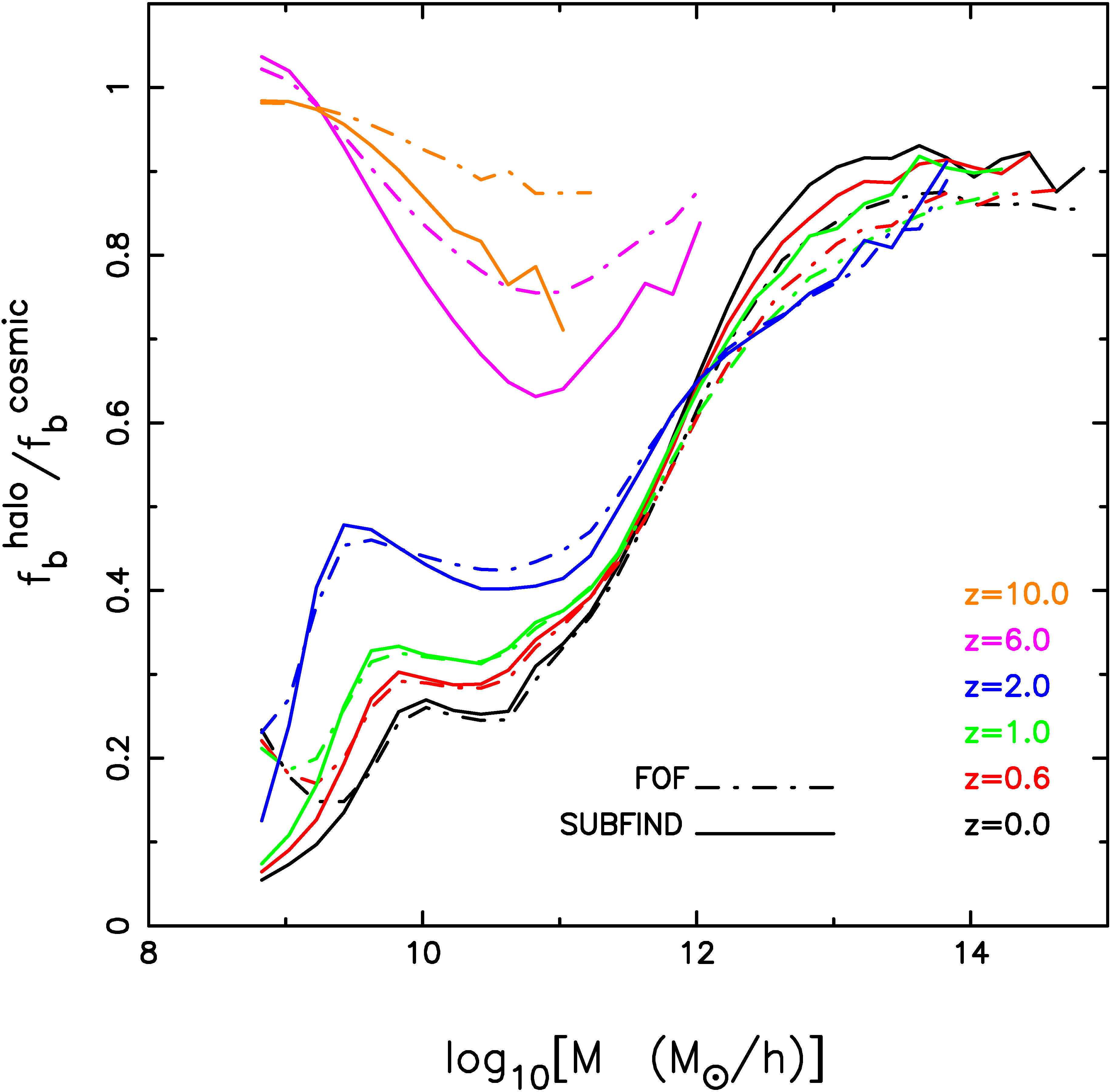} \\
  \end{tabular}
  \caption{The evolution of the baryon fraction of halos ($f_b^{halo}$) in units of the cosmic baryon fraction $f_b^{cosmic}$ as a function of halo mass.
    The solid line is for FOF halos and the dot-dashed is for halos identified with {\small SUBFIND}. The colors (black, red, green, blue, pink and orange) represent
  the baryon fraction for different redshifts ($z=0.0, 0.6, 1, 2, 6, 10$)}
  \label{fig_barfrac}
\end{figure}

We start by looking at the baryon fraction of halos in
figure~\ref{fig_barfrac} as a function of halo mass.
We plot the ratio of the baryon fraction of halos to the cosmic
baryon fraction $f_b^{halo}/f_b^{cosmic}$
where $f_b^{cosmic} = \Omega_b/\Omega_m$ and $f_b^{halo}$ is the ratio
of baryonic to total mass of the halo.
The solid line represents halos identified with the FOF algorithm
whereas the dot-dashed line
represents those identified by {\small SUBFIND}.

We find that the distinction between halos and
subhalos  has little effect on the baryon
fraction of halos below $z=6$.
At $z \geq 6$ the baryon fraction of subhalos identified with {\small SUBFIND}
 is
larger compared to objects identified with FOF although
the qualitative trend with mass is similar.
We find that the baryon fraction plateaus to around 80-90\%
around $\mhalo \geq 10^{13}\msun/h$ at low redshifts
and drops significantly below that mass scale.
At this point it is worthwhile to compare our results with
\citet{2007MNRAS.377...41C} who looked at the baryon fraction
of halos in an adiabatic resimulation of the Millennium simulation
and another smaller volume (higher resolution) simulation at $z=0$.
These authors also included
an additional simulation with a simple photoheating model, where
a gas temperature floor of $T_{floor}= 2\times 10^4$K  was assumed to mimic
the IGM temperature at mean density in the post-reionized Universe.
Star and blackhole formation and associated feedback processes
were not included.
These authors found that the baryon fraction plateaued to around
90\% for $\mhalo \geq 10^{10}\msun/h$ and dropped significantly
below that in their photoionisation model, whereas the baryon
fraction in the adiabatic simulation did not show any significant
behavior with halo mass.
Our results largely agree well the results of \citet{2007MNRAS.377...41C},
for larger masses however, we find that feedback can drastically
prevent the collapse of baryons in halos below
$\mhalo \lsim 10^{13}\msun/h$. \citet{2007MNRAS.377...41C} found that
photoionisation regulates the formation of smaller galaxies with
$\mhalo \lsim 10^{10}\msun/h$. Interestingly we find that at
$\mhalo \sim 10^{10}\msun/h$ the baryon fraction
plateaus to around 20-40\% of the cosmic mean.
It drops below that mass scale, which can be attributed to
photoheating. The suppression of the baryonic fraction
in $\mhalo = 10^{10}-10^{13}\msun/h$ can be attributed
to feedback from stars and blackholes.

Since we see that baryon effects play an important role in
the formation of halos we expect to see deviations in the halo mass function
which is the premise of the next section.

\subsection{The mass function}
\label{subsec_massfn}

We use the FOF and {\small SUBFIND} algorithms to compute the halo and subhalo
mass functions respectively. We look at the total halo/subhalo
 mass and the mass of the dark
matter component.
In section~\ref{sec_galaxy} we will look at the galaxy stellar
mass function (GSMF) and compare
them with observational constraints.
In this analysis we choose the mass bin to be $\Delta \log{M}=0.2$ which is well within the recommended bin width \citep{2007ApJ...671.1160L}
to avoid any systematic error that may arise in the estimate  of the
mass function due to large bins. We assume Poisson errors for the counts of halos in this mass bin.

It is convenient to rewrite the differential mass function,
$dn/d\log_{10}M$, in a rescaled form, $f(\sigma)$, which is independent
of redshift, power spectrum and cosmology \citep{1994MNRAS.271..676L}.
The computed differential mass function $dn/d\log_{10}M$
can be rescaled to $f(\sigma)$
\beq
\frac{dn}{d\log_{10}M} = \frac{M}{\rho}
\frac{d\ln \sigma^{-1}}{d\log_{10}M}f(\sigma)
\label{eq_fnu}
\eeq
where $M$ is the halo mass, $\rho$ is the mean matter density and the
variance in mass, smoothed with a real-space spherical top hat filter $W(k,M)$
at a scale $R(M)=(3M/4\pi\rho)^{1/3}$, is instead used as a
mass variable and is given by
\beq
\sigma^2(M,z) =  \frac{D_+(z)^2}{2\pi^2}\int_0^\infty k^3 P(k)W^2(k,M)d\log{k}
\eeq
The redshift dependence is encapsulated in the growth factor $D_+(z)$
which is normalized to $D_+(0) = 1$. $W(k,M)$.
When written in this form equation~\ref{eq_fnu}
is universal since the dependence of redshift,
power spectrum and cosmology are absorbed into the variable $\sigma(M,z)$.
Therefore $f(\sigma)$ at multiple redshifts should fall on a single curve.
The commonly and most used mass functions, namely the Press-Schecter \citep{1974ApJ...187..425P} and Sheth-Tormen \citep{1999MNRAS.308..119S}
mass functions, can then be written in a compact form:
\begin{eqnarray}
  f_{\mathrm{PS}} &=& \sqrt{\frac{2}{\pi}}\frac{\delta_c}{\sigma}\exp\left[-\frac{\delta_c^2}{2\sigma^2}\right]\\
  f_{\mathrm{ST}} &=& A\sqrt{\frac{2a}{\pi}}\left[1+\left( \frac{\sigma^2}{a\delta_c^2}\right)^p\right]
  \frac{\delta_c}{\sigma}\exp\left[-\frac{a\delta_c^2}{2\sigma^2}\right]
\end{eqnarray}
where, $\delta_c  = 1.686$ is the linearly extrapolated overdensity of a spherical top-hat density perturbation
at virialization in an Einstein-de Sitter Universe. For the Sheth-Tormen\citep{1999MNRAS.308..119S}
mass function $(A, a, p)=(0.3222, 0.707, 0.3)$ are additional parameters which better describe the shape of the mass function
when compared to simulations.

\begin{figure*}
  \begin{tabular}{cc}
    \includegraphics[width=3.2truein]{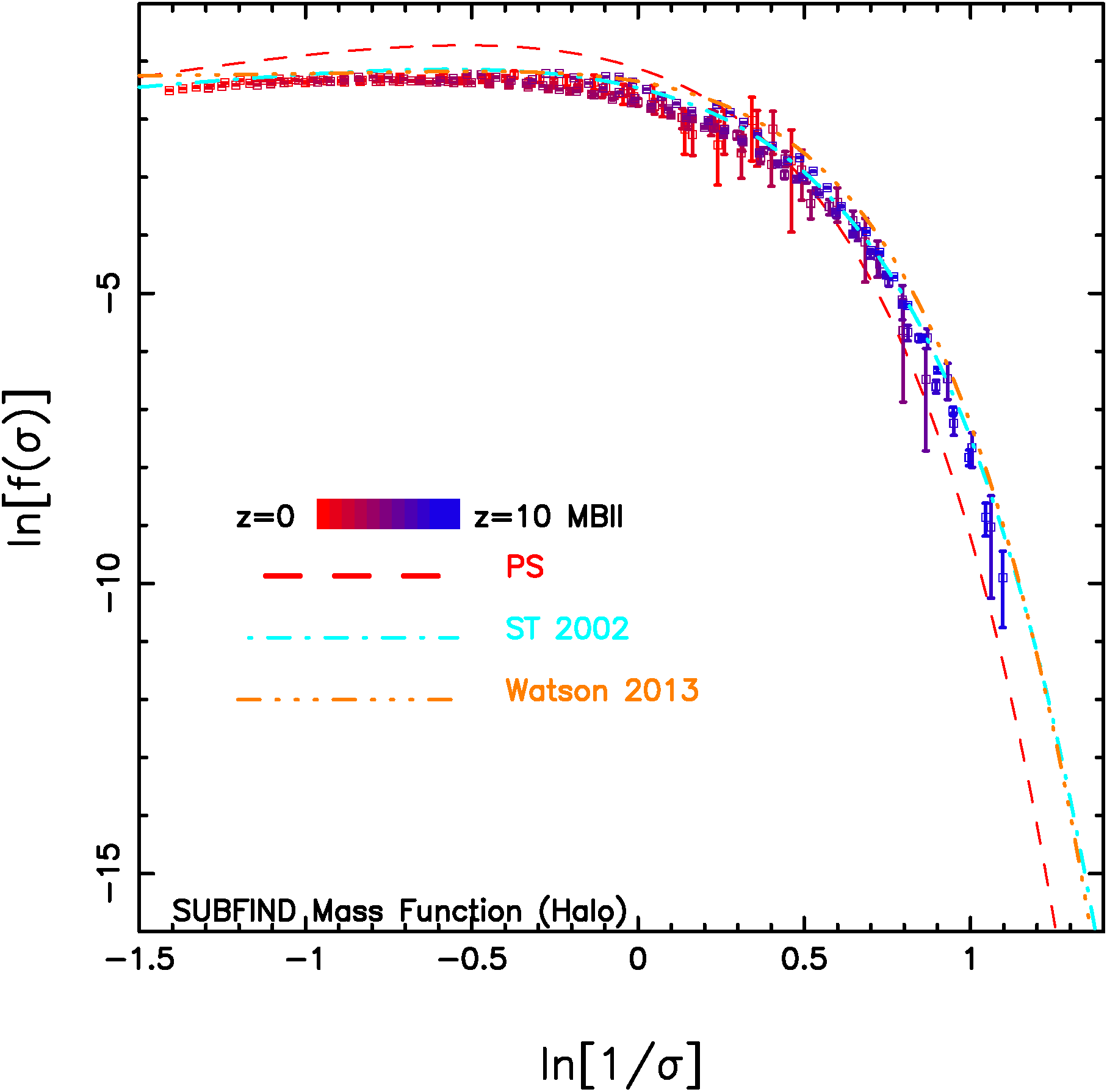}
    \includegraphics[width=3.2truein]{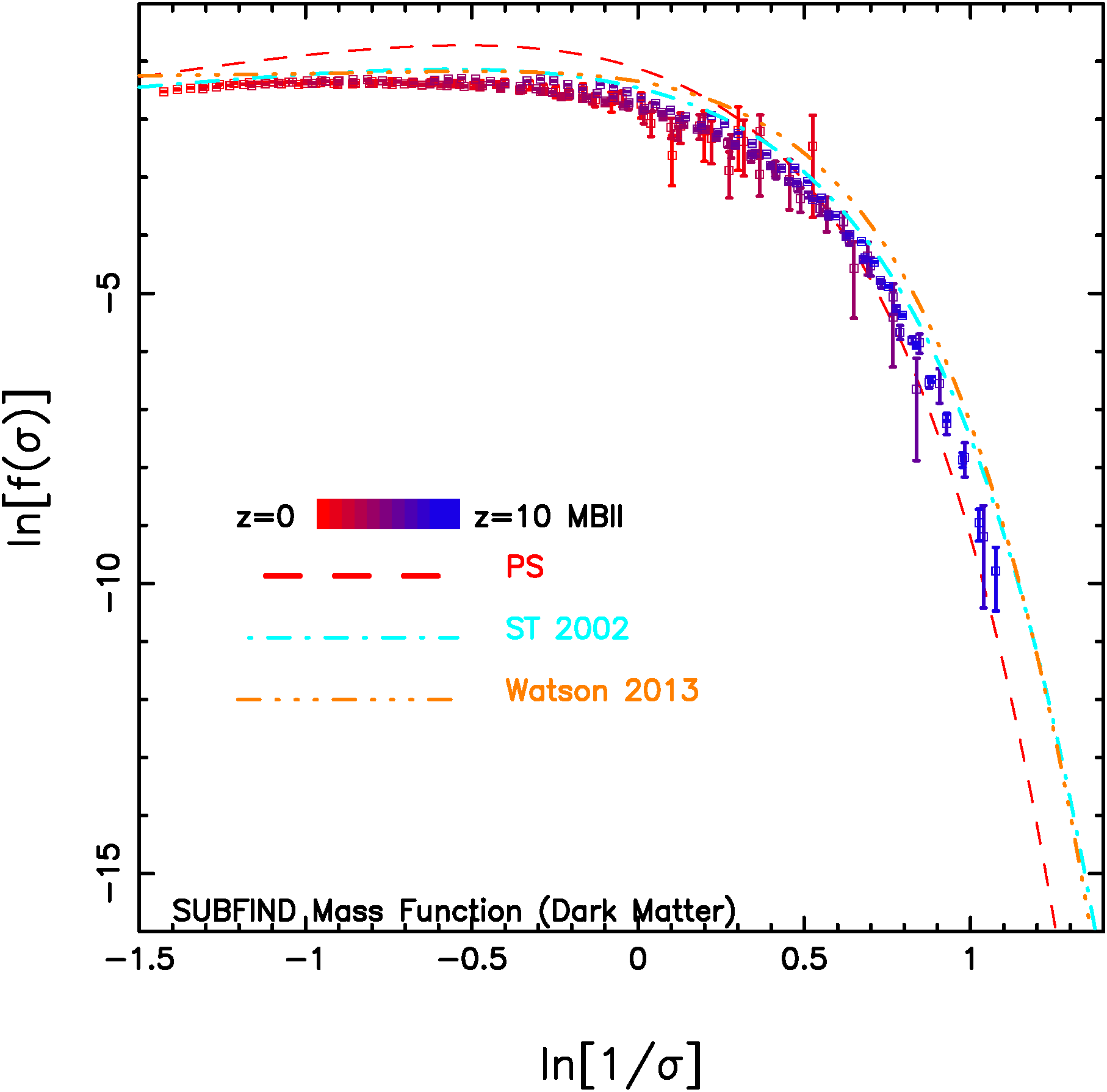} \\
    \includegraphics[width=3.2truein]{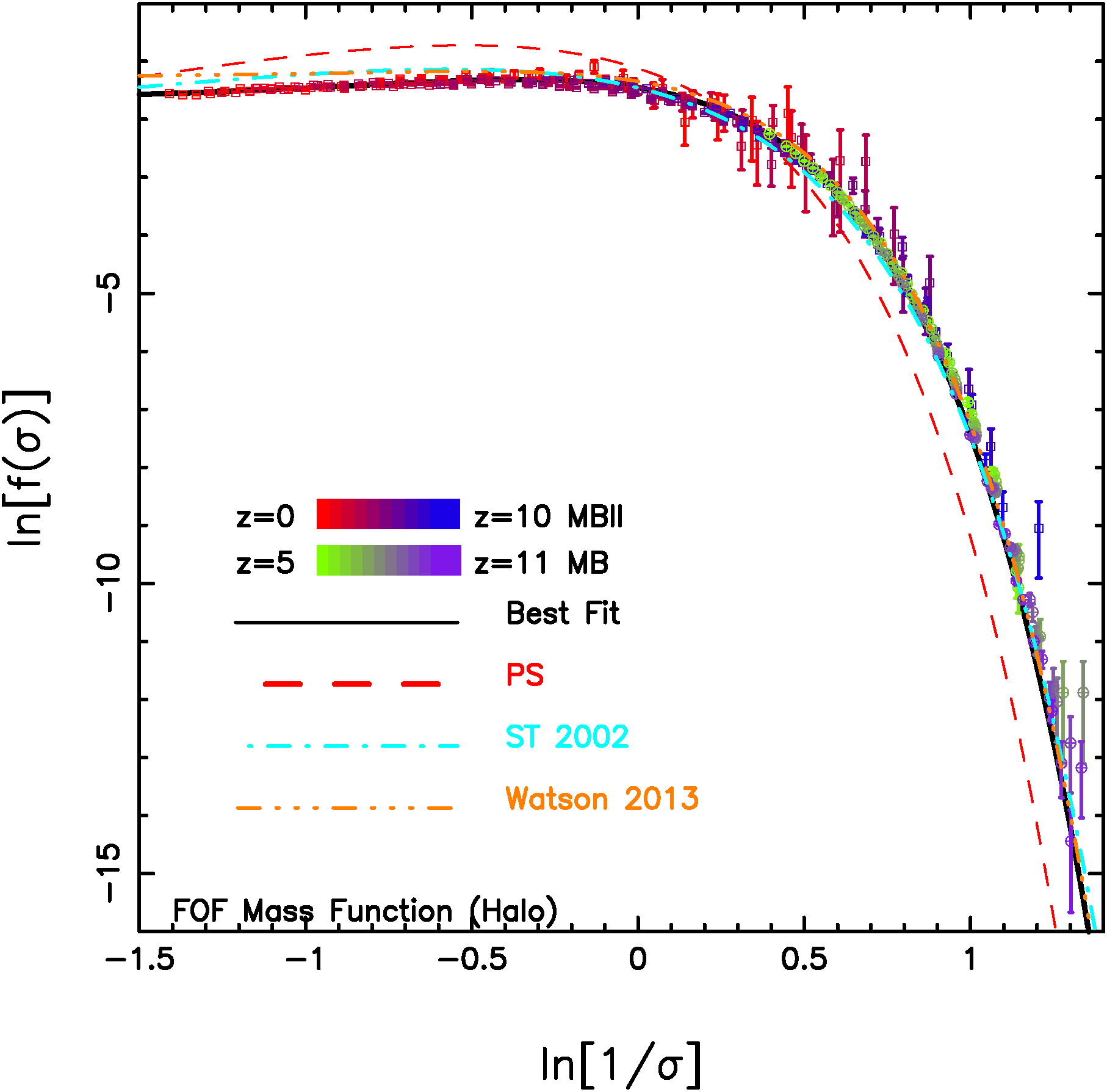}
    \includegraphics[width=3.2truein]{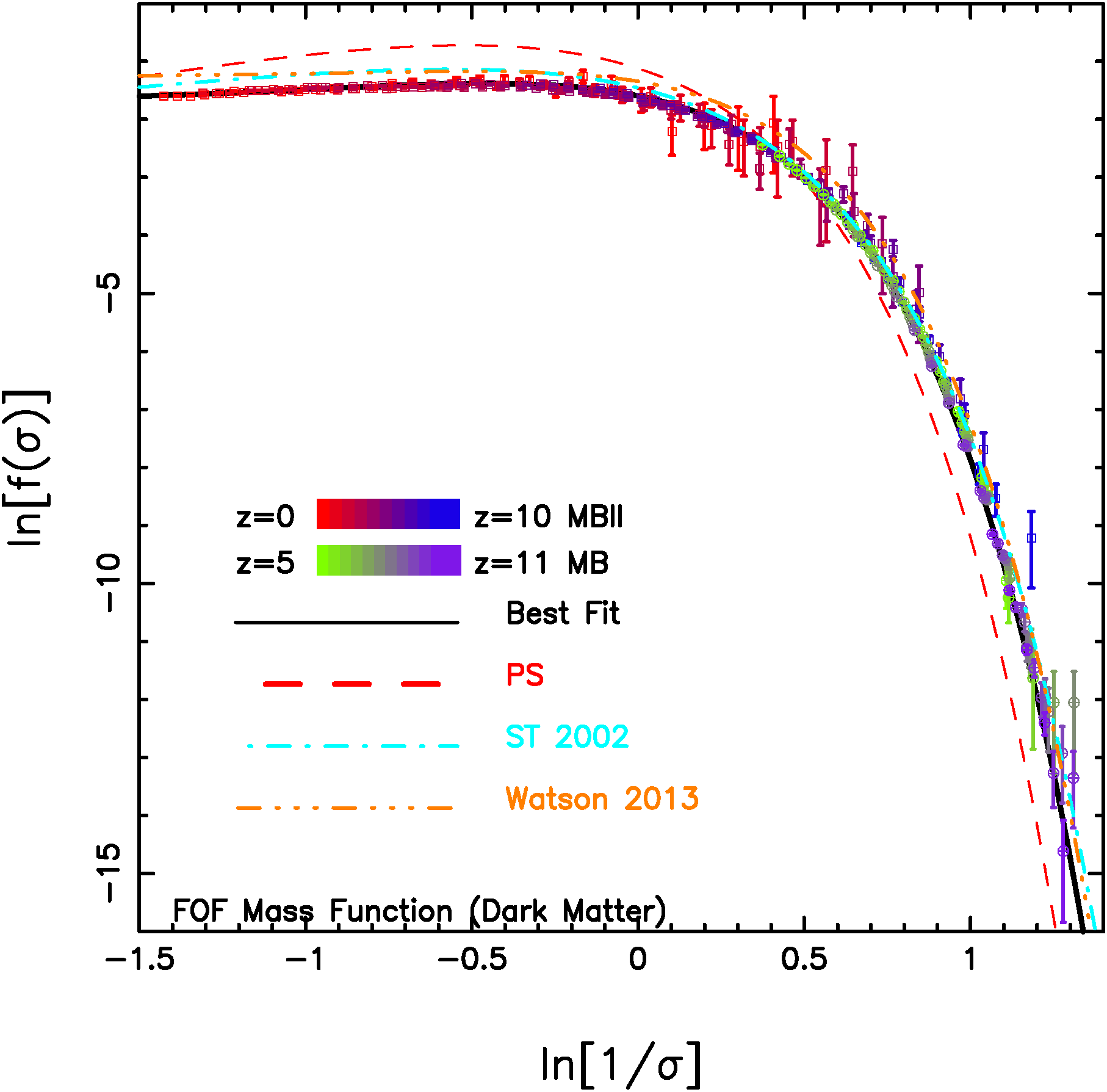} \\
  \end{tabular}
  \caption{The {\small SUBFIND} (top) and FOF (bottom) mass functions are plotted from $z=0-11$. The columns denote the mass function based on the total halo mass
    and the dark matter component of the halo mass respectively. The open squares denote data points from the MBII simulation from $z=0-10$ (red-blue).
    The open circles in the FOF mass function (bottom panels) denote data from the MB simulation from $z=5-11$ (green-purple).
    The dashed (red), dot-dashed (cyan), dot-dot-dot-dashed (orange) lines are for the mass functions from
    \citet{1974ApJ...187..425P,2002MNRAS.329...61S,2013MNRAS.433.1230W}.
    The solid (black) line in the bottom panels (FOF mass functions) denote the best fit mass function to the MB and MBII data based on the \citet{2008ApJ...688..709T}
    parametrization of the \citet{2006ApJ...646..881W} mass function.}
  \label{fig_fnu}
\end{figure*}

In figure~\ref{fig_fnu} we plot the rescaled
mass function $f(\sigma)$ for redshifts $z=10$ to $z=0$ from the MBII simulation (open squares, blue to red).
For the FOF mass functions we also add data from the MB simulation from
redshifts  $z=11$ to $z=5$ (open circles purple to green).
For any simulation the mass function data points move from right to left along a single curve
in the $f(\sigma) - \sigma$ plane.
The top row and bottom rows denote the {\small SUBFIND}
 and FOF mass functions.
The columns denote the full mass of the halo and the dark matter
mass of the halo respectively.
The dashed (red), dot-dashed (cyan), dot-dot-dot-dashed (orange) lines are for the mass functions from
\citet{1974ApJ...187..425P,2002MNRAS.329...61S,2013MNRAS.433.1230W}
The solid (black) line in the bottom panels (FOF mass functions) denote the best fit mass function to the MB and MBII data based on the \citet{2008ApJ...688..709T}
parametrization of the \citet{2006ApJ...646..881W} mass function.

The PS mass function \citep{1974ApJ...187..425P} overpredicts the abundance of
low mass halos and underpredicts the abundance
of large mass halos.
This has been seen in numerous studies \citep{1999MNRAS.308..119S,2001MNRAS.321..372J,2007ApJ...671.1160L,2007MNRAS.374....2R,2013MNRAS.433.1230W}.
This has led to a renewed effort in recent years to recalibrate
the mass function of halos based on simulations.

We find that the {\small SUBFIND} mass functions for halos and dark
matter do not fall on a single curve.
There is significant and systematic scatter across redshifts at
small masses. This is similar to the
analyses of
\citet{2001MNRAS.321..372J,2008ApJ...688..709T,2013MNRAS.433.1230W}
who found a for an object definition different from
FOF halos a less universal mass function (having a redshift dependence). We therefore do not provide a universal fit
to the {\small SUBFIND} mass functions which show a strong redshift dependence.

On the other hand the FOF mass function has been shown
to be more universal \citep{2001MNRAS.321..372J,2011ApJ...732..122B,2013MNRAS.433.1230W}
and this is also seen in the lower panels of figure~\ref{fig_fnu}.
We denote FOF and FOFD to represent the FOF halo and dark matter component of the FOF halo.
We find that the FOF and FOFD mass functions agree well with the
\citet{2002MNRAS.329...61S,2013MNRAS.433.1230W} mass functions at the larger masses. The FOFD shows a systematic shift with
respect to the FOF mass function due to a systematic shift in the halo mass, since baryon contribution has been subtracted
from the halo mass. However the mass function at small masses is
systematically underestimated in MBII. This discrepancy is larger for FOFD which is again due to a systematic shift in
the halo mass. Our results are consistent with \citet{2013MNRAS.431.1366S} who used
the set of simulations described in \citet{2009MNRAS.399.1773C}
to look at the effect of baryons on the abundance of halos. \citet{2013MNRAS.431.1366S} compared the GIMIC (which include gas, dark matter, star formation and feedback)
and the DMO (dark matter only) simulations \citep{2009MNRAS.399.1773C} which were performed by resimulating at higher resolution an $18 \hmpc$ spherical
region of the Millennium simulation \citep{2005Natur.435..629S}.  Both the GIMIC and DMO runs were done with the same initial conditions
making it possible to look at the effect of baryons on halo properties directly.
They found that both simulations agree well on large scales however objects below $\sim 10^{12}\msun$ have systematically lower masses in the GIMIC
simulation when compared to the DMO counterpart. This result translated to an overestimate of the abundance of structures in the DMO simulation, by
approximately $\sim 10\%$ at $10^{11.5}\msun$ and $\sim 30\%$ at $10^{10}\msun$. We do not have a DMO version of the MBII simulation to make a direct
comparison. We therefore make a comparison of the mass function in MBII with published results based on dark matter only simulations.

Given the binned FOF and FOFD mass functions we perform separate fits with the \citet{2008ApJ...688..709T} parametrization of the \citet{2006ApJ...646..881W} mass function:
\beq
f(\sigma) = A\left[\left(\frac{\beta}{\sigma}\right)^{\alpha} + 1\right]\exp\left[-\frac{\gamma}{\sigma^2}\right]
\label{eq_fnubf}
\eeq

The solid black line in the lower panels (FOF and FOFD) of figure~\ref{fig_fnu} represents our fits to the mass function.
For the FOF mass function we have also added data from the $\sim \times 150$ larger volume MB simulation from $z=5-11$ (open circles).
This is done to obtain a larger range at the tail of the mass function. The fitted mass function function for FOF and FOFD are good to within $\sim 13\%$ across
the full range of masses and redshifts. This means that the universality of the mass function holds for the FOF and FOFD mass functions at the $\sim 13\%$
level.  The best fit parameters are quoted in table~\ref{tab_bf}. We have also added the latest fit from dark matter simulations described in \citet{2013MNRAS.433.1230W}.
\citet{2013MNRAS.433.1230W} also find that their fit is accurate to $\sim 10\%$ across all redshifts and provide redshift dependent fits to obtain greater accuracy.
The tail of the mass function which is governed by $\gamma$ is consistent with \citet{2013MNRAS.433.1230W}. We however find that the best fit mass function
in MBII systematically underpredicts the abundance of halos in the tail of the mass function, which is dominated by high redshift data points (with large error bars).
Such a behavior is also seen in \citet{2013MNRAS.433.1230W} and can only be improved by assuming a redshift dependent fit, which we leave to a forthcoming paper.
\begin{table}
      \begin{center}
        \begin{tabular}{c|c|c|c|c}
          \hline
                         & A       & $\alpha$ & $\beta$ & $\gamma$ \\
          FOF (MB+MBII)  & 0.1897  & 1.9607 & 1.7880 & 1.2067\\
          FOFD (MB+MBII) & 0.1738  & 1.6907 & 1.8812 & 1.2104\\
          \citet{2013MNRAS.433.1230W} & 0.282  & 2.163 & 1.406 & 1.210\\
         \hline
        \end{tabular}
      \end{center}
\caption{Best fit parameters for the FOF and FOFD mass functions. The parameters are described in equation~\ref{eq_fnubf}. The last row
  is the best fit parameters from \citet{2013MNRAS.433.1230W}.}
\label{tab_bf}
\end{table}

\begin{figure}
  \begin{tabular}{c}
    \includegraphics[width=3.2truein]{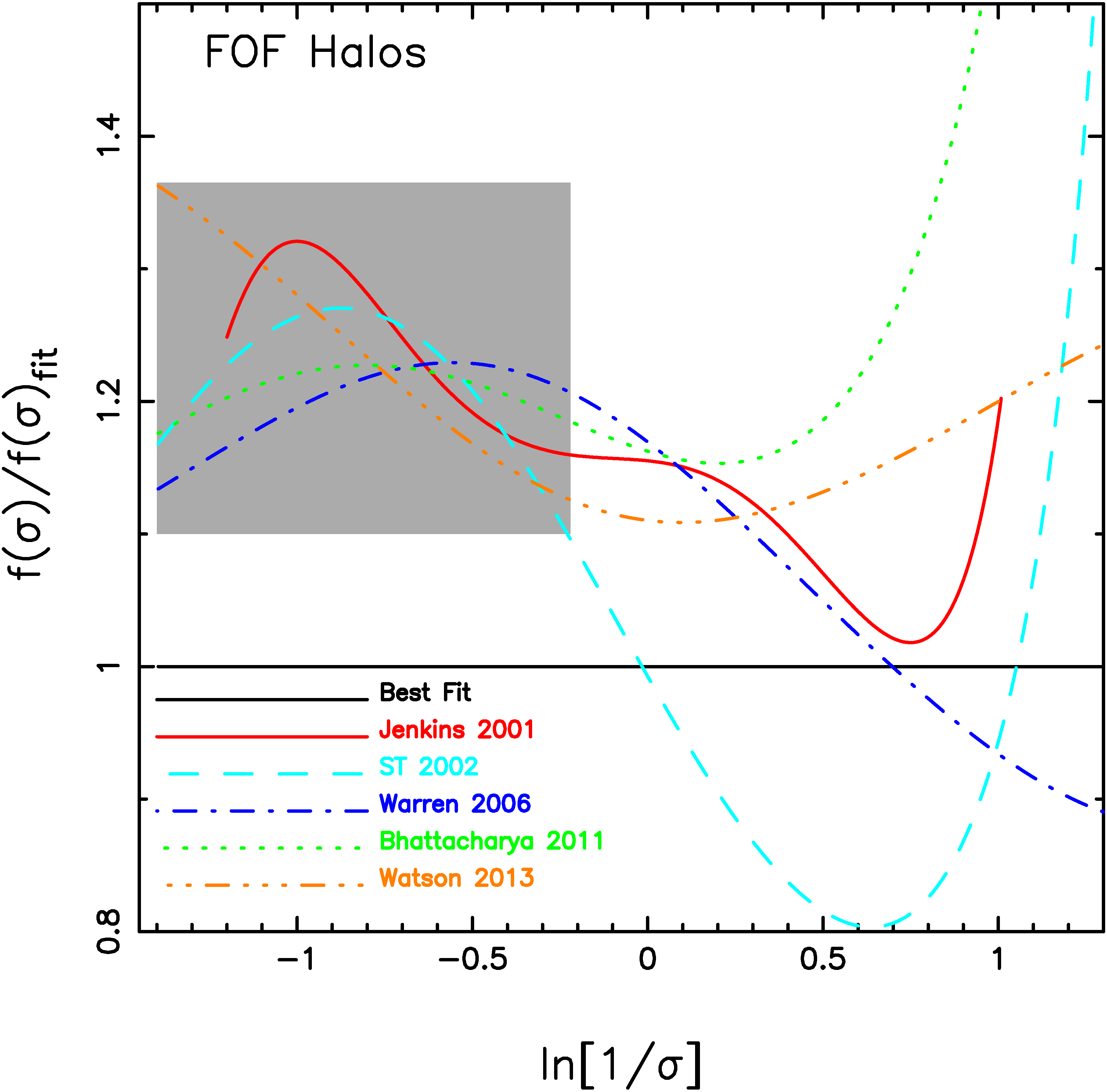} \\
  \end{tabular}
  \caption{Ratio of the best fit FOF mass function with fits based on dark matter only simulations for FOF halos.
  The solid (black) horizontal line is for MBII. The solid (red), dashed (cyan)
  , dot-dashed (blue), dotted (green) and dot-dot-dot-dashed (orange) lines
  are from \citet{2001MNRAS.321..372J,2002MNRAS.329...61S,2006ApJ...646..881W,2011ApJ...732..122B,2013MNRAS.433.1230W}. The gray shaded box highlights the region below the knee of the mass function where dark matter simulations systematically overpredict the abundance of halos.}
  \label{fig_massfn_ratio}
\end{figure}

We end this section by comparing the best FOF mass function (MBII +MB)
to earlier earlier work in figure~\ref{fig_massfn_ratio}. We plot the ratio
of the FOF mass functions in
\citet{2001MNRAS.321..372J,2002MNRAS.329...61S,2006ApJ...646..881W,2011ApJ...732..122B,2013MNRAS.433.1230W}
(solid (red), dashed (cyan), dot-dashed (blue),
dotted (green) and dot-dot-dot-dashed (orange) lines) to our fit (table~\ref{tab_bf}) and focus our attention at smaller masses, i.e.
the gray shaded box bounded by $-1.4 \leq \ln(1/\sigma) \leq -0.2$
which highlights the region below the knee of the mass function,
where dark matter simulations systematically overpredict the abundance of halos.
We find that all the fits based on the dark matter simulations overpredict
the mass function at the $20-35\%$ level at around $\ln(1/\sigma) \simeq -0.9$ when compared to our fit.
$\ln(1/\sigma) = -0.9$ corresponds to $\mhalo = 10^{11.2}\msun$/h at $z=0$ and $\mhalo = 10^{9.3}\msun$/h at $z=1$.
Even at the right edge of the shaded region, i.e. $\ln(1/\sigma) = -0.2$  which corresponds to $\mhalo = 10^{13.2}\msun$/h at $z=0$,
dark matter simulations overpredict the FOF mass function at the 10-20\% level.
To our knowledge the large effect baryonic processes have in shaping the mass function has been neglected up to this point.

\begin{figure*}
\begin{tabular}{c}
\includegraphics[width=\textwidth]{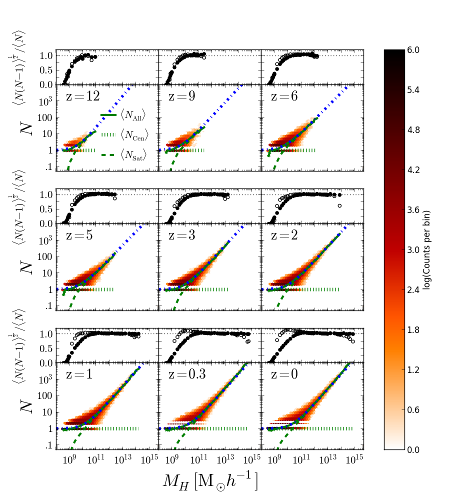}
\end{tabular}
\caption{Occupation number and scatter of sub halos as a
function of halo mass.  The color corresponds to the log
of the grid density at each occupation number and halo mass.
The green over plotted curves are the mean occupation number of
central sub halos(\textit{dotted}),
satellite sub halos (\textit{dashed}),
and all sub halos (\textit{solid}).
The blue dash-dotted curve is the best fit power law of the
occupation number through all data points.
The top panel for each is the width of the probability
distribution for all sub halos (\textit{solid circle})
and satellite sub halos (\textit{open circle}).
For a Poisson distribution, with width would be 1 which is the dotted line.}
\label{OccNum}
\end{figure*}

\section{The Halo Occupation Distribution}
\label{sec_hod}

The Halo Occupation Distribution (HOD) is a powerful theoretical formalism used for describing, predicting, and interpreting the clustering of galaxies in the large scale structure of the universe.
The HOD model describes the probability distribution $P(N|\mhalo)$ that a halo with virial mass $\mhalo$ contains $N$ galaxies.
In addition to this probability distribution, the model also describes the relative spatial distribution of galaxies within the halo.
Recently several papers have shown the robustness of the HOD by constructing and applying the model to large cosmological hydrodynamic simulations (e.g.,
\citet{2001ApJ...550L.129W, 2002ApJ...575..587B, 2003ApJ...593....1B, 2005ApJ...633..791Z} and references therein).
Like the recent hydrodynamic simulations, we will apply the HOD model to our own runs in this section.
In the HOD model, the distribution of matter within a halo is described by two main components: first, the probability distribution $P(N|\mhalo)$ that a halo with mass $\mhalo$ hosts $N$ number of galaxies, and second, the relationship between the distribution of galaxies within the halos.
In this section, we will briefly analyze and discuss these two components from our simulation.

The most important component of the HOD model is the probability distribution $P(N|\mhalo)$.
Figure~\ref{OccNum} shows the occupation number $N$ as a function of halo mass $\mhalo$ for nine snapshots.
Each halo from the snapshots has one point on the plot and the color corresponds the log of the number of points per grid space.
It is easy to see the power-law tail for high halo mass greater than $\mhalo\sim10^{13}\mathrm{M_\odot}h^{-1}$ at almost all snap shots.
However for high redshift, i.e. $z=12$, there are few halos with $\mhalo>10^{13}\mathrm{M_\odot}h^{-1}$, so the power-law tail isn't observed.
The color plot corresponds to the total number of galaxies $N_\mathrm{All}$ as a function of halo mass, but the literature has shown that is is perhaps more robust to explore the occupation number of the central galaxy $N_\mathrm{Cen}$ and satellite galaxies $N_\mathrm{Sat}$ separately \citep{2005ApJ...633..791Z}.
Previous studies have shown that a halo above a certain mass threshold will host one central galaxy, while halos below this mass threshold will not.
Therefore, $N_\mathrm{Cen}$ can be modeled as a step function
\begin{equation}
N_\mathrm{Cen}(\mhalo)=
\begin{cases}
0 & \mhalo<M_\mathrm{min} \\
1 & \mhalo\ge M_\mathrm{min}
\end{cases}
\label{CenOcc}
\end{equation}
where $M_\mathrm{min}$ is the minimum mass of a halo which hosts a central galaxy.
The green dotted lines in figure~\ref{OccNum} is the average number of central galaxies $\langle N_\mathrm{Cen}\rangle$.
For all snapshots $\langle N_\mathrm{Cen}\rangle$ plateaus at $N=1$ quickly because nearly all halos identified with the FOF algorithm host one central galaxy.
The halos that host a central galaxy can also be populated by satellite galaxies.
The dashed green lines in figure~\ref{OccNum} show the average number of satellite galaxies $\langle N_\mathrm{Sat}\rangle$.
Like the occupation number of all galaxies, $\langle N_\mathrm{Sat}\rangle$ follows a power-law for halo masses greater than $\mhalo\sim M_1$, where $M_1$ is the mass of a halo that on average hosts one satellite galaxy.
In other words, the average occupation number of satellite galaxies follows a power law proportional to $\langle N_\mathrm{Sat}\rangle\propto\left(\mhalo/M_1\right)^\alpha$.
The solid green line in figure~\ref{OccNum} is the average occupation number of both central and satellite galaxies: $\langle N_\mathrm{All}\rangle\equiv\langle N_\mathrm{Cen}\rangle+\langle N_\mathrm{Sat}\rangle$.
For low halo masses, $\langle N_\mathrm{All}\rangle$ is closely related to $\langle N_\mathrm{Cen}\rangle$; however, for higher $\mhalo$, as the occupation number of satellite galaxies increases, $\langle N_\mathrm{All}\rangle$ is dominated by $\langle N_\mathrm{Sat}\rangle$.

\begin{figure}
\centering
\includegraphics[width=\columnwidth]{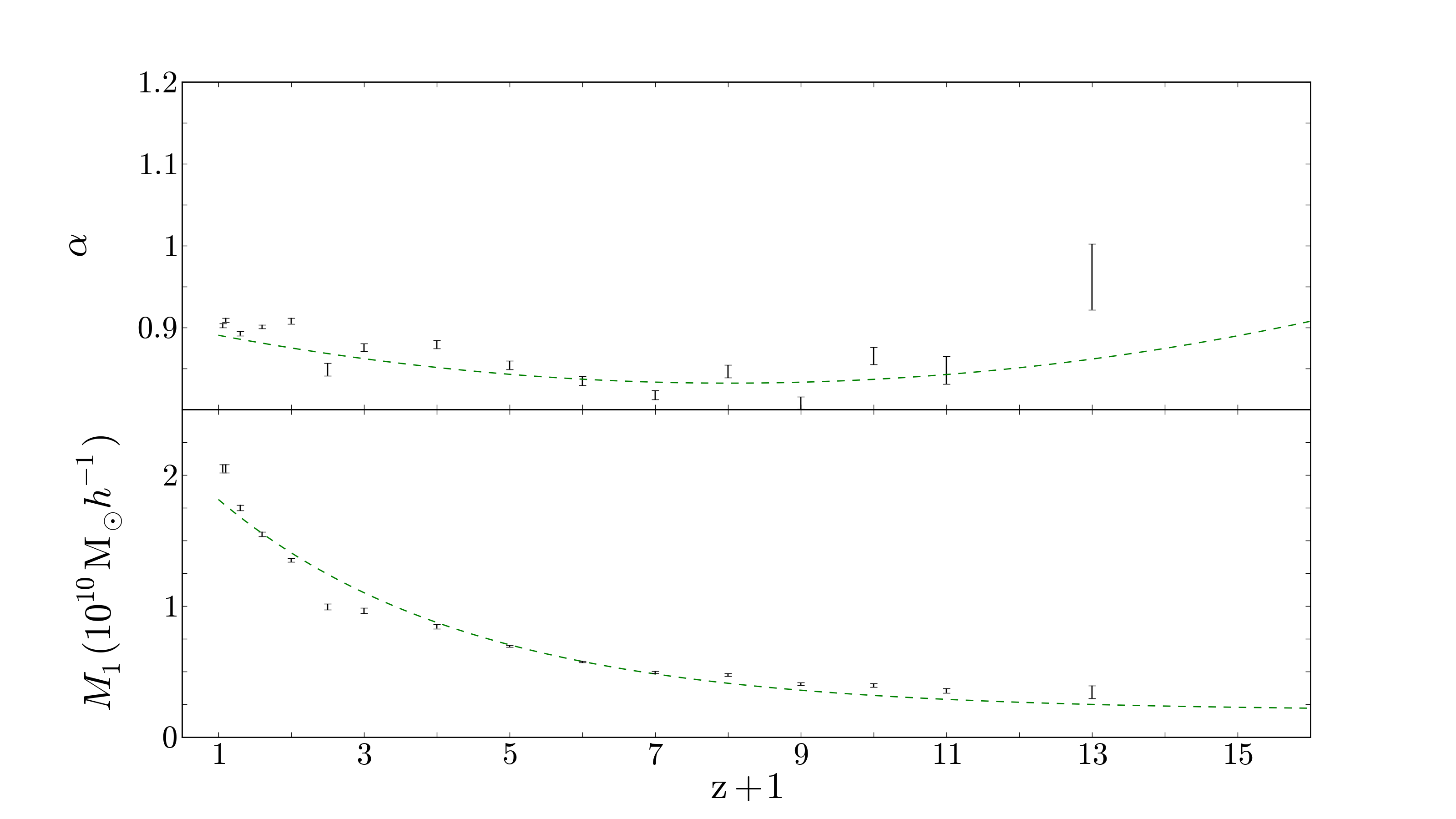}
\caption{Summary plot of the best fit parameters $\alpha$ and $M_1$
as a function of redshift.  The green curves are the best fits through each.
Table \ref{ParamTable} shows the fit functions used and the best fit
parameters through these fits.}
\label{ParamFit}
\end{figure}

In order to fit the halo occupation number, we consider a function that is a combination of $\langle N_\mathrm{Cen}\rangle$ and $\langle N_\mathrm{Sat}\rangle$.
Because the occupation number will be zero if $N_\mathrm{Cen}=0$, we only fit halos that host a central galaxy; additionally, the fit function must have a power-law tail for large $\mhalo$.
Therefore, we used a fit function of the form:
\begin{equation}
\langle N(\mhalo)\rangle=1+\left(\mhalo/M_1\right)^\alpha
\label{OccFit}
\end{equation}
where $M_1$ is defined by $N_\mathrm{Sat}(M_1)=1$, and $\alpha$ is the power-law index of the distribution.
These fits are shown as the blue dot-dash curves on figure~\ref{OccNum}.
The two fit parameters $\alpha$ and $M_1$ exhibit an evolution with redshift, which is shown in figure~\ref{ParamFit}.
The bottom panel of figure~\ref{ParamFit} shows the evolution of the normalization mass $M_1$, which exhibits a clear exponential decay with increasing redshift.
The green dashed line is the best fit curve, and the fitting parameters and function are given in Table \ref{ParamTable}.
In the top panel of figure~\ref{ParamFit} is the evolution of the slope of the power-law tail $\alpha$.
There seems to be a slight parabolic evolution with redshift for $\alpha$, which is fitted as the green dashed curve; however, this evolution is very slight and may just be an artifact of the simulation.
Nevertheless, the fit function along with the best fit parameters for $\alpha$ are also shown in Table \ref{ParamTable}

\begin{table}
\begin{center}
\begin{tabular}{clclclc}
\hline
Function&A&B&C\\
\hline\hline
$\alpha=A+B(\mathrm{z}+1-C)^2$&$0.84(2)$&0.003(1)&5.4(1.3)\\
$M_1=A+Be^{-C(\mathrm{z}+1)}$&0.2(1)&2.4(2)&0.34(3)\\
\hline
\end{tabular}
\end{center}
\caption{Best fits for the evolution of the power law index $\alpha$
and normalization mass $M_1$ for increasing redshift.}
\label{ParamTable}
\end{table}

\begin{figure*}
\begin{tabular}{c}
\includegraphics[width=\textwidth]{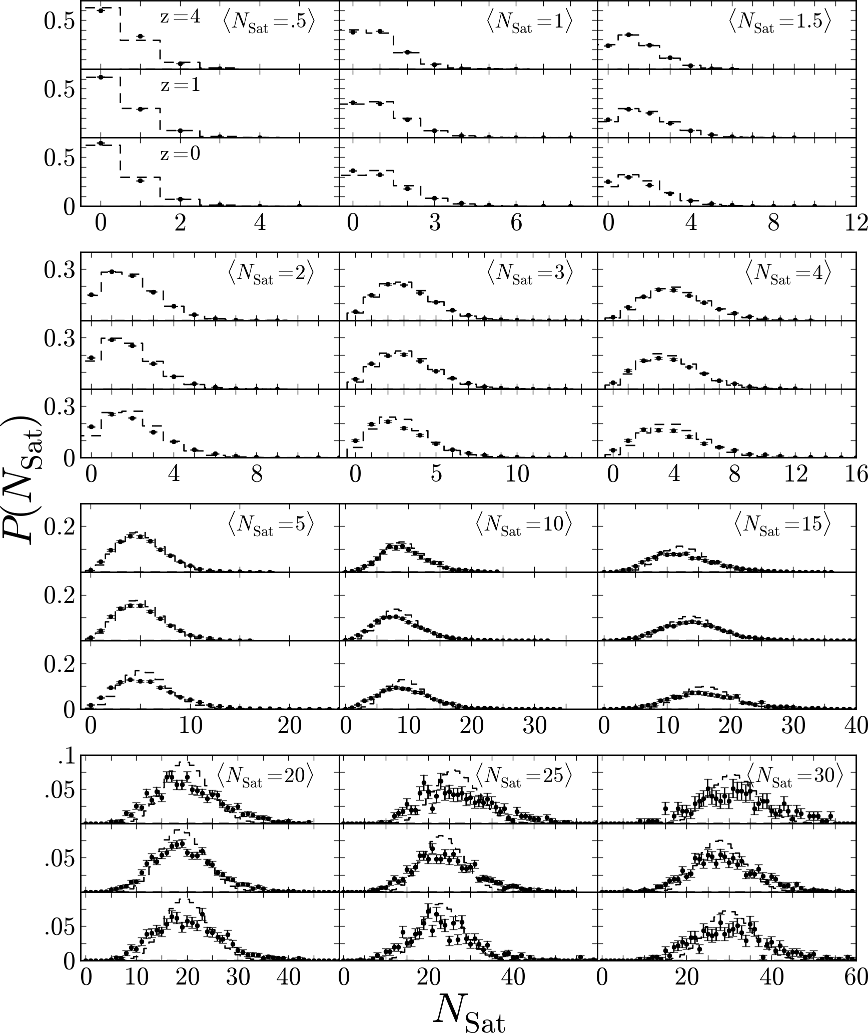}
\end{tabular}
\caption{Probability distribution $P(N_\mathrm{Sat}|\langle N_\mathrm{Sat}\rangle)$ for a halo with virial mass $\mhalo$ and an average occupation number of $\langle N_\mathrm{Sat}(\mhalo)\rangle$ will host $N_\mathrm{Sat}$ galaxies.  Each plot is the distribution about a different value for $\langle N_\mathrm{Sat}\rangle$ which is given in the top right, while the three panels correspond to the three redshifts given in the first plot.  The error bars shown are Poisson error bars.  For comparison the dotted histogram is the corresponding Poisson distribution centered about $N_\mathrm{Sat}$.  Each of the distributions can be very accurately approximated as a Poisson distribution.}
\label{ProbDist}
\end{figure*}

\begin{figure*}
\centering
\includegraphics[width=\textwidth]{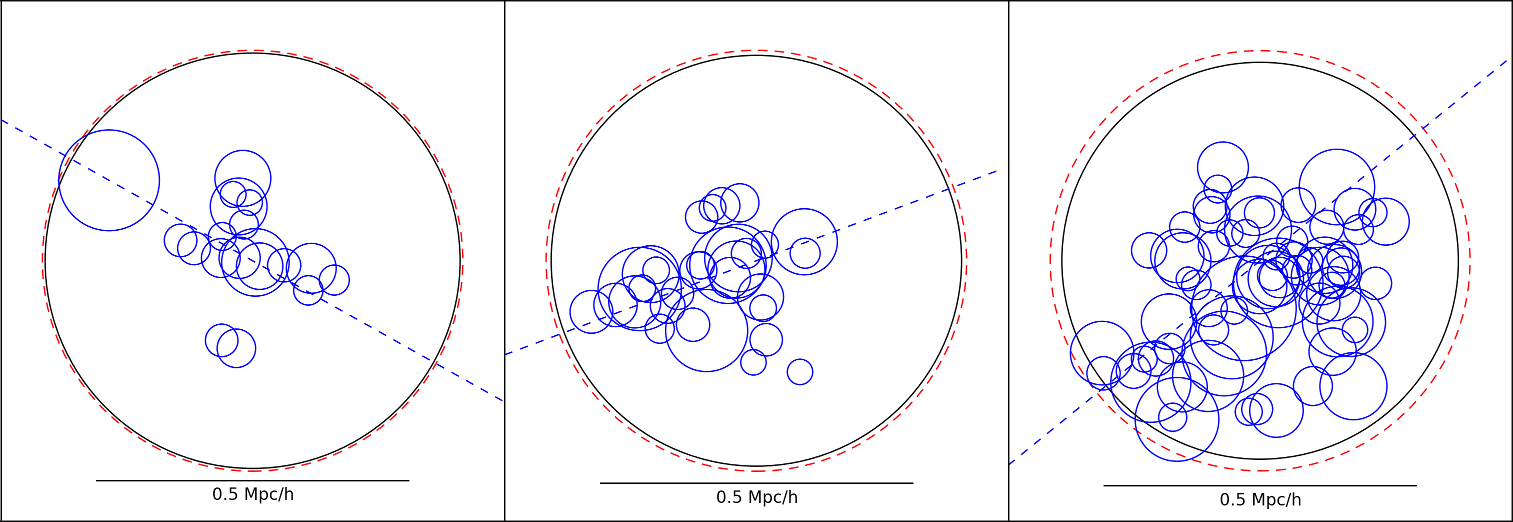}
\caption{Plotted here is a simplistic schematic of the distribution of sub halos within three groups of mass $\mhalo\sim10^{12}M_\odot h^{-1}$.  The left panel shows a group with the fewest number of subhalos from our simulation, the center panel shows a group with an occupation number equal to $\langle N_\mathrm{Sat}\rangle$ given by Eq. \ref{OccFit}, and the right panel shows a group with the largest number of subhalos from our distribution.  The red dotted circles map out $R_{Vir}$ of the parent halo, while the black and blue circles show $R_{Vir}$ of the central and satellite subhalos, respectively.  Subhalos are believed to live upon filamentary structures within groups so the blue dotted line roughly shows the filaments of each group.}
\label{HaloSchematic}
\end{figure*}

\begin{figure*}
\includegraphics[width=\textwidth]{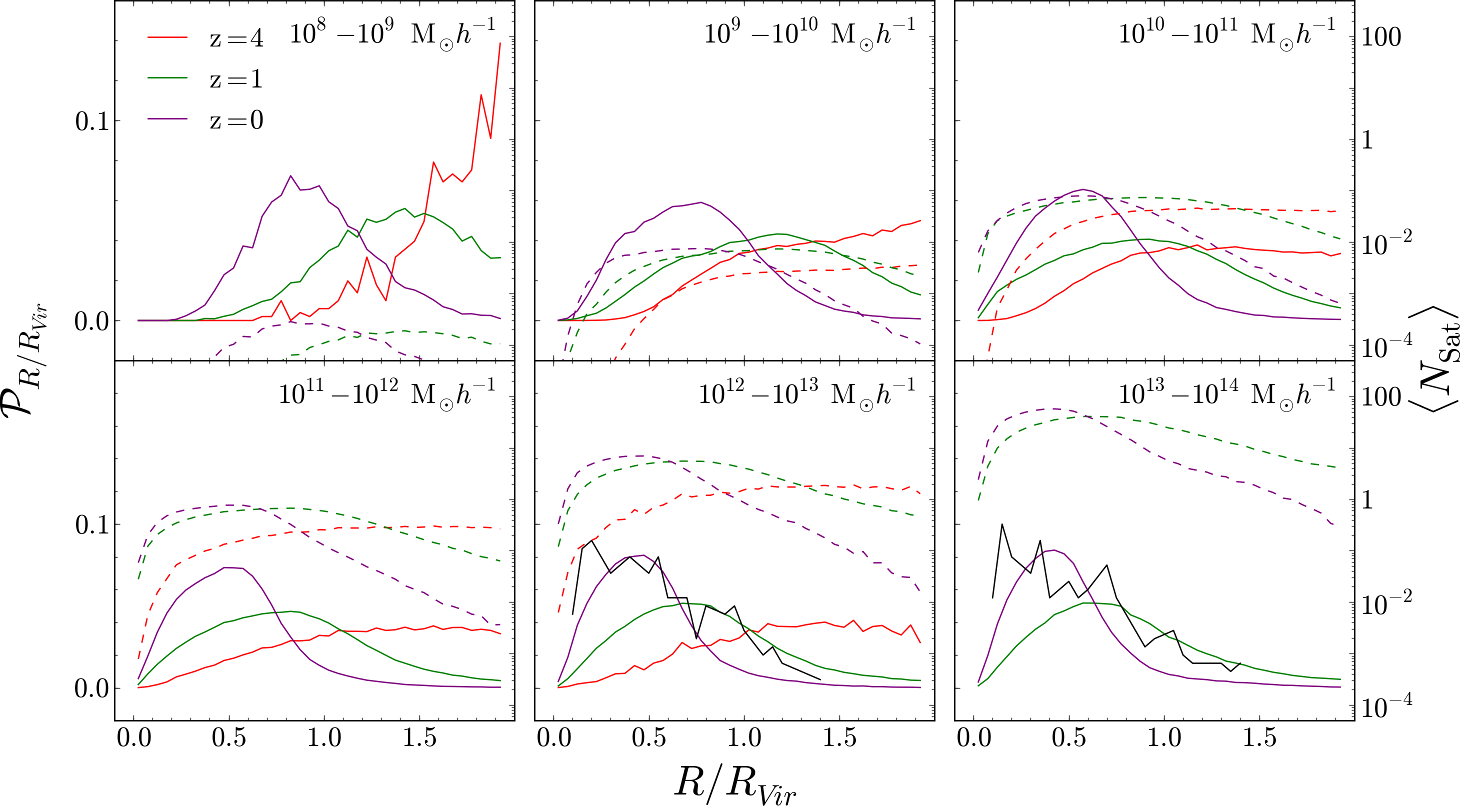}
\caption{The solid curves here are the probability distribution $\mathcal{P}_{R/R_{Vir}}$ that a satellite subhalo will be located at a radial distance $R$ from the center of the parent halo in units of the parent halo's $R_{Vir}$.  Each panel corresponds to a parent halo mass bin which is given in the right hand corner of each panel.  The colors of the curves correspond to a specific redshift which is given in the top left corner of the first (top left) panel.  The dotted curves correspond to the occupation number $\langle N_\mathrm{Sat}\rangle$ at a certain radial distance from the center of the parent halo.  In the last two panels, the black curves show the SPH results from \citet{2003ApJ...593....1B}.}
\label{RadDistr}
\end{figure*}

With the best fit functions given in Table \ref{ParamTable}, one could determine the normalization mass $M_1$ and power law index $\alpha$ at a given redshift, then use Eq. \ref{OccFit} to determine $\langle N\rangle$ in order to populate a given halo of mass $\mhalo$ for a simulation.
However, in order to fully populate said halos, one must also understand the spread of the sub halo population at a certain value of $\langle N\rangle$.
The best technique for this analysis is to compare the probability distribution of sub halos $N$ at a given average occupation number $\langle N\rangle$, $P(N|\langle N\rangle)$, with that of a well defined probability distribution, more specifically a Poisson distribution (e.g. \citet{2005ApJ...633..791Z, 2003ApJ...593....1B}).
If $P(N|\langle N\rangle)$ follows that of a Poisson distribution, then one could use Poisson statistics to quantize the spread of occupation number from the mean.
The top panels of figure~\ref{OccNum} show the width of the distribution; for a Poisson distribution the width is $\langle N(N-1)\rangle=\langle N\rangle^2$ which is shown as the dotted line at unity.
The solid circles show the width of the probability distribution as a function of $\mhalo$ for all galaxies (both central and satellite).
At all redshifts, this distribution is sub Poisson for low values of $\mhalo$, however, it quickly approaches Poissonian at larger $\mhalo$.
Because the number of central galaxies follows a step function, there is very little spread in this value, i.e. it is either $0$ or $1$, so instead of exploring the probability for all sub halos we focused on the probability distribution of satellite sub halos $P(N_\mathrm{Sat}|\langle N_\mathrm{Sat}\rangle)$.
The open circles in the top panels of figure~\ref{OccNum} show the width of the probability distribution as a function of $\mhalo$.  Again for low halo masses, $P(N_\mathrm{Sat})$ is sub Poisson, but approaches a Poissonian width for higher masses.
On the other hand, for extremely high halo masses at low redshifts, $P(N_\mathrm{Sat})$ again is sub Poisson.
From the width of the distribution, it appears that $P(N_\mathrm{Sat})$ is very close to a Poisson distribution.
In figure~\ref{ProbDist} we plotted $P(N_\mathrm{Sat}|\langle N_\mathrm{Sat}\rangle)$ for three different snapshots.
Each plot shows the probability that a halo $\mhalo$ will host $N_\mathrm{Sat}$ satellite galaxies if on average the halo occupation number is $\langle N_\mathrm{Sat}(\mhalo)\rangle$.
In figure~\ref{ProbDist} the error bars are Poisson error bars and the dotted histogram is the Poisson distribution centered at $\langle N_\mathrm{Sat}\rangle$ which is given in the top right of each set of plots.
From figure~\ref{ProbDist} and the top panels of figure~\ref{OccNum}, it is clear that $P(N_\mathrm{Sat}|\langle N_Sat\rangle)$ can be approximated as a Poisson distribution; therefore, one can adequately use Poisson statistics to quantify the spread of sub halos from the best fit mean of Eq. \ref{OccFit}

Now that we understand the mean occupation number for a halo at a certain redshift, we must also analyze the relative spatial distribution of the sub halos within their parent halo.
As we discussed above, all sub halos can be classified as either central or satellite; furthermore, each halo will host either one or none central sub halos, so studying the relative spatial distribution of a central sub halo is a bit trivial, i.e. if the halo hosts a central sub halo it will be located in the center of the halo.
Therefore, we will focus on analyzing the spatial distribution of only satellite sub halos.
Figure~\ref{HaloSchematic} shows a simplified 2-D schematic of three different halos of mass $\mhalo\sim10^{12}\ \mathrm{M_\odot}h^{-1}$ at $\mathrm{z}=0$ which is comparable to the Milky Way.
The middle panel shows a halo with an occupation number equal to the mean at $\mhalo$ (see Eq. \ref{OccFit}), while the left and right panels show, respectively, the halos with the minimum and maximum occupation number at $\mhalo$.
The red dotted circle shows the virial radius of the parent halo, the black solid circle shows the virial radius of the central sub halo, and the blue circles show the satellite sub halos.
Because the virial radius of a galaxy is depended on the mass of the galaxy, the size of the circle also represents the mass of each galaxy.
This simplified representation easily shows the distribution of mass within each halo: the majority of the mass is located in one large sub halo located at the center of the group with many less massive galaxies scattered around the central sub halo.
Furthermore, this simple schematic shows that the sub halos prefer to populate their parents halos along filaments which is roughly shown as the blue dotted line in figure~\ref{HaloSchematic}.

In figure~\ref{RadDistr} we have plotted (\textit{solid curves}) the probability density $\mathcal{P}_{R/R_{Vir}}$ that a satellite sub halo will be located
a radial distance $R$ from the the center of the group in units of the halo's $R_{fir}$.
We chose to only investigate satellite galaxies here because we have already shown that halos can only host one central galaxy which would lead to a trivial analysis.
We plotted halos from three different snapshots while each panel corresponds to different parent halo mass bins.
In the last two panels (the high mass panels), there a no halos with mass $10^{12}<\mhalo<10^{14}$ at $\mathrm{z}=4$, so there is no data from that snapshot plotted in these panels.
For completeness, plotted on the right hand axes are the corresponding dotted curves of $\langle N_\mathrm{Sat}\rangle$
as a function of radial distance scaled by $R_{200}$ for each snapshot.
Here, $\langle N_\mathrm{Sat}\rangle$ is the average number of satellite sub halos per group per radial bin.
In other words this value corresponds to the average number of satellite sub halos within a group \textit{at} a certain radial distance $R$,
which is \textit{not} the average number of satellite sub halos \textit{within} $R$.
Also plotted as the the black curve plotted in the last two panels are SPH data from \citet{2003ApJ...593....1B}. Our distribution follows
the same general form as \citet{2003ApJ...593....1B}. As can be seen in figure~\ref{RadDistr} the peak and width of the
radial distribution of satellite galaxies decreases with decreasing redshift, irrespective of the mass of the parent halo.
This suggests that with time satellite galaxies cluster strongly around the central galaxy and that mergers dominate over the accretion
of new satellites. Additionally we see a mild variation
for the peak of the distribution with the mass of the host halo at a given redshift. The relative
location of the peak (with respect to $R_{\mathrm{vir}}$) decreases with increasing halo mass which suggests that
the clustering of satellite galaxies is stronger in more massive halos.

\section{Galaxy clustering}
\label{sec_gal_clustering}

The MBII simulation is of a large enough volume that galaxy
clustering can be studied meaningfully. The sheer number of galaxies
in the MBII (particularly for low mass selection thresholds) means that
clustering
measures can be computed with a high signal to noise level and
consequently subtle features be noticed and analyzed.  In this
section we concentrate on two-point correlation functions of the
galaxies and dark matter, including the cross-correlation of the two.

\subsection{Two point correlation functions}

We analyze 15 snapshots
of the simulation between redshifts $z=10$ and $z=0.06$. For each
snapshot, we compute the two point autocorrelation of dark matter particles
and also the two point autocorrelation function of  subhalos.
For the latter, we measure this quantity for
several subsamples defined by a lower limit on the subhalo mass:
$m_{\rm tot} > 10^{9} \msun,  10^{10} \msun,
 10^{11} \msun,  10^{12} \msun$. We also
do the same for subsamples defined by lower limits on the stellar mass
of subhalos:
$m_{*} >  10^{8} \msun,  10^{9} \msun,
 10^{10} \msun,  10^{11} \msun$.
We also compute the cross-correlation of dark matter and subhalos
for subsamples defined by the above mass bins.

We note that
before computing the correlation functions for any sample in the simulation,
if the number of elements (dark matter particles
or subhalos) is greater than $256^{3}$, for speed
we randomly subsample
down to this number, as shot noise errors on the scales we are interested
in will be negligible at the sampling density.

\begin{figure}
  \begin{tabular}{c}
    \includegraphics[width=3.2truein]{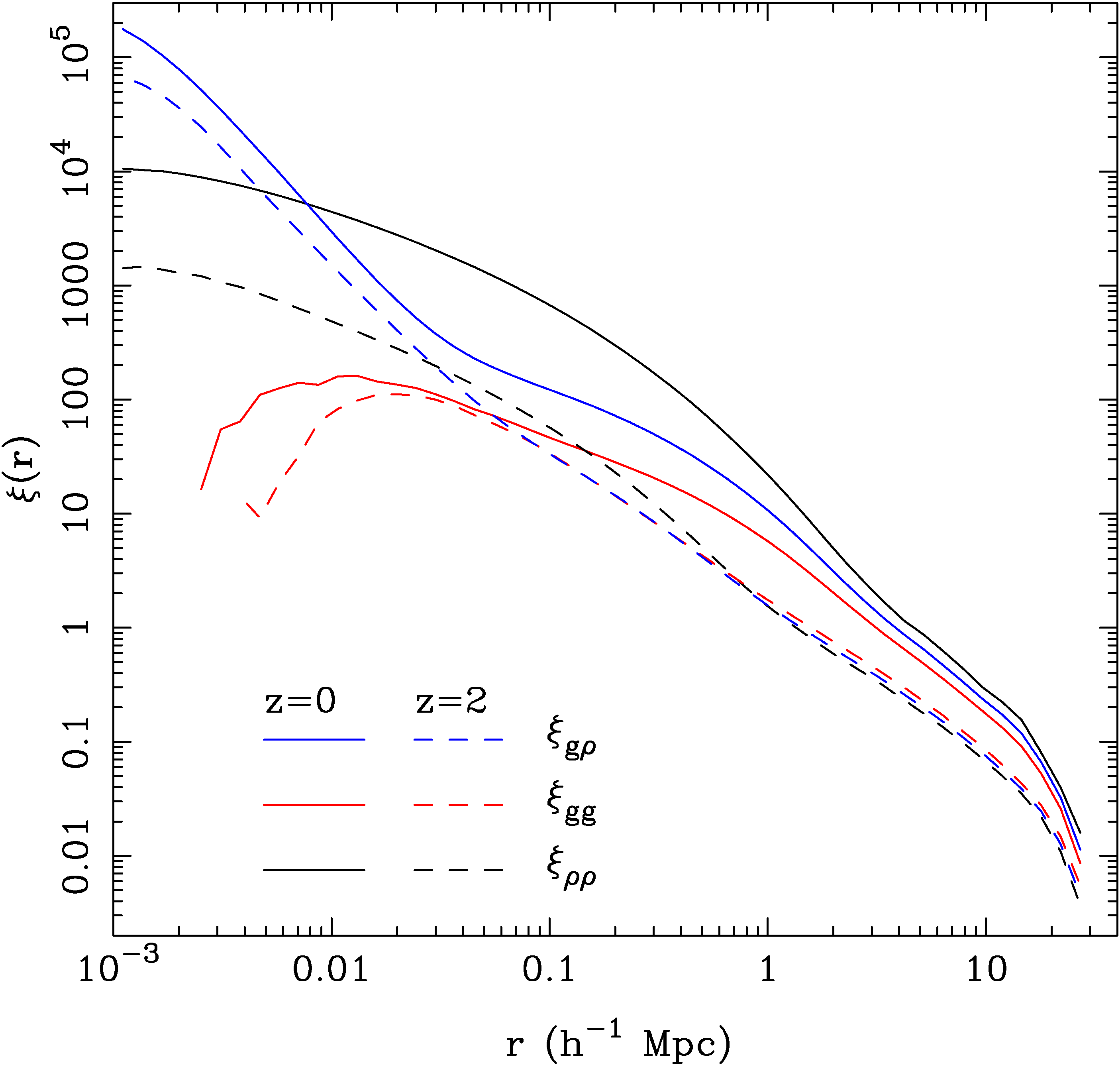}
  \end{tabular}
  \caption{The two-point auto-correlation function of dark matter (black),
    galaxies (red) and the two-point cross-correlation function
    of dark matter and galaxies (blue) in the MBII simulation.
    We show results at two redshifts $z=0$ and $z=2$. The galaxies were selected
    to be those above a (total) mass threshold of $10^{9}\msun$.}
  \label{xi}
\end{figure}

In Figure~\ref{xi}, we show  examples of the autocorrelations
and crosscorrelations for galaxies and dark matter
in the MBII at two redshifts, $z=0$ and $z=2$. In this example,
the galaxies used to compute the clustering were selected above a total
mass threshold of $10^{9} \msun$. There were $1.65\times 10^{6}$
galaxies in the subsample at $z=0$ and $1.89\times 10^{6}$ at
$z=2$.
In Figure~\ref{xi} we can see that the dark matter autocorrelation
(referred to as $\xi_{\rho\rho}$ although it is not the autocorrelation
of the total density)
at $z=0$ has the pronounced dip at $r\sim 1 {\rm h^{1} Mpc}$
indicating the transition between one-halo and
two-halo terms (\citealt{2002PhR...372....1C}).
At redshift $z=2$, the autocorrelation of
galaxies and of dark matter particles  have similar shapes and ampitudes
on scales $r> 20 $kpc/h, but at $z=0$ the galaxies (which have low
mass) are significantly anti-biased with respect to the dark matter on
all scales (see e.g., \citealt{2007MNRAS.378..641A}).
The cross-correlation, $\xi_{g\rho}$
has a second dip in it at $r\sim 20 $kpc/h. The galaxies at the two redshifts
have pretty similar $\xi_{g\rho}$  (dark matter profiles) interior to this,
 as expected, because they were selected above the same threshold mass.

\begin{figure}
\begin{tabular}{c}
  \includegraphics[width=3.2truein]{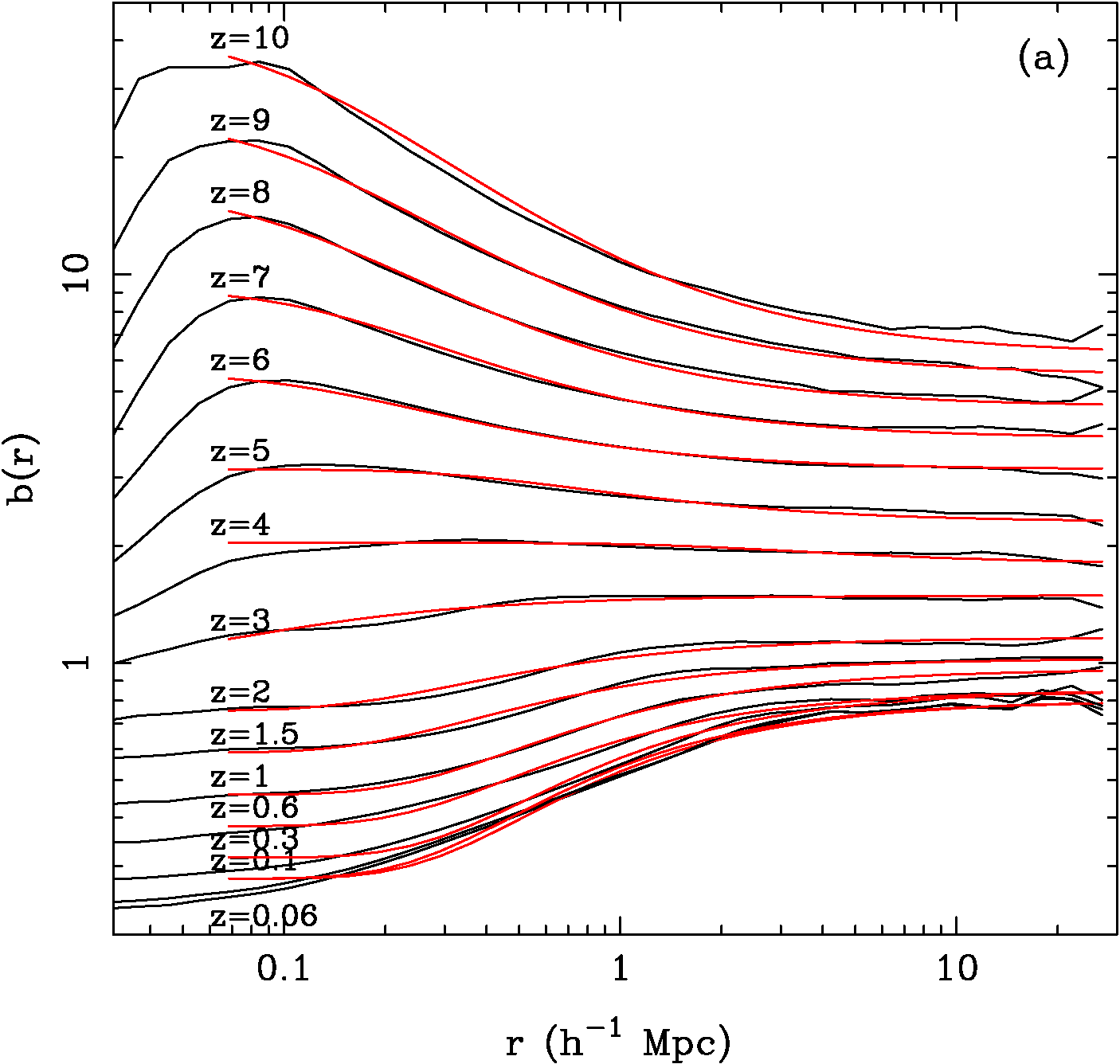} \\
  \includegraphics[width=3.2truein]{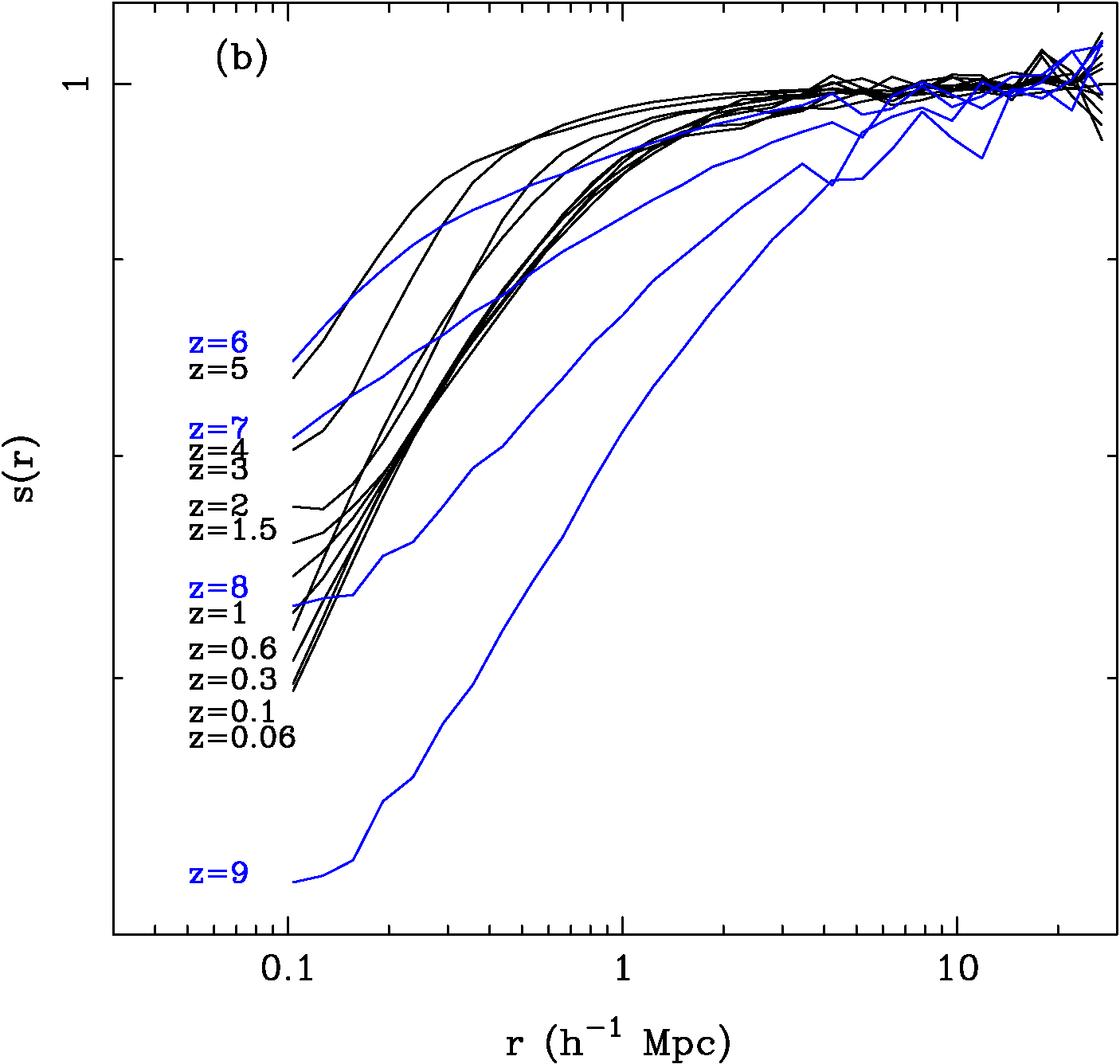}
\end{tabular}
\caption{(a)\emph{Top panel}: Bias vs scale for
galaxies in the MBII simulation,
at 15 different redshifts. The lower
threshold mass of the MBII galaxy sample was chosen to be $10^9 \msun$.
The simulation results are shown as solid black
lines, and a simple parametric fit (Equation \ref{bfiteq}) to
the results for each redshift is plotted as a solid red line.
(b)\emph{Bottom panel}: Stochasticity of galaxy clustering
(Equation \ref{sfiteq}) as a function of
scale for MBII galaxies. The same threshold mass ($10^{9} \msun$) was
used as in the top panel and the same redshift snapshots. In order
to make the redshift progression clearer, we plot results at redshifts
above $z=5$ with blue lines and the low redshift results in black.}
\label{biasstochvsscale}
\end{figure}

\subsection{Bias and stochasticity in MBII}

\subsubsection{Bias}

That the ratio of dark matter and galaxy correlation functions can
vary as a function of scale is obvious from Figure~\ref{xi}.
In Figure~\ref{biasstochvsscale} we plot
$b(r)=\sqrt{\xi_{gg}(r)/\xi_{\rho\rho}}$
for the same lower total mass threshold $10^{9}\msun$ as was used in
Figure~\ref{xi}, but for redshifts between $z\sim 0$ and $z \sim 10$.
The $b(r)$ function is approximately flat for separations $r > 2-5 {\rm h^{1} Mpc}$ ,
depending on the redshift, reaching the limit on large scales
usually referred to as linear bias (see e.g., \citealt{1998ApJ...504..607S}).
On smaller scales, the bias is scale-dependent,
with bias decreasing as $r$ becomes smaller for redshifts $z<4$ and
increasing for redshifts $z >4$. For the mass threshold plotted,
bias is approximately scale independent at $z=4$ down to
$r\sim 0.1 {\rm h^{1} Mpc}$ . The scale dependence at late times
is presumably due at least partly to non-linear effects such as
merging of galaxies reducing the number of pairs on small scales,
as well as  halo exclusion.
In the halo model framework, bias is scale dependent with a
change of slope at the transition scale between the one and the two halo terms.
At earlier times, because we are using a fixed threshold mass, the galaxies
become rarer and so are likely to lie in primary halos (i.e. they are
not in subhalos of larger halos).

\begin{figure}
  \begin{tabular}{c}
    \includegraphics[width=3.2truein]{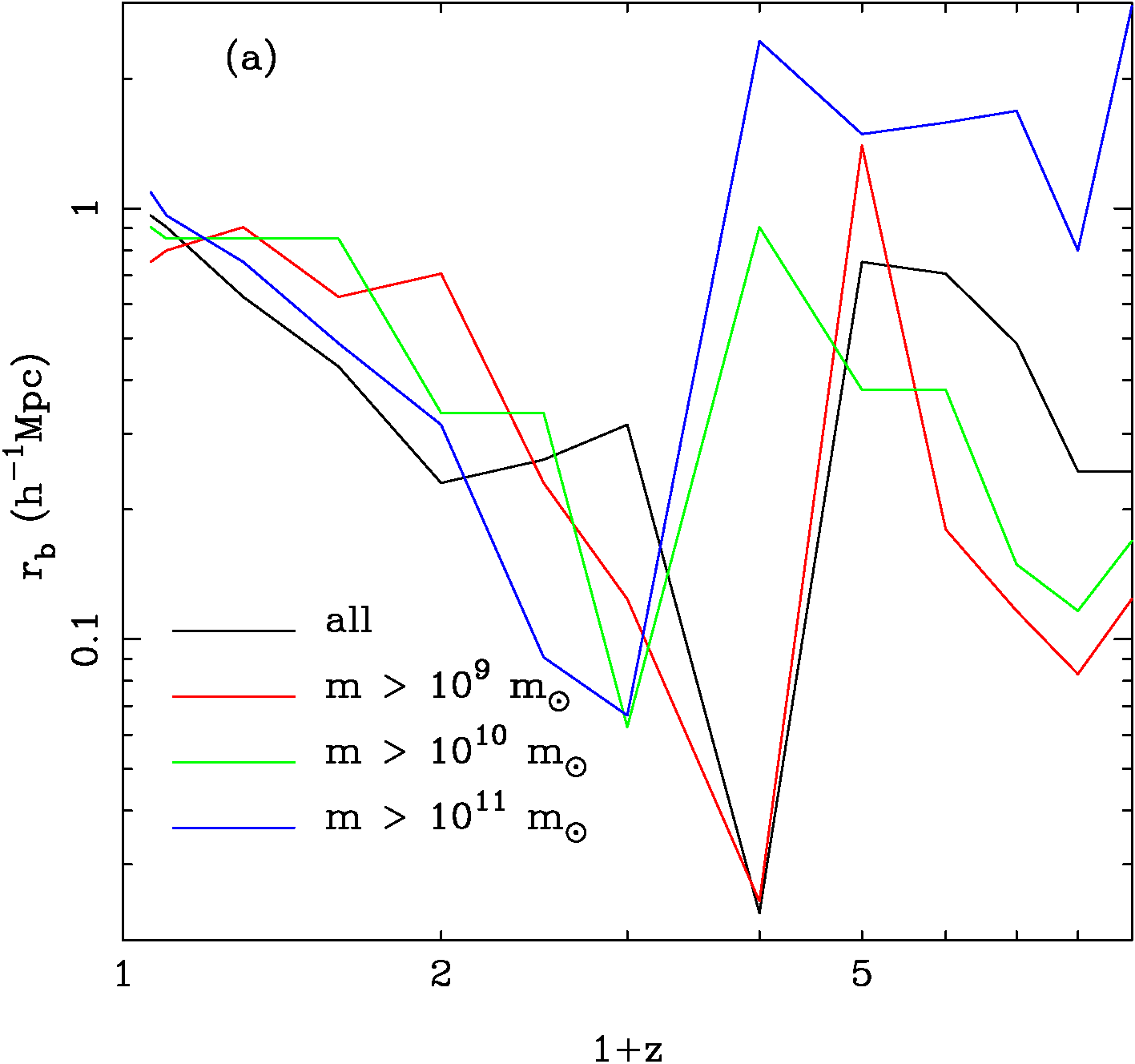}\\
    \includegraphics[width=3.2truein]{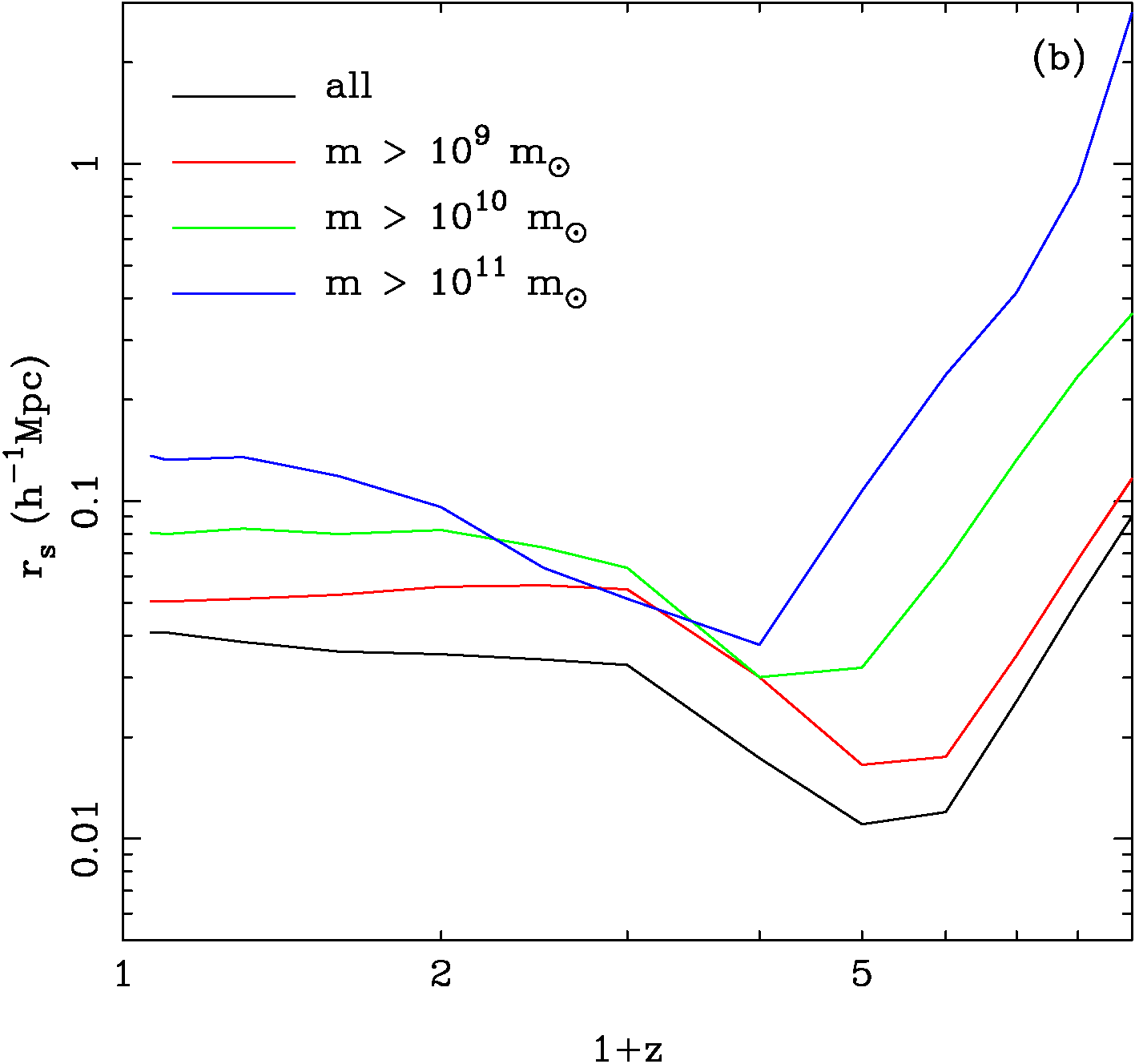}
  \end{tabular}
  \caption{(a)Top panel: The non-linear scale in
the bias (parameter $r_{b}$ in
    Equation \ref{bfiteq}) of MBII galaxies as a
function of redshift. We show results
    for four different lower mass thresholds as different colored lines.
    (b) Bottom panel: The non-linear scale in the stochasticity (parameter
    $r_{s}$ in Equation \ref{sfiteq}) of MBII galaxies as a function of redshift. We show
    results for the same  four different lower mass thresholds
    as the top panel.}
  \label{biasstochvsz}
\end{figure}

\begin{figure}
  \begin{tabular}{c}
    \includegraphics[width=3.2truein]{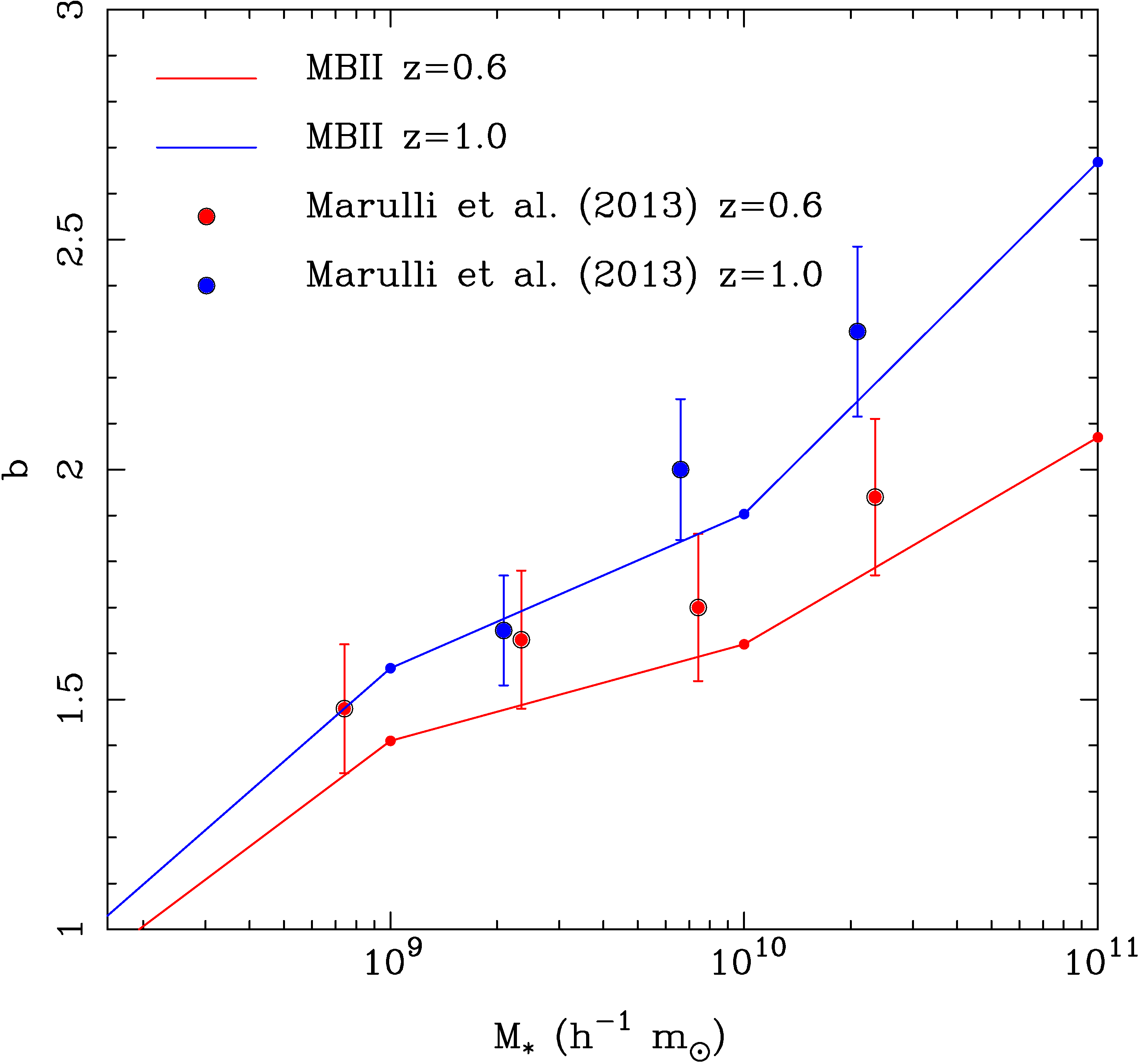} \\
  \end{tabular}
  \caption{Top panel: Linear bias versus threshold galaxy stellar mass.
    Results for galaxies in the VIPERS survey (\citealt{2013A&A...557A..17M}), are
    shown as points  with error bars, at two different redshifts,
    $z=0.6$ and $z=1$. We show results from the MBII simulation at the
    same redshifts as solid lines.}
\label{vipers}
\end{figure}

In order to further see how this trend evolves, we have plotted in
Figure~\ref{biasstochvsz} a simple fitting function for $b(r)$:

\beq
b_{\rm fit}(r)=(b_{\rm large}-b_{\rm small})e^{-(r_{b}/r)}+b_{\rm small}
\label{bfiteq}
\eeq

where $b_{\rm large}$ and $b_{\rm small}$ are fitting parameters
corresponding to the large-scale asymptote of the bias parameter and
a value of bias on small scales, respectively. The
parameter $r_{b}$ corresponds to the exponential length scale
over which the bias changes from its large to small scale value.
This parameter $r_{b}$ can therefore be considered to represent a type
of non-linear scale parameter for the bias.
We fit this function to
the $b(r)$ curves for points with $r > 0.25 {\rm h^{1} Mpc}$ , so that
we avoid the downturn of $b(r)$ on small scales.
The corresponding fits
are shown as red lines in Figure
\ref{biasstochvsscale}(a).

We note that in  Figure \ref{biasstochvsscale}(a) that for
the subhalo mass sample we are plotting the values of $b_{\rm large}$
and $b_{\rm small}$ in the fits will change relative to one another
as we change redshift. As a result there will be a redshift (for this
mass subsample it is around redshift $z=3-4$) where the bias will be
almost linear ($b_{\rm large} \simeq b_{\rm small}$). There is also likely to
be a  transition in the non-linear scale parameter $r_{b}$
at around this redshift.

In Figure \ref{biasstochvsz}(a) we plot the behavior of $r_{b}$ vs redshift,
with  results for several different mass bins on the same plot.
At low redshifts
$z < 2$
we can see a gradual increase in $r_{b}$
with scale for all mass subsamples. This non-linear scale reaches a maximum
of $1 \hmpc$ at $z=0$. This can be compared to the scale at which matter
clustering becomes non-linear (the matter clustering deviates from the linear
extrapolation), which is $\sim 5 \hmpc$ at this redshift
(\citealt{2001ApJ...558L...1G}). Galaxies therefore
trace the mass to scales significantly smaller than the non-linear mass
clustering scale in this simulation.

The minimum value for $r_{b}$ is reached at redshift $z=2$ and is between
$r_{b}=0.01-0.05 \hmpc$ with the smaller halos being at the lower end of
this range. At earlier redshifts there is then a switch to a much larger value
for $r_{b}$.

\subsubsection{Stochasticity}

Another quantity of interest is the stochasticity of clustering for which
we use the correlation coefficient   (see e.g., \citealt{2013PhRvD..87l3523S},
we use the symbol
$s(r)$ in order to avoid confusion with length scale $r$.)
\beq
s(r)=\xi_{g\rho}/\sqrt{\xi_{gg}\xi_{\rho\rho}}
\label{stocheq}
\eeq
We show the $s(r)$ curves for the same redshifts and lower mass threshold
($10^{9} \msun$) as used for Figure \ref{biasstochvsz}(a) in Figure
\ref{biasstochvsz}(b). On large scales, $r> 1-15 {\rm h^{1} Mpc}$ , depending
on the redshift, the $s(r)$ reaches unity, indicating that the
galaxy and dark matter fluctuations trace each other deterministically.
For clarity, we have not plotted the smallest scales
( $r< 0.1 {\rm h^{1} Mpc}$ ) on this plot, but all curves eventually rise
again and go above $s=1$ (this can be seen in Figure \ref{jullo}
below). The $s(r)$ increases for the smallest $r$
because galaxies cannot be closer together than the sum of
their radii. This causes $\xi_{gg}$ in the denominator of Equation
\ref{stocheq} to be very small (see e.g. Figure \ref{xi}) and
so $s(r)$ to increase.

Again in order to explore a wide range of masses and redshifts
in one plot, we have fit a simple curve to the $s(r)$ results.

\beq
s_{\rm fit}(r)=e^{-(r_{s}/r)}
\label{sfiteq}
\eeq

where $r_{s}$ is a parameter which determines the
scale at which the stochasticity $s(r)$ deviates from $s=1$. The results for
different mass bins are shown as a function of redshift in
Figure \ref{biasstochvsz}(b). We can see that halos with larger masses have
systematically higher values of $r_{s}$ at almost all redshifts. Curves for
all masses also have a trend of $r_{s}$ with redshift which is somewhat
similar to the $r_{b}$ parameter in Figure \ref{biasstochvsscale}. As
the density field evolves below redshifts $z=3-4$, the scale at which
stochasticity becomes important increases. Unlike the bias parameter
$r_{b}$ it does appear to level off at the lowest redshifts, however.

\begin{figure}
  \begin{tabular}{c}
    \includegraphics[width=3.2truein]{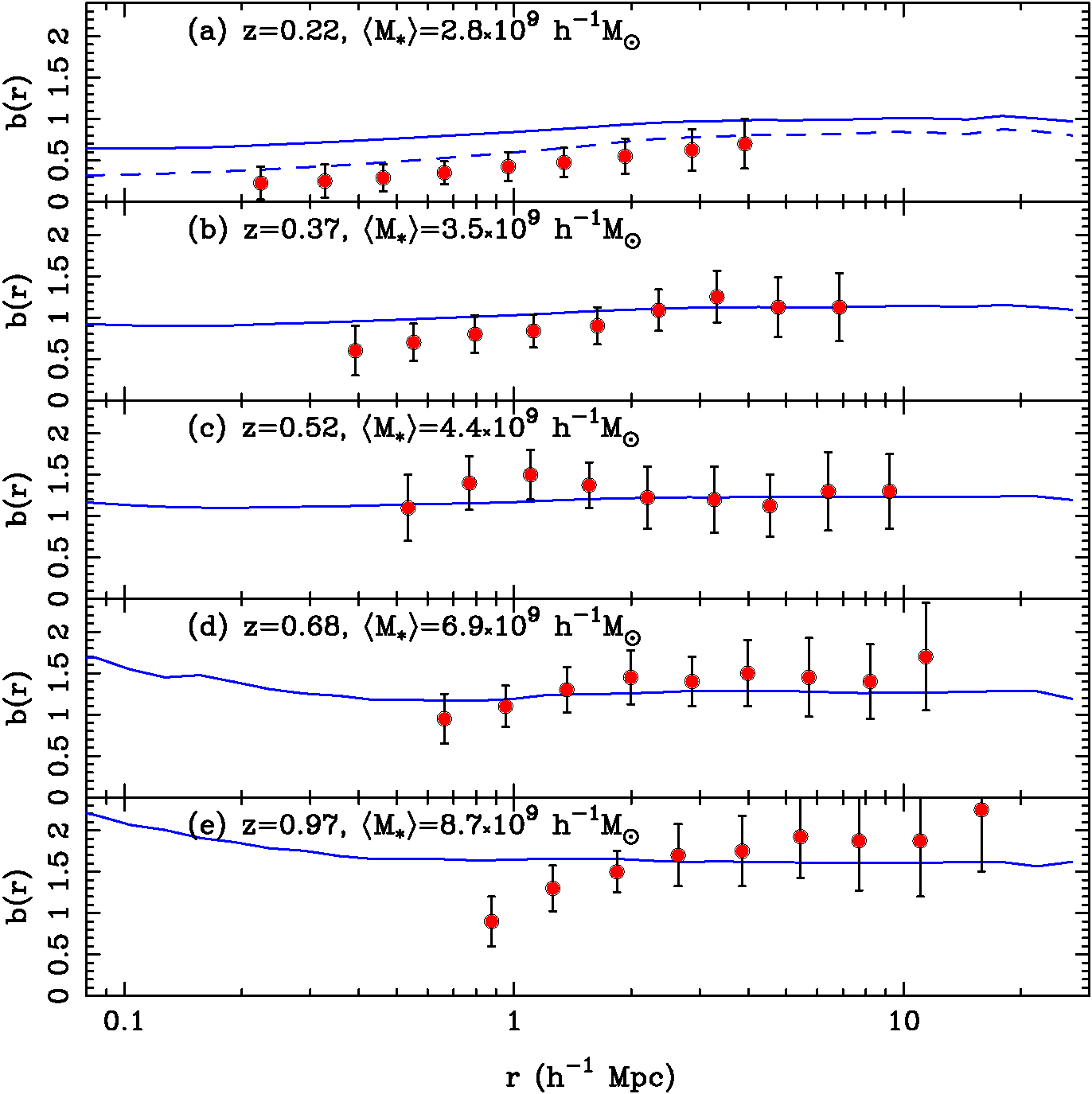}
  \end{tabular}
  \caption{The scale dependence of bias in the MBII simulation (blue
    lines) compared to the observational data of \citet{2012ApJ...750...37J}
    (points with error bars)
    for various samples with different mean stellar mass
    and redshift (given in the panels). The dashed line in panel (a) shows
    the results for an MBII sample with a
    mean stellar mass of $1.5 \times 10^{9} \msun$.}
\label{xij}
\end{figure}

\subsection{Comparison with observations}

We compare first
to the galaxy autocorrelation function published by the VIMOS Public
Extragalactic Redshift Survey
(VIPERS)
team (\citealt{2013A&A...557A..17M}).
The VIPERS survey is an ongoing deep and
well sampled
spectroscopic survey of 100,000 galaxies in the redshift range $z=0.5-1.2$
(see \citealt{2013arXiv1303.2623G} for details).
We use two of the three redshift bin measurements
published by \citet{2013A&A...557A..17M},
centered at $z=0.6$ and $z=1.0$. \citet{2013A&A...557A..17M}
give the linear bias parameter
for each redshift in a series of bins in
 stellar mass threshold (their table 3). We plot these
linear bias vs stellar mass points in Figure~\ref{vipers}.
Results from the simulation
are shown as lines
($b$ in this case is
the parameter $b_{\rm large}$ fit using equation \ref{bfiteq}).
From Figure~\ref{vipers} we can see that
the simulation and observations show the expected trend of increasing
bias with increasing galaxy stellar mass, and that the simulations are
consistent with the observational results.

Observationally
analyses such as \cite{2012ApJ...750...37J} have been able to
probe scale dependence of bias by comparing weak-lensing measurements
of the matter distribution with galaxy clustering.
In Figure \ref{xij} we show a quantitative comparison between the MBII
simulation and the data from figure 11 of \cite{2012ApJ...750...37J}
The \cite{2012ApJ...750...37J} measurements  were for 5 different
subsamples of observed galaxies
with different mean stellar masses and mean redshifts.
In order to match the
mean stellar masses
and redshifts of the \cite{2012ApJ...750...37J}
data samples we carried out a quadrilinear
interpolation in log mass and in redshift between the correlation
function results measured from the MBII simulation for different redshift
snapshots and mass bins.
The relevant redshifts and mean stellar masses for the different samples
are given in the panels of Figure
 \ref{xij}.

\begin{figure}
\begin{tabular}{c}
  \includegraphics[width=3.2truein]{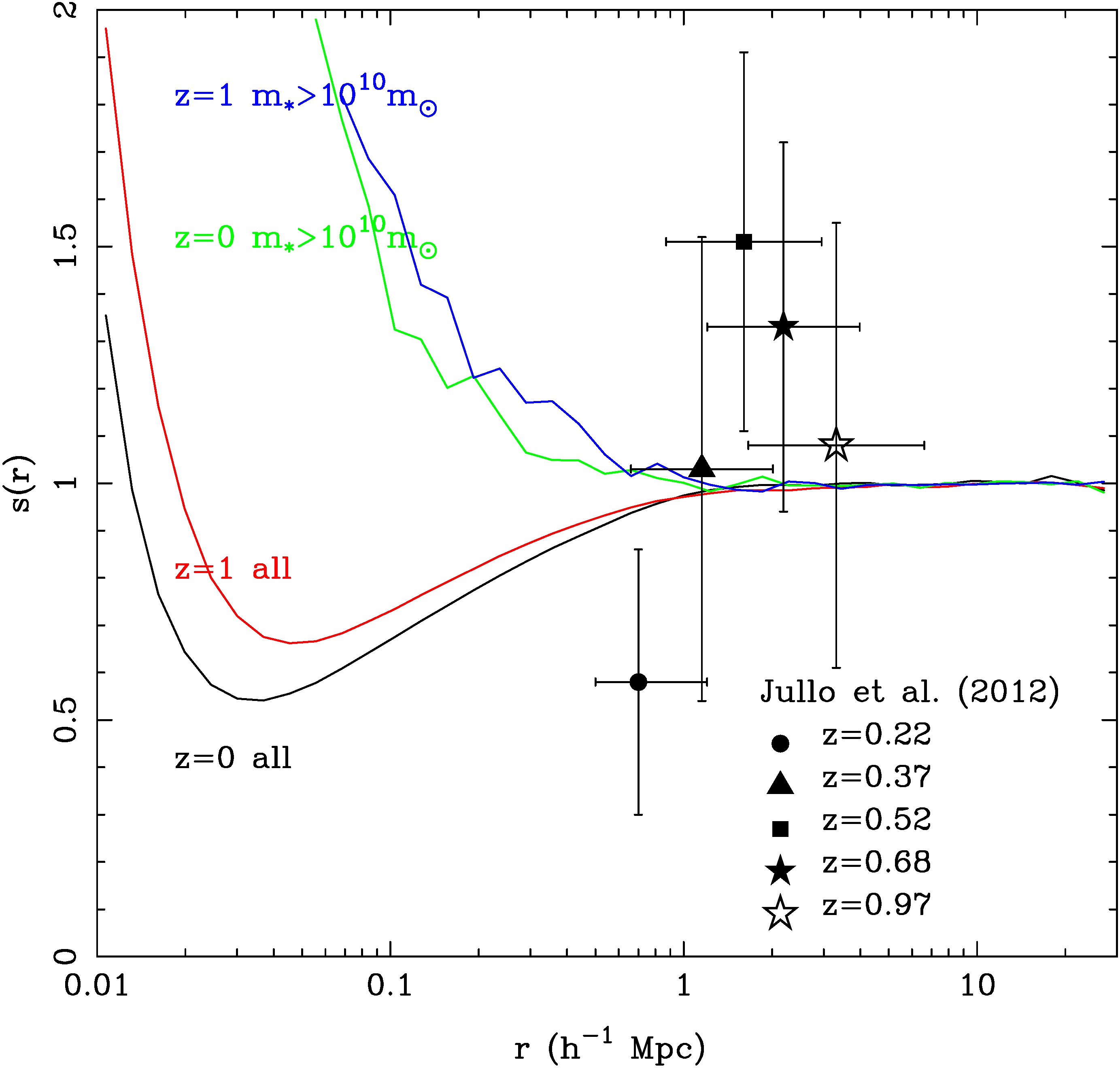}
\end{tabular}
\caption{Stochasticity versus radius. Results for different redshifts
and mass thresholds in the MBII simulation are shown
as colored lines. The points with error bars denote observational
determinations for galaxy stochasticity by \citet{2012ApJ...750...37J}.
The simulation redshifts and mass bins shown bracket those of the
observational data.
}
\label{jullo}
\end{figure}

Looking at Figure \ref{xij} we can see that observed data does show a
pronounced antibiasing ($b<1$) of galaxies with respect to dark matter
on small scales for many of the galaxy subsamples.
For the lowest redshift subsample (top panel) this is particularly
significant, given the small error bars. The MBII simulation data
exhibits this trend also, for the bins with low mass and low redshift.
The bias for the lowest redshift bin is systematically higher at
all scales in the simulations compared to the observations, however. In
order to show quantitatively how this relates to the mean
stellar mass of the subsample, for this panel, (a), we have also plotted the
results for a subsample with approximately half the mean stellar mass
($1.5 \times 10^{9} \msun$).

\begin{figure*}
\begin{tabular}{ccc}
  \includegraphics[width=2.33truein]{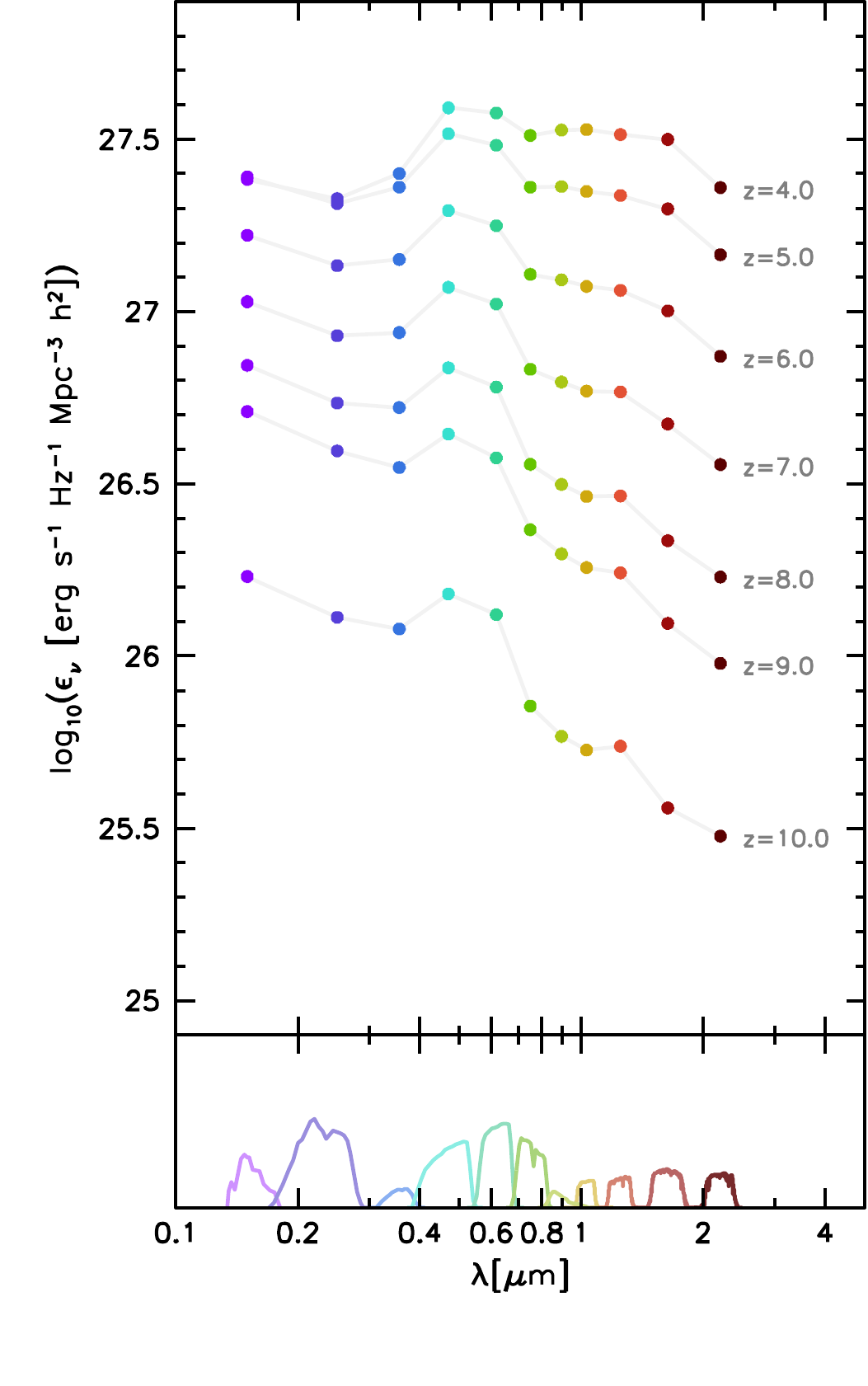}
  \includegraphics[width=2.33truein]{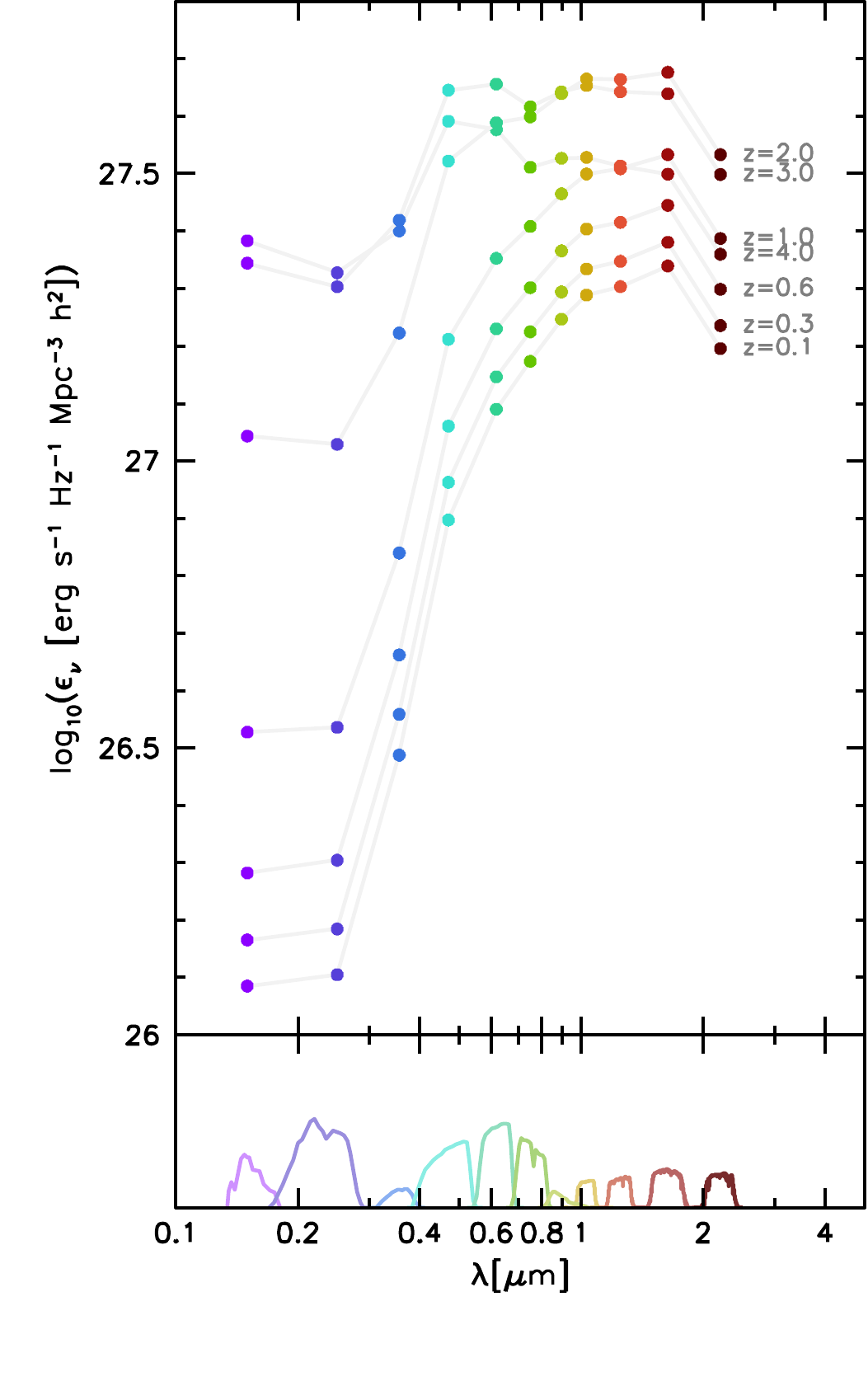}
  \includegraphics[width=2.33truein]{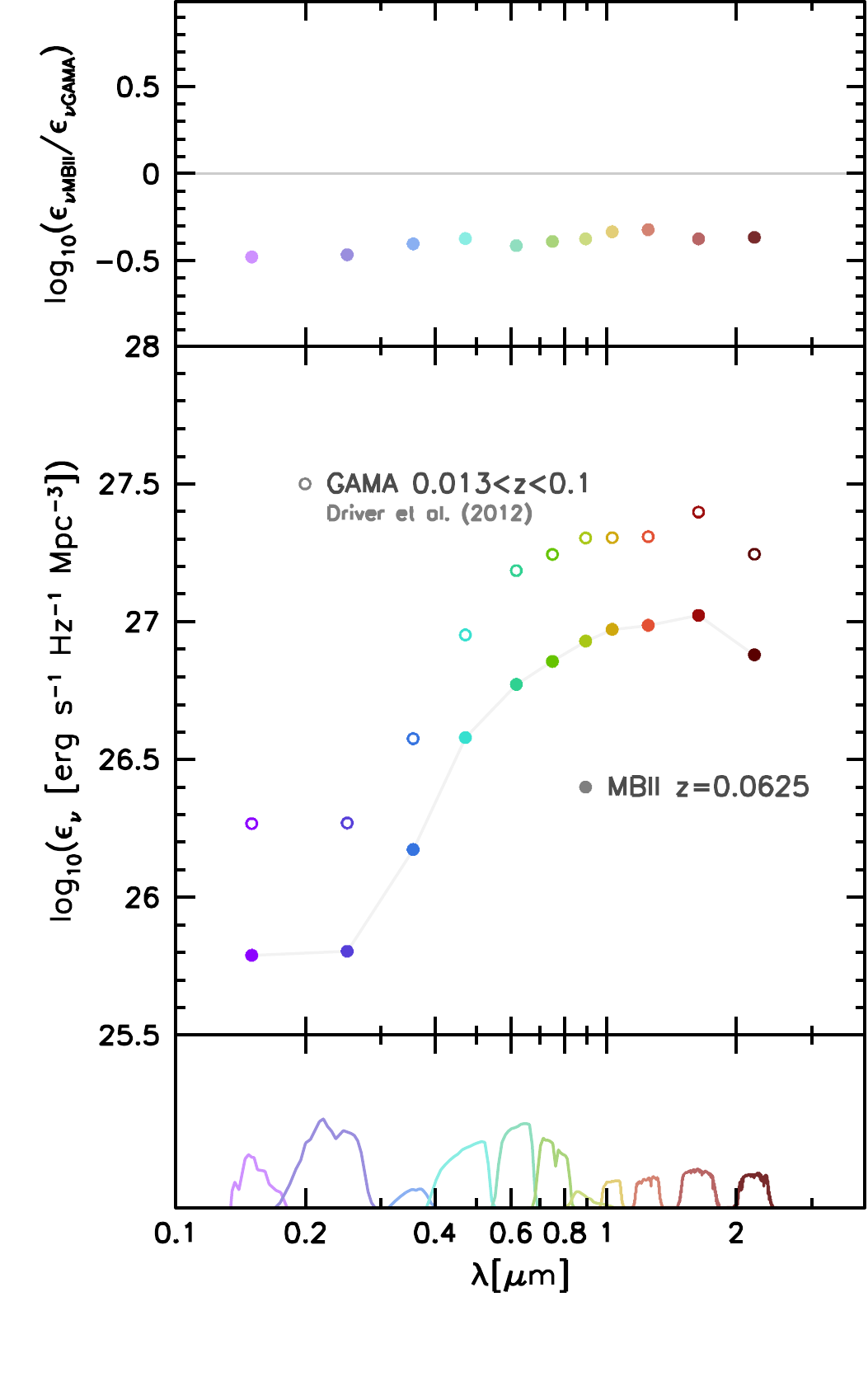} \\
\end{tabular}
\caption{The Evolution of the Cosmic Spectral Energy Distribution in MBII (left and center panels. Comparison is made at $z=0.0625$ with the GAMA survey at $z=0.05$ \citet{2012MNRAS.427.3244D}.}
\label{fig_csed}
\end{figure*}

\cite{2012ApJ...750...37J} have also searched for
for stochasticity in clustering by comparing weak-lensing measurements
of the matter distribution with galaxy clustering. \cite{2012ApJ...750...37J}
find
no significant amount of stochasticity on scales between $r=0.2 {\rm h^{1} Mpc}$
and $r=15 {\rm h^{1} Mpc}$  at redshifts between $z=0.2$ and $z=1$.

A comparison
between the MBII simulation results and those of
\cite{2012ApJ...750...37J} is shown in Figure \ref{jullo}. The observational
points are for a range of redshifts, with each redshift's measurement
being reliably inferred over a particular range of scales
(shown by the horizontal
error bars).  The different observational points also came from
different flux limited samples, which have a mean stellar mass
varying between $6\times 10^{9} \msun$/h and $1.8\times10^{10} \msun$/h.
We can see that the observational results are all
consistent with no stochasticity ($s=1$) at at least the $1.5 \sigma$
level. The simulation results are shown for redshifts and stellar mass
ranges which bracket the observational results. We can see that the
observational and MBII results are consistent, but at the scales $r>0.4 {\rm h^{1} Mpc}$
that are probed by \cite{2012ApJ...750...37J} we expect no significant deviation
from $s=1$. On smaller scales for the MBII,
we see  differences between the two samples
of different masses. The galaxy-exclusion effect mentioned above
means that $s(r)$ goes above 1 at smaller scales for smaller galaxies.

\section{Properties of galaxies}
\label{sec_galaxy}

The MB and MBII simulations have been very successful in reproducing the observed properties of galaxies in the high redshift Universe.
\citet{2013MNRAS.429.2098W} showed that the galaxy stellar mass function (GSMF) predicted by MB and MBII at $z \geq 5$
could be reconciled with observations if one assumed that the mass-to-light ratio (as predicted in MB and MBII) of these galaxies
was evolving with redshift. \citet{2012MNRAS.423.2397K} showed that the MB simulation reproduced the observed
properties of galaxies hosting the highest redshift quasars \citep{2007ApJ...666L...9C, 2010ApJ...714..699W, 2011AJ....142..101W}.
In this section we focus our attention on the properties of galaxies in the MBII simulation at $z < 4$.
We will compare general properties of galaxies with observations and leave a detailed analysis to future
publications.

We start by looking at the Cosmic Spectral Energy Distribution (CSED) in MBII.
We select subhalos using {\small SUBFIND} and consider only those
which have more than 100 dark matter particles.
We refer to these subhalos as galaxies for the rest of this section.
The spectral energy distribution (SED) of a galaxy is generated by summing the SEDs of each star particle in the galaxy.
The left and right panels of figure~\ref{fig_csed} show the evolution of the CSED in the MBII simulation from
$z=10$ to $z=0$. We find that the amplitude of CSED in all bands increases rapidly with decreasing redshift to $z=4$ with little change in shape.
This is expected and is in line with the behavior of the observed cosmic star formation rate (CSFR) (see figure~\ref{fig:SFRBHdensity}).
Observations find that the CSFR plateaus around $z\sim 3-4$ and declines rapidly at lower redshifts.

\begin{figure*}
  \begin{tabular}{cc}
    \includegraphics[width=3.2truein]{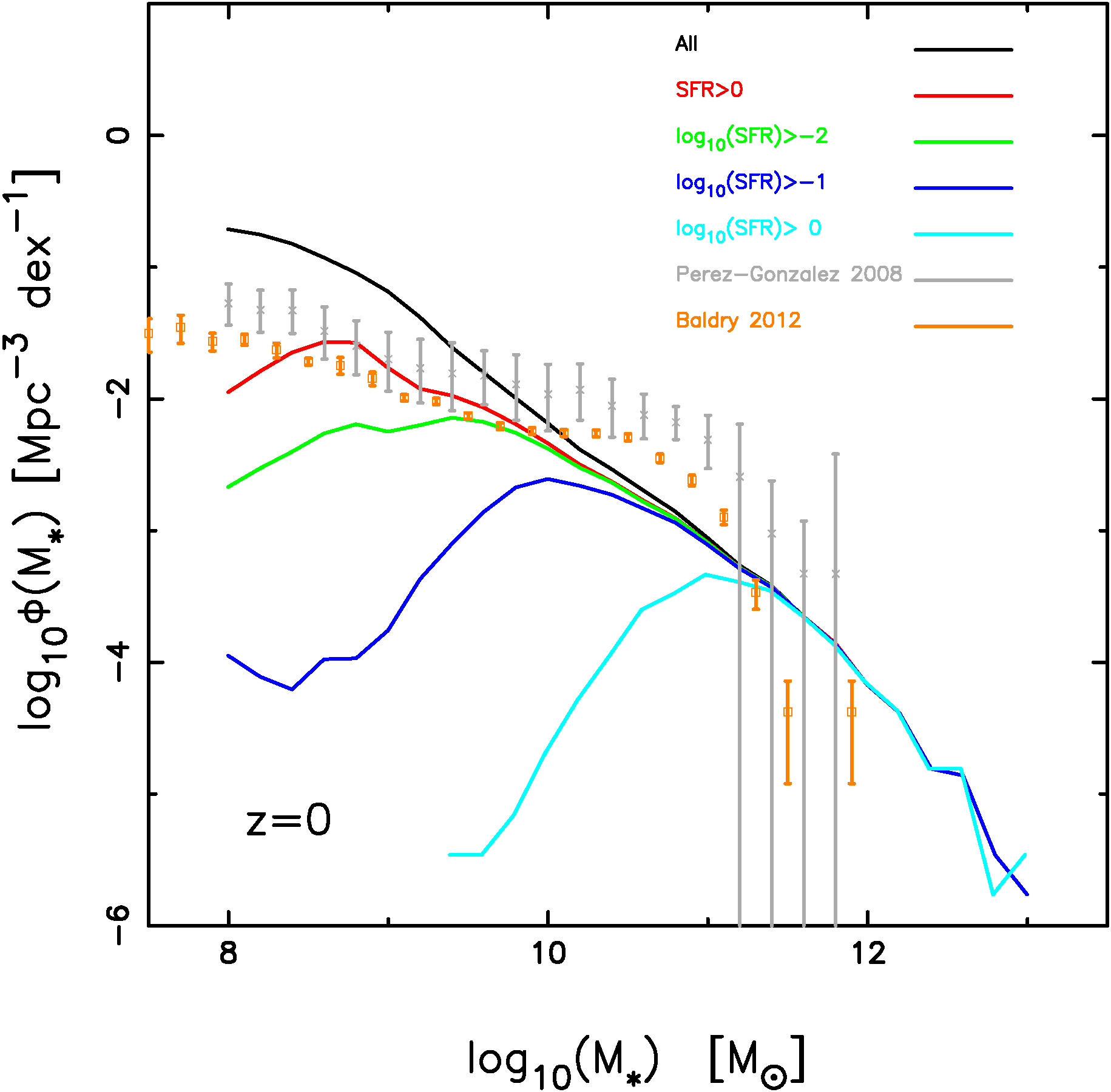}
    \includegraphics[width=3.2truein]{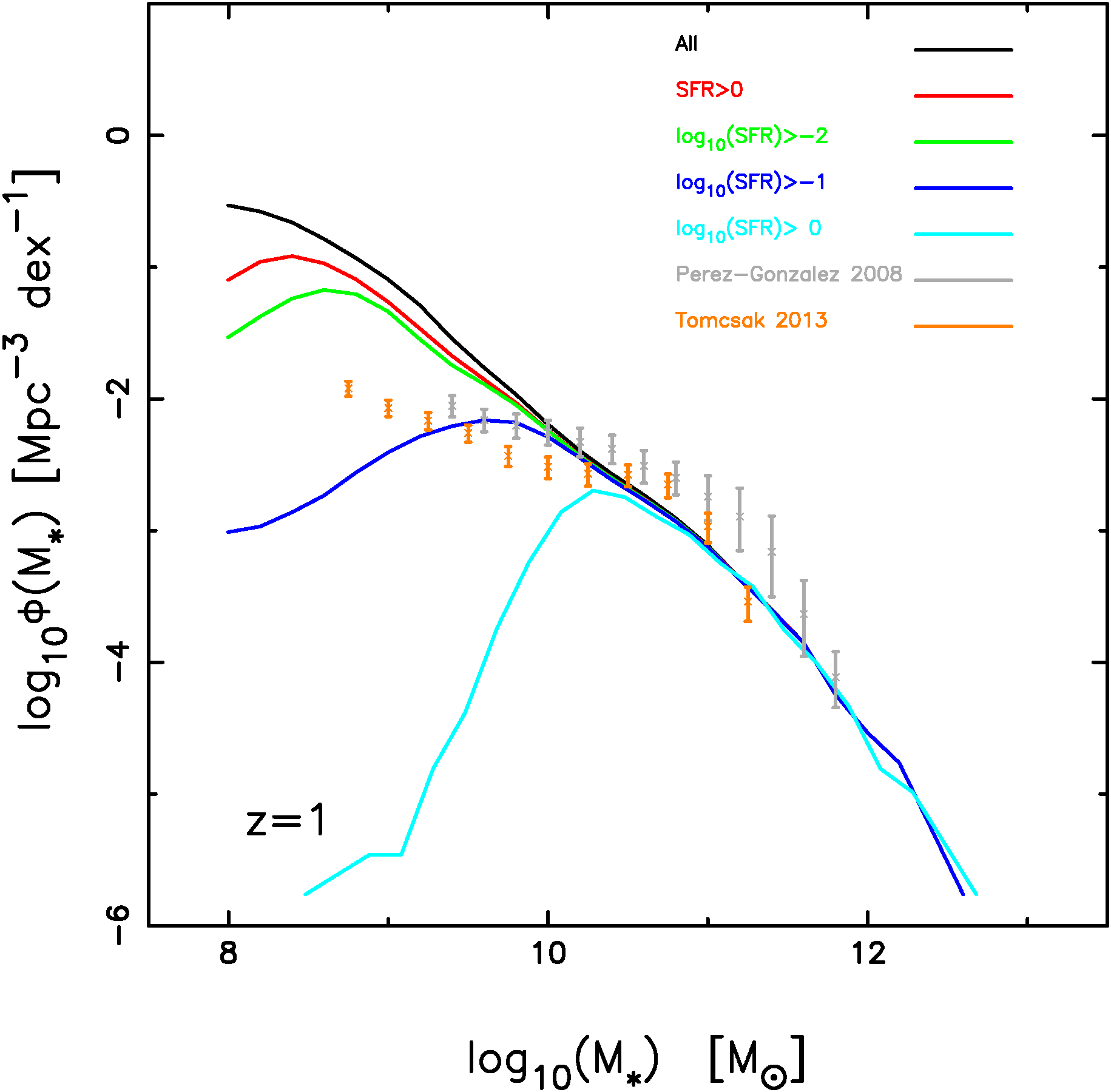} \\
    \includegraphics[width=3.2truein]{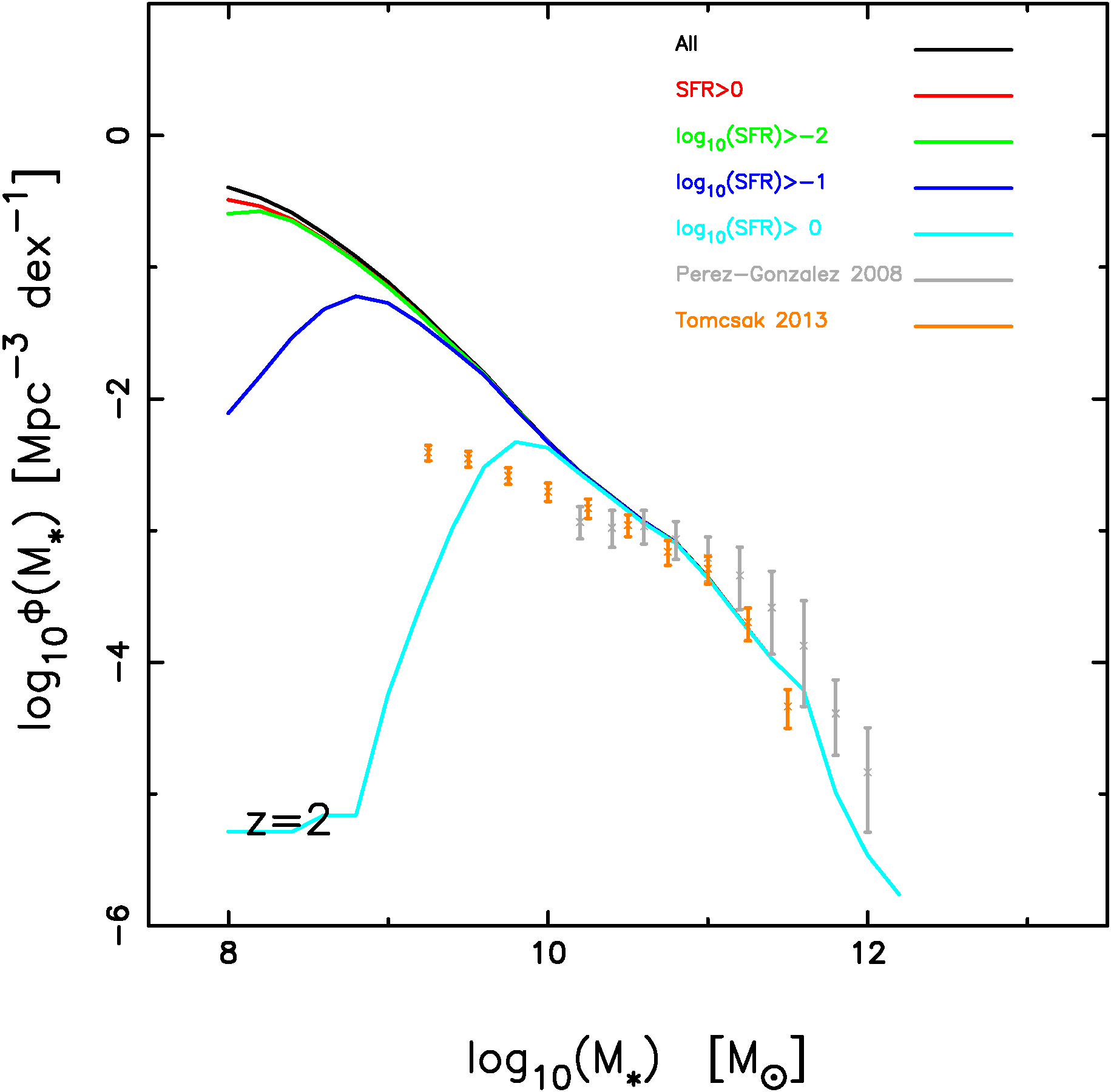}
    \includegraphics[width=3.2truein]{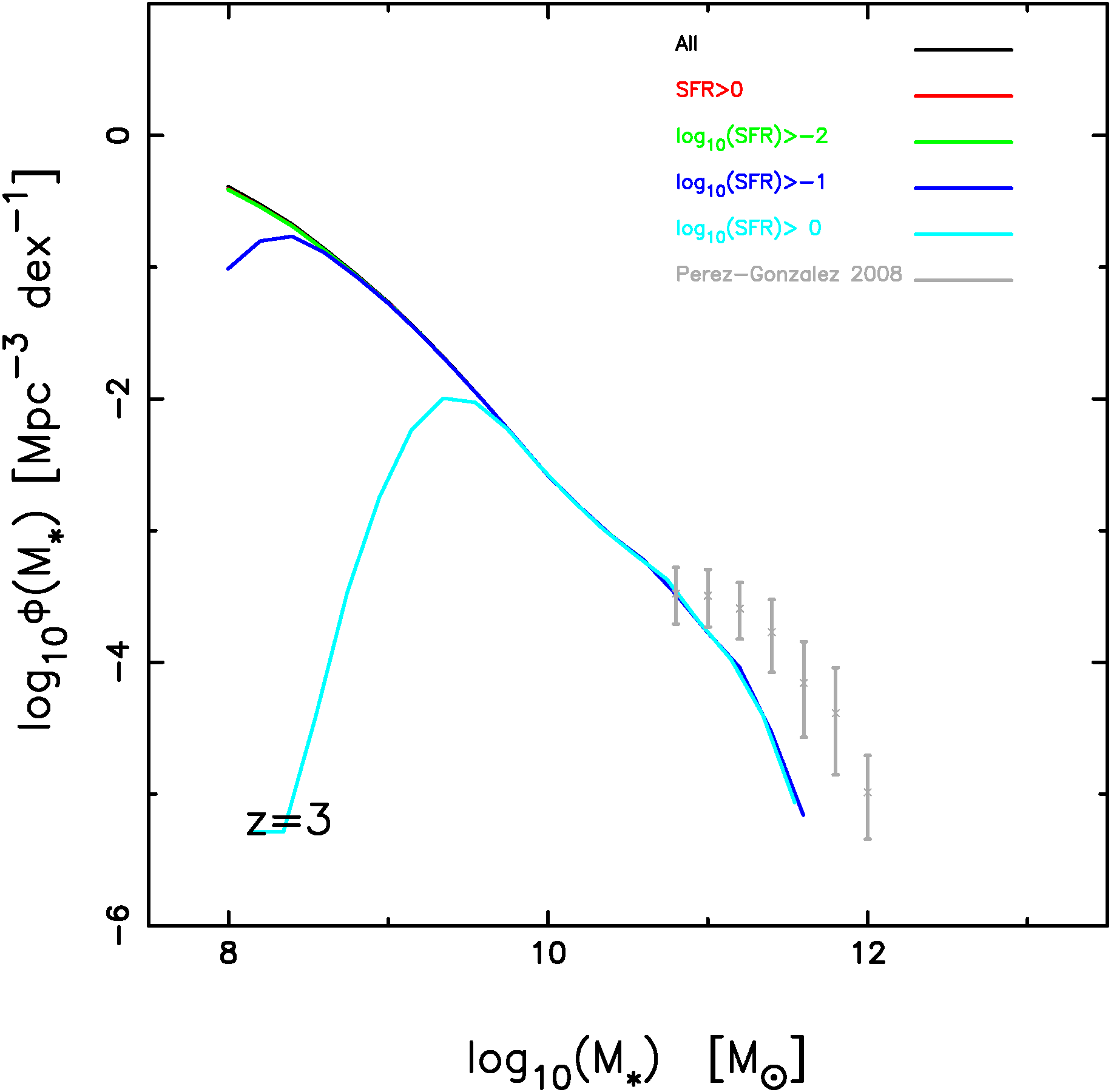} \\
  \end{tabular}
  \caption{The GSMF from $z=0-3$ in MBII.
    Comparison is made with observational estimates of \citet{2008ApJ...675..234P}
    (gray data points with error bars), \citet{2012MNRAS.421..621B} (z=0, orange data points with error bars).
    For $z=2$ and $z=3$ (orange data points with error bars) the observational estimates are taken
    the CANDELS survey \citep{2013arXiv1309.5972T}.
    The black line represents the GSMF in MBII when we consider all galaxies.
    The red, green, blue and cyan lines denote the population of
    galaxies which have a SFR greater than 0, 0.01, 0.1. and 1 respectively (in units of $\msun/$yr.)}
  \label{fig_gsmf_sfrcut}
\end{figure*}

The shape of CSED evolves dramatically below $z\lsim 3$.
We find that the bluer part of the CSED, which strongly correlates with the CSFR starts to decline below $z\lsim 3$
consistent with the observational trend, whereas the redder part of the CSED increases very slowly with decreasing redshift to $z\sim 1$ and declines thereafter.
This is because at these redshifts galaxies are forming new stars at much reduced rates and are passively evolving into a redder population.

In the right-hand panel of figure~\ref{fig_csed}
we compare the intrinsic CSED at $z=0.0625$ with dust-corrected
observations from the GAMA survey at $0.013<z<0.1$ \citep{2012MNRAS.427.3244D}.
We find that the shape of the CSED compares well with the observational
results. However, the amplitude of the CSED
predicted by MBII does not match that of observations,
falling systematically below it.
This will be in part caused by incompleteness as
MBII does not resolve all galaxies (particularly those at low stellar masses).
This discrepancy is also sensitive to the
initial mass function (IMF) assumed in the
processing of MBII.
Assuming an IMF in which a larger fraction of the mass
is converted into high-mass stars
\citep[such as those proposed by][]{2001MNRAS.322..231K,2003PASP..115..763C}
would increase the luminosity density bringing MBII more closely inline with the observations

\begin{figure}
  \begin{tabular}{c}
    \includegraphics[width=3.2truein]{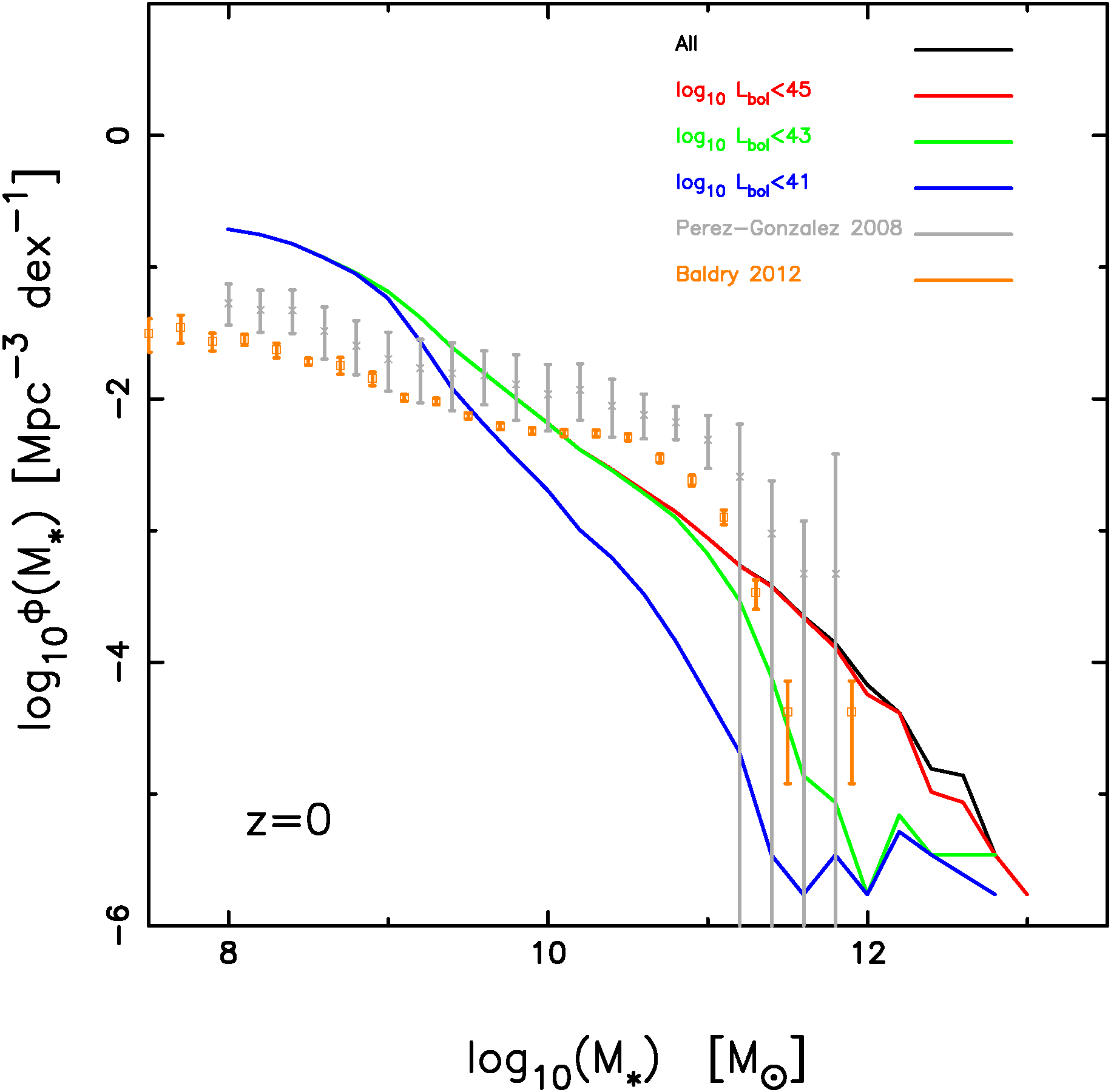}
  \end{tabular}
  \caption{Same as in figure~\ref{fig_gsmf_sfrcut} but we now consider the
    the population of galaxies at $z=0$ which may host an AGN
    with bolometric luminosity, $\log_{10}(L_{bol}) < 45$ (red),   $\log_{10}(L_{bol}) < 43$ (green)
    and  $\log_{10}(L_{bol}) < 41$ (blue). }
  \label{fig_gsmf_lbolcut}
\end{figure}

We now look at the galaxy stellar mass function (GSMF) predicted in MBII in figures~\ref{fig_gsmf_sfrcut} and \ref{fig_gsmf_lbolcut}.
In figure~\ref{fig_gsmf_sfrcut} the GSMF is compared with observational estimates.
The black line represents the GSMF in MBII when we consider all galaxies.
The red, green, blue and cyan lines denote the population of
galaxies which have a SFR greater than 0, 0.01, 0.1. and 1  respectively (in units of $\msun/$yr.)
Comparison is made with observations of \citet{2008ApJ...675..234P} (gray data points),
\citet{2012MNRAS.421..621B} (orange data points (z=0)).
For $z=2$ and $z=3$ (orange data points) the observational estimates are taken
the CANDELS survey \citep{2013arXiv1309.5972T}.
At $z=0$ MBII overpredicts the GSMF both at the high and low mass ends.
However the GSMF agrees well with the observations for $\mstar \geq 10^{10}\msun$/h for $z=1$ and $z=2$
At $z=3$ MBII underpredicts the abundance of larger mass galaxies.
This is most likely due to the finite volume of MBII.

We find that the amplitude of the GSMF of \citep{2013arXiv1308.0333H} is larger
compared to MBII although the shape seems to
be in reasonable agreement. The boxsize, star formation and feedback model in
MBII and \citet{2013arXiv1308.0333H} are similar. Therefore given that the mass resolution of
\citep{2013arXiv1308.0333H} is $\sim \times  60 $ lower than MBII
and based on the resolution tests carried out in \citet{2014MNRAS.tmp...38T} the amplitude
of the GSMF in \citep{2013arXiv1308.0333H} should be lower compared to MBII.

One of the striking feature at all redshifts is the steep slope in the GSMF in MBII at $\mstar \leq 10^{10}\msun$/h.
These galaxies are less affected by AGN feedback and are therefore more sensitive to the star formation
and stellar feedback model. This feature is seen across all redshifts in figure~\ref{fig_gsmf_sfrcut}.
For example in MBII we find that
there are many more lower mass galaxies which have zero star formation (i.e. the difference between the black and red lines) at $z=0$
as compared to higher redshifts, the difference between the two decreasing with increasing redshift.
We therefore need to understand why do small galaxies form rapidly so early and why do they stop forming stars later.
A better treatment of the star formation  and stellar feedback model is therefore required
in order to suppress the overproduction of lower mass galaxies.
For example our model does not include the treatment of molecular gas \citep{2011ApJ...729...36K}
which would tend to suppress star formation rates in lower mass galaxies. Alternately one may need to assume
a feedback model which is dependent on the mass of the galaxy
\citep{2006MNRAS.373.1265O,2011MNRAS.415...11D,2013MNRAS.436.3031V, 2014MNRAS.tmp...38T}.
The variable wind model which is dependent on the galaxy velocity dispersion
is described in \citet{2006MNRAS.373.1265O} and indeed flattens the GSMF at $z=0$ better reproducing observations. However
\citet{2014MNRAS.tmp...38T} find that the GSMF is still steep at higher redshifts and additional modeling
may be required to suppress the production of stars in low mass galaxies.
Interestingly we find that if we account for those galaxies which have non-zero star formation at $z=0$
the lower mass end of the GSMF is in better agreement with observations.

In figure~\ref{fig_gsmf_lbolcut} we look how the AGN population affects the GSMF at $z=0$.  We consider the
the population of galaxies which may host an AGN with bolometric luminosity in units of erg/sec,
$\log_{10}(L_{bol}) < 45$ (red),   $\log_{10}(L_{bol}) < 43$ (green) and  $\log_{10}(L_{bol}) < 41$ (blue)
We focus our attention on larger mass galaxies at $z=0$ since they are most affected by AGN feedback.
We find that the tail of the GSMF is in reasonably good agreement with observations down to $z\geq 1$ (see figure~\ref{fig_gsmf_sfrcut})
and one does not need to consider a subsample of galaxies without bright AGNs (green and blue curves) to match observations.
We find that the stellar mass in galaxies that host AGNs brighter than $\log_{10}(L_{bol}) = 43$ is overpredicted in MBII suggesting
insufficient quenching / AGN feedback. This is also seen in the results of \citet{2013arXiv1308.0333H}.
\citet{2014MNRAS.tmp...38T} on the other hand reproduce the tail of the GSMF reasonably well although it maybe a result
of missing cluster sized halos in their simulation volume of $L_{box} = 25$Mpc/h. \citet{2014MNRAS.tmp...38T} also define
the galaxy stellar mass to be the sum of the stellar mass within twice the half mass radius. Such a definition affects only
larger objects as it gets rid of the stellar mass in the diffuse intracluster medium and is not traditionally counted
as contributing to the central galaxy's mass. Such a definition may help in better bringing in line our results
with observations but as we will discuss in the next section the bright end of the QLF is still overestimated due to insufficient
AGN feedback.

\section{Blackholes}
\label{sec_bh}

In this section we present some basic properties of our simulated black hole
population and their relation to the galaxies in MBII. In particular
we show overall history of the black hole mass assembly and
look at the relation between black hole and stellar mass in galaxies.
We look at the predictions for the bolometric luminosity function
and clustering strength as a function of luminosity for
the AGN population in MBII. More detailed analysis of the blackhole-galaxy
relations and comparisons with observational constraints will be presented in
a separate paper.

\begin{figure*}
\centering
\hbox{
\includegraphics[width=3.2truein]{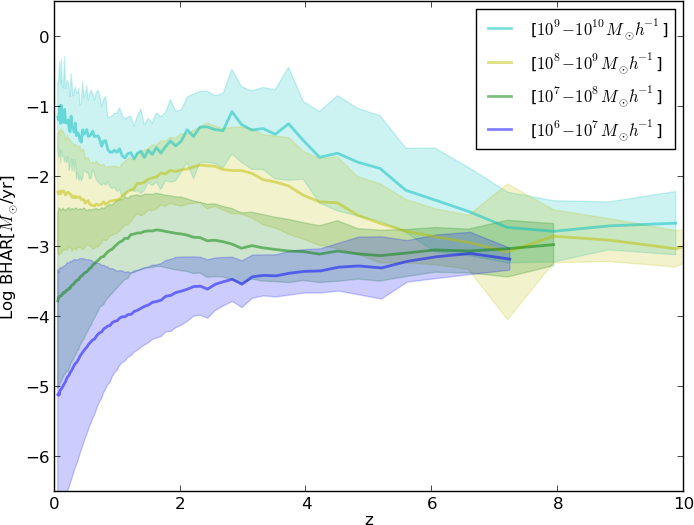}
\hspace{0.4cm}
\includegraphics[width=3.2truein]{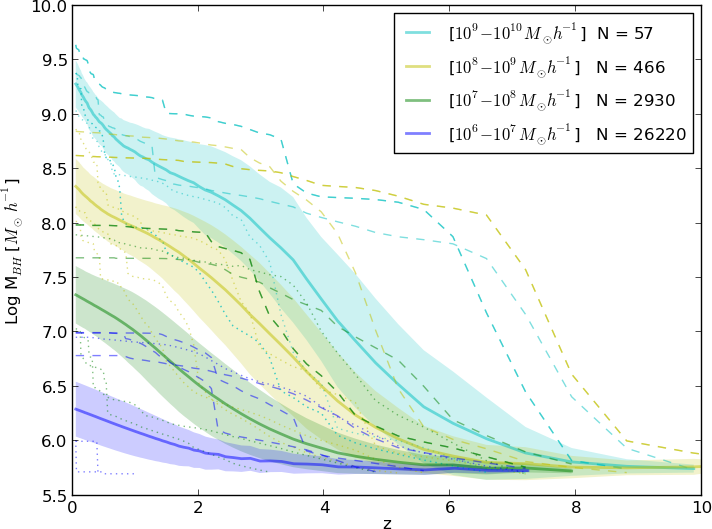}}
\caption{The mean black hole accretion rate (left panel)
and the mean black hole mass assembly history for the main progenitor
 (right panel)  black holes ending up in different mass ranges
(as labeled) at $z=0$. $N$
indicates the number of black holes in each of the mass bins. The dashed
and dotted lines show a sample of single main progenitor histories.}
\label{fig:bhmass_history}
\vspace{-0.5cm}
\end{figure*}

Every single black hole in our simulations accretes and grows
according to local gas properties so as an outcome of our black hole
model, each single black hole that has a lightcurve and an associated mass
history over the cosmic time since it is seeded.  MBII
contains tens of thousands of blackholes and
Figure~\ref{fig:bhmass_history}, we show the mean (and associated dispersion,
in the corresponding colored areas) blackhole mass assembly history for
blackholes that, at $z=0$, end up in different mass bins, from the the lowest
masses ($10^6 \msun/{\mathrm{h}} < M_{BH} < 10^7 \msun/{\mathrm{h}}$)
to the highest mass bin ($10^9
\msun/{\mathrm{h}} < M_{BH} , 10^{10} \msun/{\mathrm{h}}$).
In addition to the mean histories for
 the population
we also specifically show a sample of single, main progenitor mass assembly
histories. This is to illustrate how the most massive,
and earliest growing black holes, within the different mass bins, depart from
the mean (dominated by the more numerous lower mass blackholes).
In general we see that the
largest black holes $z=0$ are largely assembled at high redshifts. The
dispersion in mass assembly histories is also typically larger for the high
mass black holes. This is because there is a much larger variety of assembly
histories than for the lower masses. For example, some black holes form early
and grow to large masses quickly at high-$z$, but their mass assembly history
remains flat subsequently. We also note an upturn in the growth histories of
massive black hole at $z < 0.2$. This is likely the result
of insufficient AGN feedback (in the form that
is modeled in MBII) in massive galaxy hosts at low redshift.
Similar results were also found by recent work of \citet{2013arXiv1308.0333H}.

\begin{figure*}
\begin{tabular}{cccc}
  \includegraphics[width=1.6truein]{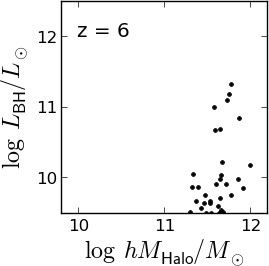}
  \includegraphics[width=1.6truein]{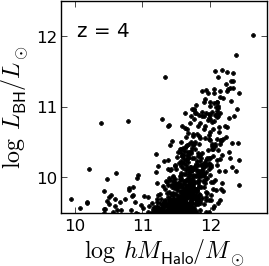}
  \includegraphics[width=1.6truein]{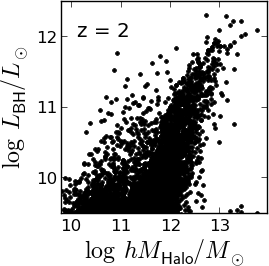}
  \includegraphics[width=1.6truein]{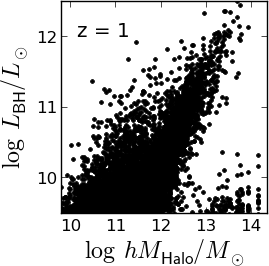}\\
  \includegraphics[width=1.6truein]{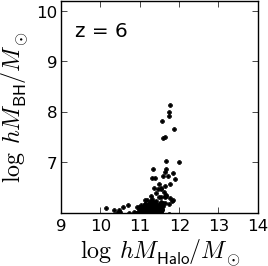}
  \includegraphics[width=1.6truein]{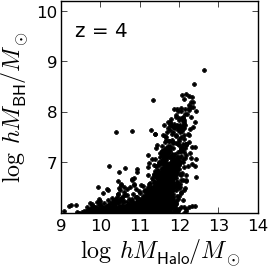}
  \includegraphics[width=1.6truein]{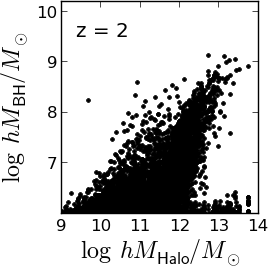}
  \includegraphics[width=1.6truein]{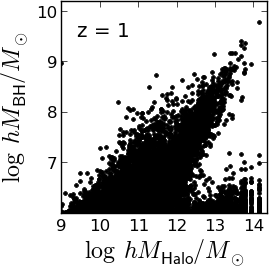}\\
\end{tabular}
\caption{The relation between blackhole luminosity and host halo mass (top)
  and blackhole mass and host halo mass (bottom).}
\label{fig_bhmass_bhlum}
\end{figure*}

To illustrate the range of black hole masses and luminosities in MBII we show
these two quantities as a function of total halo mass in our galaxies in
Figure~\ref{fig_bhmass_bhlum}. Although there is an overall correlation
between both quasar luminosity and black hole mass with halo mass the scatter
in both these relations is rather large indicating the halo mass is an
extremely rough proxy to black hole properties. Complex hydrodynamic and
associated feedback effects play an important role in the central
region of galaxies.

\begin{figure*}
\centering
\hbox{
\includegraphics[width=3.2truein]{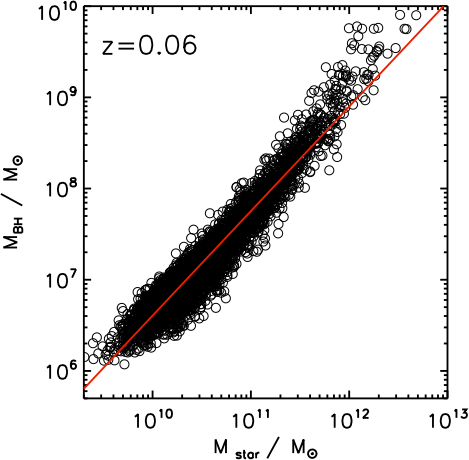}
\hspace{0.4cm}
\includegraphics[width=3.2truein]{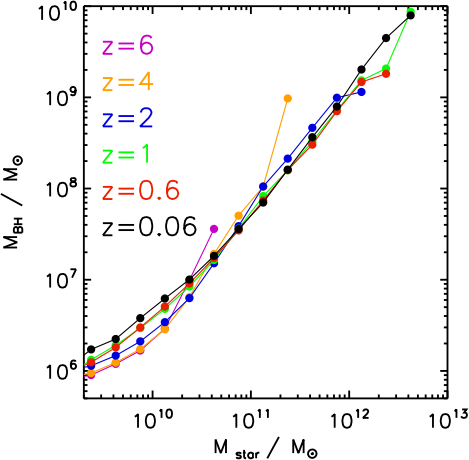}}
\caption{Left panel: present day blackhole mass stellar mass relation in MBII.
Right panel: the evolution of the relation from $z=0$ to $z=6$. }
\label{fig:mmstar}
\vspace{-0.5cm}
\end{figure*}

In Figure~\ref{fig:mmstar} we show the prediction for the
 relationship between black hole
mass and stellar mass in galaxies in our simulation. In the right panel, the
relation is shown for the $z=0$ blackhole population. For simplicity here we
use the total stellar mass and not only the bulge mass which is what is
normally used in the local universe (and hence show only groups with $M_{*} >
10^{10} \msun$; in future work we will look at the associated black hole-
stellar velocity dispersion relation).  In
 the left panel of Figure~\ref{fig:mmstar}
we show the evolution of the mean relation
derived from a number of snapshots from MBII. As in previous work
\citep{2008ApJ...676...33D, 2011MNRAS.413.1158B, 2013arXiv1308.0333H}
we find good agreement between
simulations and observations. Although we do not find strong evolution in the
relation with redshift the relation appears to steepen slightly toward higher
redshift. This is overall consistent with previous findings and
observation constraints \citep{2010ApJ...708.1507B, 2010ApJ...708..137M}
and we defer detailed comparison with different observational
constraints as a function of redshift to a future paper \citep{DeGraf14}.

\begin{figure}
\centering
\includegraphics[width=3.2truein]{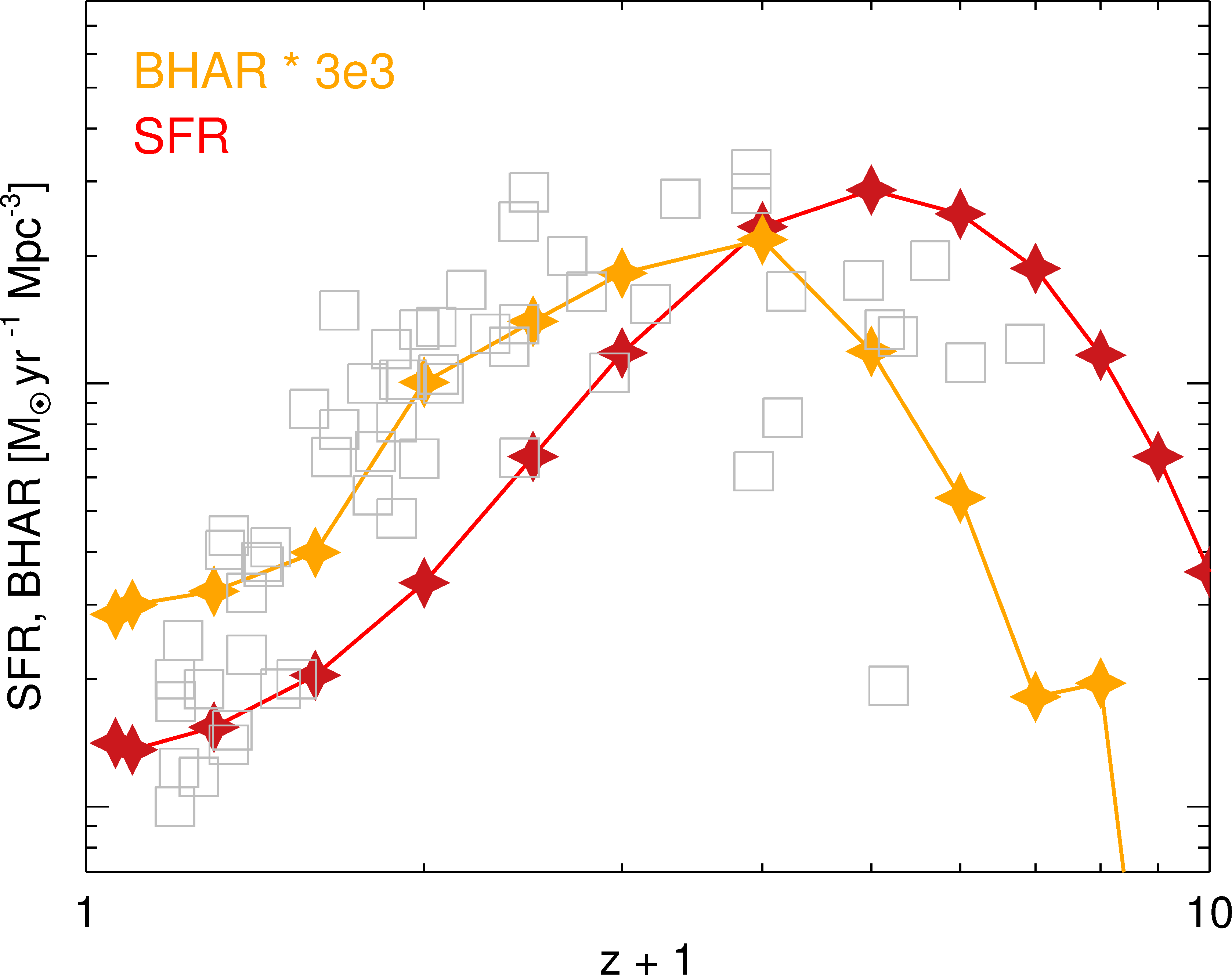}
\caption{The star formation rate (SFR, in red) and black hole accretion rate
  (BHAR, in orange, multiplied by $3 \times 10^3$ for comparison) density
  prediction from MBII. The observation constraints for the SFR density from
  \citet{2006ApJ...651..142H} are shown in gray squares (without error bars for
  clarity).}
\label{fig:SFRBHdensity}
\vspace{-0.5cm}
\end{figure}

The global star formation rate and black hole accretion rate density
(multiplied by a factor of $3\times 10^{3}$)  is shown
in Figure ~\ref{fig:SFRBHdensity} together with the observational
compilation of \citet{2006ApJ...651..142H} for the star formation rate density
derived from different wavebands. Overall the star formation rate and black
hole accretion rate density have similar shapes and track each, with the peak
of star formation proceeding that of BHAR density by roughly a unit redshift.
Both the star formation rate density (and also the BHAR density) show a
flattening below $z \sim 1$. This is likely a result of too inefficient
feedback to quench both SFR and BHAR in low redshift high mass halos.

\begin{figure*}
\begin{tabular}{c}
  \includegraphics[width=6.4truein]{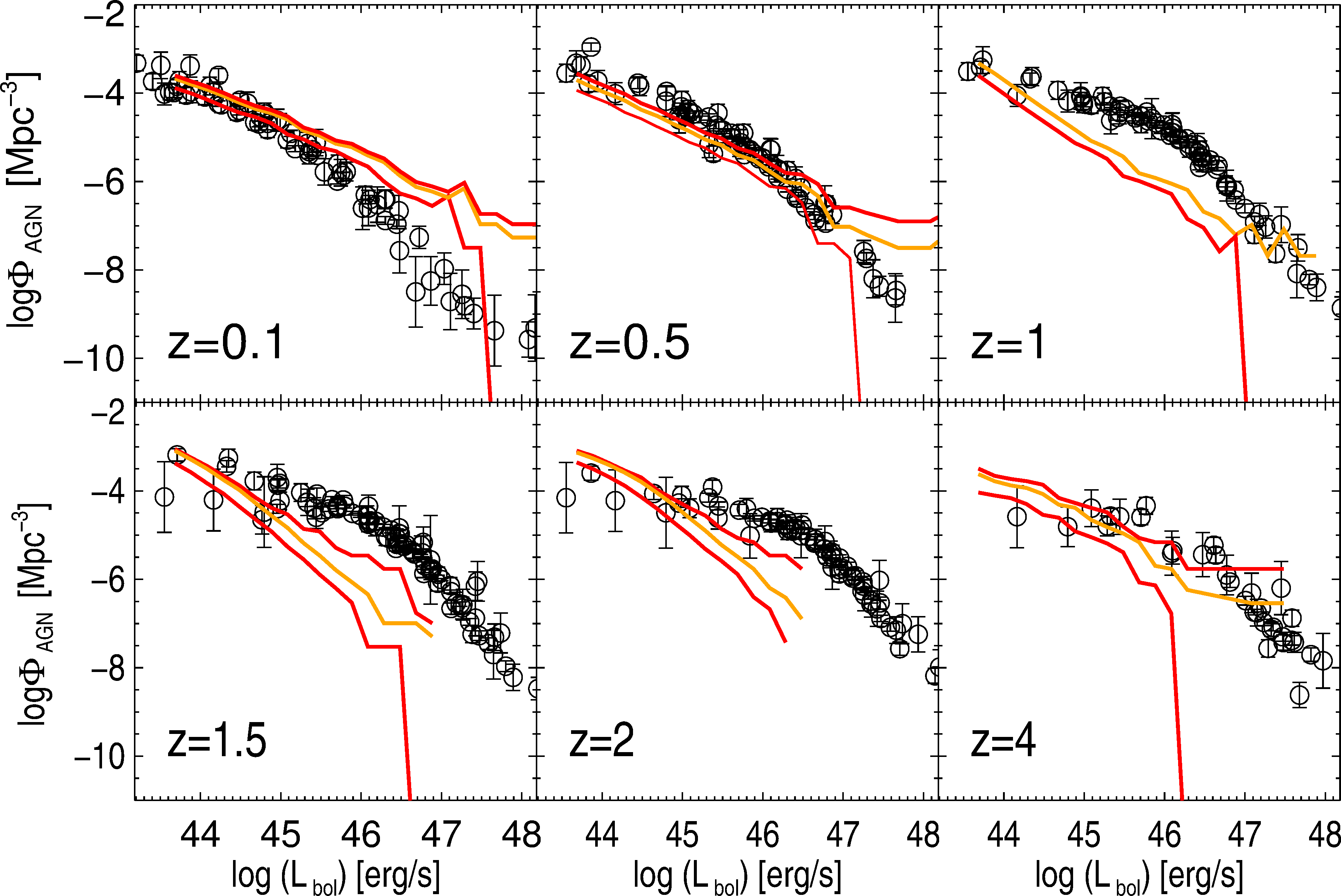}
\end{tabular}
\caption{MBII Predictions for the quasar luminosity function (QLF) compared to
  the \citet{2007ApJ...654..731H} data compilation at different redshifts.}
\label{fig_qsolf}
\end{figure*}

In figure~\ref{fig_qsolf} we show the bolometric quasar luminosity function
(QLF) compared to the compilation of data from \citet{2007ApJ...654..731H},
at $z=0.1,
0.5, 1, 1.5, 2$ and $4$. We note that $z > 5$ predictions from MBII (and MB)
are presented in \citet{2012MNRAS.424.1892D, 2013ApJ...768..105M}
and predictions for $z=2.0, 2.4, 3.2$
compared to the most up-to-date BOSS QLF in \citet{2013ApJ...773...14R}.
In general, we find overall good agreement
between the simulations and observations at
$z>2$ (here and in
\citet{2012MNRAS.424.1892D, 2013ApJ...768..105M, 2013ApJ...773...14R}
and $0.5 < z < 1$.
At the lowest redshifts ($z=0.1$), however, the bright end of the QLF
is overestimated by the simulations. As we discussed earlier in the Section,
this indicates insufficient quenching / AGN feedback at these redshifts.
Conversely, in the redshift range between $z=1.5 -2$ the bright end of the QLF
is underestimated \citep[see also][]{2013ApJ...773...14R}.
This would indicate that our
peak of the BHAR density occurs somewhat too early (see
Fig.~\ref{fig:SFRBHdensity}). Again this may have to do with the details AGN
feedback. Given the simple model adopted here, these results suggest that a
constant $f$ (our feedback energy parameter) may be too simplistic a model for
AGN feedback.
A redshift evolution of $f$ as also implemented by \citet{2013arXiv1308.0333H}
may alleviate some of these issues (however, even in those simulations
the bright end (at low-$z$) is overestimated indicating that some more
complicated modeling may be necessary).

\begin{figure}
\begin{tabular}{c}
  \includegraphics[width=3.2truein]{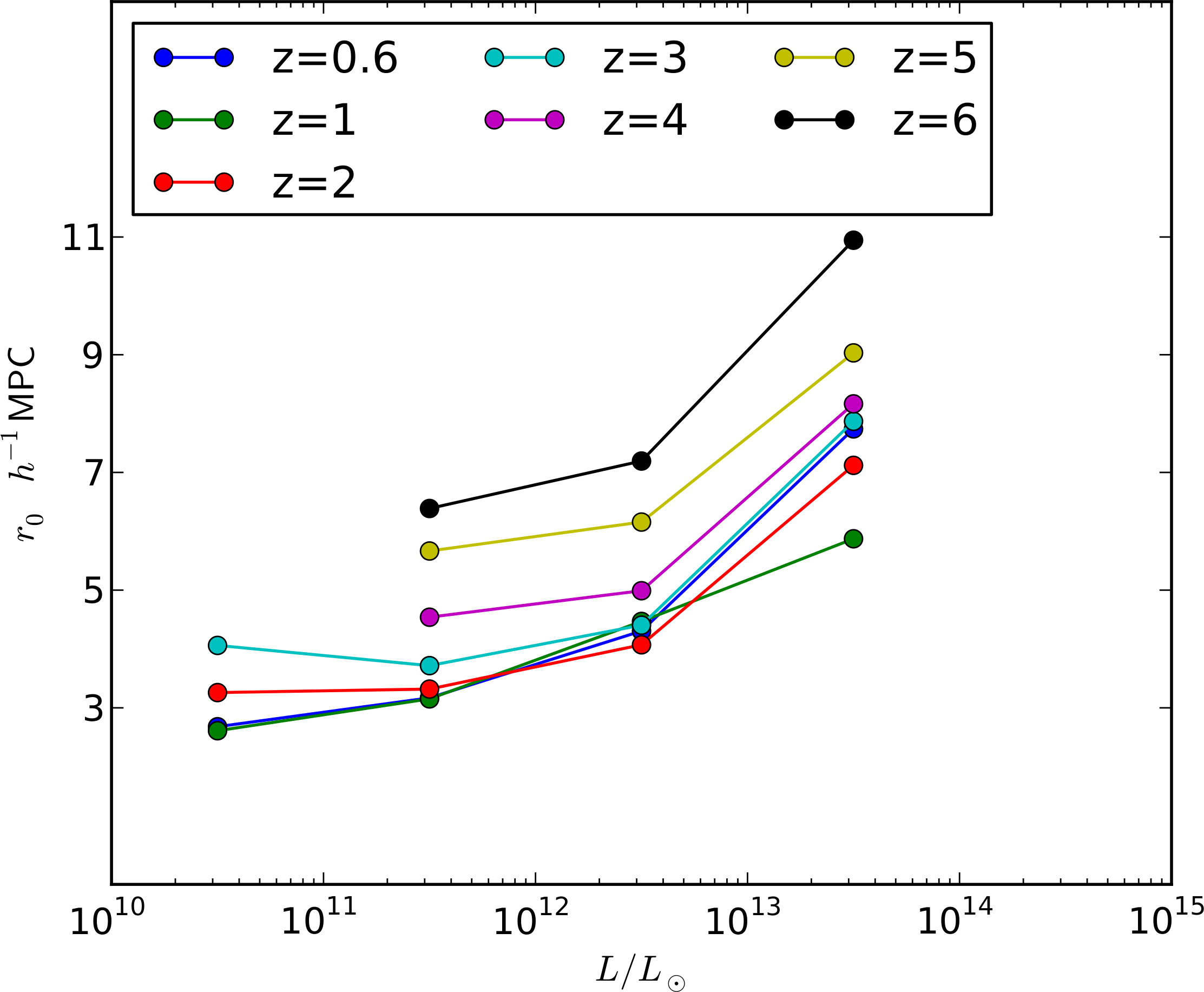}
\end{tabular}
\caption{The correlation length, $r_{0}$ measured for the quasar population as
  a function of quasar luminosity plotted for different redshifts (as
  labeled).}
\label{fig_qsor0vsz}
\end{figure}

Finally we briefly present the MBII prediction for the correlation length,
$r_0$ of AGN as a function of redshift and luminosity in
Figure~\ref{fig_qsor0vsz}.
In general the correlation lengths are in agreement with observational
constraints~\citep{2009ApJ...697.1656S}.
The correlation length, at constant luminosity,
increases as a function of redshift. This is expected if same luminosity AGN
are hosted in similar mass halos at different redshifts. In addition, there is
a luminosity dependence in the clustering and $r_{0}$ increases by a factor of
$2-3$ at the highest luminosities.

\section{Conclusions}
\label{sec_conc}

In this paper we have examined a variety of standard predictions
using the recently completed MBII simulation. In particular, the results presented
here range from the clustering of halos and galaxies, HOD, the mass function of
halos to basic properties of galaxies and AGNs.

Our main conclusions are:

\begin{itemize}

\item We find that halos with masses $M_{halo} \geq 10^{13}\msun/h$
have a baryon fraction close to 80\%-90\% of the cosmic mean. The baryon
fraction decreases steadily with decreasing halo mass.
Our results are broadly consistent with \cite{2007MNRAS.377...41C}
for cluster sized halos but differ significantly for lower mass halos.
The discrepancy can be attributed to feedback both from AGN
and supernovae which were not been modeled in \citet{2007MNRAS.377...41C}.

\item We find that the FOF halo mass function
(where halo mass includes both dark matter and baryonic
components) in MBII can be fit with a universal
form (valid for all redshifts) at the $\sim 13\%$ level.

\item One of the most striking results predicted by MBII is
the behavior of the halo mass function which shows a strong suppression
in the abundance of halos below the knee of the mass function ($\ln(1/\sigma)
\lsim -0.2$ which corresponds to $\mhalo \sim 10^{13.2}\msun$/h at $z=0$ and
$\mhalo \sim 10^{9.5}\msun$/h at $z=3$)
when compared to dark matter only predictions of the halo mass function.
This is due to the significant impact of
baryonic processes which tend to suppress
the mass of the halo by up to 30\% \cite{2013MNRAS.431.1366S}.

\item Fits to the mass function from dark matter only
simulations overpredict the  mass function at the 20-35\% level below the knee of the mass function
at $\ln(1/\sigma) \lsim 0.2$.

\item We have quantified the scale dependence of bias and
stochasticity in the simulation. We find that scale of the
deviation from a linear fit for both bias and stochasticity
reaches a minimum at $z\sim 3-4$. The galaxy and mass density field
can in this sense be said to trace each other most closely at this redshift.

\item The MBII overall matches most observed measurements of galaxy
clustering with stellar mass. There are however some discrepancies
in the bias at the lowest masses and redshifts probed. The simulation
is consistent with observed measurements of stochasticity from
combinations of galaxy clustering and weak lensing.

\item We find that the HOD is well described
by a power law behavior (see equation~\ref{OccFit}). We find a modest
evolution for the power law slope however the normalization mass for
the distribution exhibits an exponentially decaying behavior
with redshift.

\item The location of the peak and the width for the radial
distribution of satellite galaxies decrease with decreasing redshift
irrespective of the host mass of the halo, showing that satellite galaxies
cluster more strongly around the central galaxy with time.

\item We find that the shape of the CSED in MBII is consistent with
observed data. The amplitude is however much lower and can be attributed to
incompleteness at low galaxy masses in MBII.

\item The GSMF predicted by MBII is consistent with observations out to $z=2$.
At lower redshifts however the  MBII GSMF is much steeper at lower masses
whereas MBII over produces stars at larger masses at $z=0$.

\item We find black hole mass and AGN luminosity to be broadly
  correlated to host $M_{\rm halo}$ but with very large scatter,
  indicating a wide range of black hole properties for a given halo
  mass.  However, the local $M_{\rm BH} - M_{*}$ relation is tighter
  and consistent the observed one. Our results also imply very
  moderate evolution for the $M_{\rm BH} - M_{*}$ from $z\sim 2$.  The
  global star formation rate density and black hole accretion rate
  density are also similar, but with a peak for the latter shifted to
  later times.

\item The bolometric luminosity function of the simulated AGN
  population is broadly consistent with observational constraints
  \citep[][see also]{2013ApJ...773...14R}. We note however that simple thermal
  coupling for AGN feedback appears to quench the bright end of the
  AGN LF too fast around $z \sim 2$ whilst not being sufficient to
  fully quench the brightest AGN at $z=0$. The best agreement with
  observations is at $z > 2$ and $z \sim 1$. Our results show a weak
  dependence of AGN clustering with luminosity. (Note that we defer
  to future work we to carry out detailed analysis of AGN population).
\end{itemize}

The MBII simulation is the largest simulation of its kind run to
date with sufficient resolution to resolve $10^9\msun$/h halos.  We
have found that the properties, such as HOD and clustering are
consistent with previous work and observations. However we find that
feedback from AGNs is still not sufficient at lower redshifts to properly
account for the properties of large galaxies and AGNs since it is
unable to quench star formation in massive galaxies and the
abundance of luminous quasars.  However it describes the current
state of standard SPH simulations of galaxy formation and should be
used as a testbed for improving models of galaxy formation.  The
parameters used in MBII were based on much smaller SPH simulations
which missed these large halos in their volumes and thus could not
find discrepancies with observations at the large mass end.

At smaller masses the GSMF in MBII is very steep compared to observational constraints.
A variable wind model based on the galaxy velocity dispersion
\citep{2006MNRAS.373.1265O,2011MNRAS.415...11D, 2013MNRAS.436.3031V, 2014MNRAS.tmp...38T}
has been shown to flatten the GSMF at smaller masses to better reproduce observational constraints at $z=0$.
However at higher redshifts these models still predict a steep slope.
On the other hand we find that AGN feedback is still insufficient to quench
star formation in the most massive galaxies. The effect of AGNs is important
in regulating the growth of massive galaxies;
e.g. the results of \citet{2011MNRAS.415...11D} (which do not include AGN feedback)  and those of
\citet{2013MNRAS.436.3031V, 2014MNRAS.tmp...38T} (which include AGN feedback) clearly illustrate
how AGNs affect the stellar content of massive galaxies. However these simulations are still
small in volume and miss out on massive halos. It would be interesting to see their predictions when run
on boxes similar in size to MBII. In subsequent work we will look at improving
on the feedback models employed in MBII.

One of the important effects we have quantified here is the effect of baryonic
processes on the halo mass function. Although we find that the universality of
the mass function in MBII holds at the 13\% level consistent with previous work,
we find considerable differences in the abundance of halos below
the knee of the mass function when compared to dark matter simulations.
One of the natural questions which arise are: how do different
models for star formation, black hole growth and feedback affect
the mass function? How strong is the redshift dependence?
What can be said of halo abundance matching techniques and their predictions
when done with mass functions such as MBII?
These questions can only be addressed with simulations
with different galaxy formation models and corresponding
dark matter simulations  which we will address in future work.

\section*{Acknowledgments}
The simulations were run on the Cray XT5 supercomputer
Kraken at the National Institute for Computational Sciences. This
research has been funded by the National Science Foundation (NSF)
PetaApps programme, OCI-0749212 and by NSF AST-1009781. NK would
like to aknowledge useful discussions with Michael Boylan-Kolchin and Erin Sheldon.

\label{lastpage}

\end{document}